\documentclass[12pt]{article}
\pdfoutput=1
\usepackage{amssymb, amsmath,amsfonts}
\usepackage{multirow}
\usepackage{mathrsfs}
\usepackage{array}
\usepackage{cite}
\usepackage{booktabs}
\usepackage{tikz}
\usepackage{pgfplots}
\usepackage{float}
\usepackage{subfigure, graphicx}
\usepackage[pdftex, bookmarks=true,colorlinks,linkcolor=red,urlcolor=blue,citecolor=blue]{hyperref}
\usepackage{cite}
\usepackage{slashed}
\usepackage{tabularx}

\makeatletter\@addtoreset{equation}{section}\makeatother

\def\be{\begin{equation}}
\def\ee{\end{equation}}
\def\bea{\begin{eqnarray}}
\def\eea{\end{eqnarray}}

\newcommand{\tabincell}[2]{\begin{tabular}{@{}#1@{}}#2\end{tabular}}

\def\Dslash{\,\,{\raise.15ex\hbox{/}\mkern-12mu D}}
\def\Dbarslash{\,\,{\raise.15ex\hbox{/}\mkern-12mu {\bar D}}}
\def\delslash{\,\,{\raise.15ex\hbox{/}\mkern-9mu \partial}}
\def\delbarslash{\,\,{\raise.15ex\hbox{/}\mkern-9mu {\bar\partial}}}
\def\pslash{\,\,{\raise.15ex\hbox{/}\mkern-9mu p}}
\def\calDslash{\,\,{\raise.15ex\hbox{/}\mkern-12mu {\cal D}}}

\makeatletter\@addtoreset{equation}{section}\makeatother

\renewcommand{\title}[1]{\vbox{\center\LARGE{#1}}\vspace{5mm}}
\renewcommand{\author}[1]{\vbox{\center#1}\vspace{5mm}}
\newcommand{\address}[1]{\vbox{\center\em#1}}

\def\arXiv#1{\href{http://arxiv.org/abs/#1}{arXiv:#1}}
\def\arXiv#1#2{\href{http://arxiv.org/abs/#1}{arXiv:#1}}

\def\doi#1#2{\href{http://doi.org/#1}{#2}}

\begin{document}

\unitlength = .8mm

\begin{titlepage}
\vspace{.5cm}
 
\begin{center}
\hfill \\
\hfill \\
\vskip 1cm

\title{Breakdown of hydrodynamics from holographic \\pole collision
}
\vskip 0.5cm
{Yan Liu}\footnote{Email: {\tt yanliu@buaa.edu.cn}} and
 {Xin-Meng Wu}\footnote{Email: {\tt wu\_xm@buaa.edu.cn}} 

\address{Center for Gravitational Physics, Department of Space Science\\ and International Research Institute
of Multidisciplinary Science, \\Beihang University, Xueyuan Road 37, Beijing 100191, China}

\end{center}
\vskip 1.5cm

\abstract{We study the breakdown of diffusive hydrodynamics in holographic systems dual to neutral dilatonic black holes with extremal near horizon geometries conformal to AdS$_2\,\times\,$R$^2$. We find that at low temperatures by tuning the effective gauge coupling constant in the infra-red, the lowest non-hydrodynamic mode, which collides with the charge diffusive mode and sets the scales at which diffusive hydrodynamics breaks  down, could be either an infra-red mode or a slow mode, resulting in different scaling behaviors of the local equilibrium scales. We confirm that the upper bound for the charge diffusion constant is always satisfied using the velocity and timescale of local equilibration from the pole collision. We also examine the breakdown of hydrodynamics at general temperature and find that the convergence radius has nontrivial dependence on temperature, in addition to the effective gauge coupling constant. 
}
\vfill

\end{titlepage}

\begingroup 
\hypersetup{linkcolor=black}
\tableofcontents
\endgroup


\newpage
\section{Introduction}

Hydrodynamics is a powerful and universal effective theory for a variety of physical systems at large distances and long time, capturing the dynamics of the interacting system towards thermal equilibrium  \cite{Kovtun:2012rj}. The dynamical equations of motion for hydrodynamics are local conserved equations  for the densities of the conserved charges and the corresponding currents. The currents can be expressed in terms of derivative expansions of densities (or the conjugate quantities, e.g. temperature, fluid velocity), namely the constitutive equations.\footnote{Note that we focus on the classical hydrodynamics and neglect the effects of statistical fluctuations of the hydrodynamic system, which is suppressed in the large $N$ limit \cite{Kovtun:2003vj}.} Then the evolution of the system can be solved from the conservation equations and the constitutive equations under proper boundary conditions. 

Recently the convergence of derivative expansions in hydrodynamics has been explored from the perspective of the dispersion relations of hydrodynamic modes \cite{Withers:2018srf, Grozdanov:2019kge, Grozdanov:2019uhi, Heller:2020uuy}.\footnote{The convergence of derivative expansions in hydrodynamics has been studied earlier for boost invariant flow in e.g. \cite{Heller:2013fn, Heller:2015dha}. There are also studies on all order linearized hydrodynamics using  fluid/gravity 
correspondence see e.g. \cite{Bu:2014ena}.}
The hydrodynamic modes are poles of retarded Green's function of densities with gapless dispersion relation satisfying $\mathop{\text{lim}}\limits_{k\rightarrow 0 }\omega(k)=0$ at small frequency and momentum, e.g. shear modes, sound modes and diffusion modes.  
The derivative expansions of the constitutive equations predict that the dispersion relations of hydrodynamic modes are series expansions of $\omega$ in $k$. Therefore the convergence of derivative expansions in constitutive equations could be studied from the convergence of the dispersion relations \cite{Grozdanov:2019kge, Grozdanov:2019uhi, Heller:2020uuy}. It was proposed  in \cite{Withers:2018srf, Grozdanov:2019kge, Grozdanov:2019uhi, Heller:2020uuy} that, by viewing the hydrodynamic mode as complex spectral curve in  $\mathbb{C}^2$ of complexified frequency and momentum, the convergence radii $(k_{eq}, \omega_{eq})$ of the hydrodynamic dispersion series is set by the absolute value of the complex momentum and complex frequency where the hydrodynamic pole collides with the first non-hydrodynamic gapped pole, namely ``pole collision". Along this line, the convergence of hydrodynamic dispersion series has been investigated  using kinetic theory in \cite{Heller:2020hnq}, using field theory in \cite{Choi:2020tdj, Baggioli:2020loj} and using holographic duality in \cite{ Abbasi:2020ykq, Jansen:2020hfd,  Arean:2020eus, Wu:2021mkk, Grozdanov:2021gzh, Jeong:2021zsv, Abbasi:2020xli, upper-new}.

Physically, the breakdown of hydrodynamic dispersion series is due to the presence of non-hydrodynamic gapped degrees of freedom in the system. In general, for strongly interacting quantum field theories, in addition to the hydrodynamic mode a large amount of excitations with shorter lifetime exist 
which are also captured by the poles in the retarded Green's function, i.e. the quasi-normal modes (QNM) in the gauge/gravity duality \cite{book1,book2,book3}. In the hydrodynamic limit, the higher energy excitations decay quickly with time and only hydrodynamic mode remains at late time. 
However, as the absolute value of complexified  momentum increases larger and larger to a special scale $k\sim k_{eq}$, or equivalently the distance becomes shorter and shorter, the lifetime of the hydrodynamic mode and the first non-hydrodynamic mode are of same order $1/\omega_{eq}$. In this scenario, the higher energy excitations cannot be ignored and  hydrodynamics breaks down.  

Certainly the scale at which hydrodynamics breaks down depends on the details of the microscopic dynamics. 
The convergence radii are different for hydrodynamics of field theories with different 't Hooft couplings or gauge couplings \cite{Jansen:2020hfd, Baggioli:2020loj, Choi:2020tdj, Grozdanov:2021gzh}. Meanwhile, the origins of the first non-hydrodynamic modes are different for different systems, for example, it could be a slow mode due to  symmetry breaking \cite{Davison:2018ofp, Davison:2018nxm, Davison:2014lua} or an infra-red (IR) mode for hydrodynamics at low temperature \cite{Arean:2020eus}. In this respect it is extremely interesting to extract the possible universality of the breakdown of hydrodynamics 
in particular solvable hydrodynamic systems which have holographic dual descriptions. Given the fact that the  transport physics in the quantum critical phase is universally governed by the quantum critical groundstate, it is  expected that for the breakdown of the hydrodynamics near the quantum critical groundstate there is perhaps some universality inherited from the critical groundstate.
Recently it was found in \cite{Arean:2020eus} that when the hydrodynamic system is dual to black hole geometry with extremal near horizon geometry of AdS$_2\,\times\,$R$^2$, at low temperature the non-hydrodynamic modes of the diffusive hydrodynamics seem to be universally the poles related to the near horizon geometry, namely IR modes. The generalizations to different hydrodynamic systems with exactly the same IR geometry as \cite{Arean:2020eus} have been made in \cite{Wu:2021mkk,Jeong:2021zsv} and similar universal results on non-hydrodynamic modes were obtained. Therefore, it is natural to ask, at the low temperature, if the non-hydrodynamic modes in systems with a holography dual encoding a quantum critical groundstate are really universally determined by the poles of the emergent IR critical state.  This motivates us to study the breakdown of hydrodynamics at low temperature in different quantum critical phases.

In this work, we study the breakdown of hydrodynamics near the  quantum critical state of a different semi-local quantum liquid from the one dual to AdS$_2\,\times\,$R$^2$. We consider the strongly interacting hydrodynamic systems at zero density that are holographically described by dilatonic black holes with extremal near horizon geometries conformal to AdS$_2\,\times\,$R$^2$, realized in the generalized Gubser-Rocha model with linear axion fields \cite{Gubser:2009qt, Andrade:2013gsa, Zhou:2015qui,Kim:2017dgz}. 
In the case of AdS$_2\,\times\,$R$^2$ quantum critical state  \cite{Arean:2020eus,Wu:2021mkk,Jeong:2021zsv}, there is a nonzero entropy at zero temperature and might be unstable \cite{Iqbal:2011in}, while in our case the holographic system has zero entropy at zero temperature and is expected to be a stable groundstate. We focus on the neutral hydrodynamic systems where the particle-hole symmetry allows the charge and energy to diffuse  separately, and study the breakdown of the charge diffusive hydrodynamics at low temperature near the quantum critical state. To uncover possible universality of the breakdown of hydrodynamics, we  discuss the origin of the first non-hydrodynamic mode by tuning the IR effective gauge coupling constant. In particular, we show that depending on the IR effective gauge coupling constant, the first non-hydrodynamic mode which collides with the hydrodynamic mode could be either an IR mode or a slow mode, resulting in different scaling behaviors of the local equilibrium scales. Here the IR mode is a quasi-normal mode from the near horizon geometry, while the slow mode is a long-lived gapped non-hydrodynamic mode with lifetime much longer than the Planckian time $\tau_{pl}=\hbar/(k_B T)$ and the slow mode is related to the whole geometry from the horizon to the boundary. We will also study the upper bound for the charge diffusion constant \cite{Hartman:2017hhp} and find that it is always satisfied if the velocity and timescale are defined from pole collision which sets the convergent radii of diffusive hydrodynamics, following the proposal in \cite{Arean:2020eus}. We will also comment on the  effects of temperature as well as the IR effective gauge coupling constant in the convergence radii of hydrodynamics. 

This paper is organized as follows.  We first introduce the hydrodynamic system under study and show the IR geometry at low temperature as well as the IR Green's function in Sec. \ref{sec:2}. In Sec. \ref{sec:3} we study the breakdown of the diffusive hydrodynamics from holographic pole collision and discuss related physics, including the properties of non-hydrodynamic modes, convergent radii, diffusion upper bound. Sec. \ref{sec:4} is devoted to conclusions and discussions. Some details on the  equations, calculations and figures during the discussion are collected in appendices.

\section{The generalized Gubser-Rocha model}
\label{sec:2}
We consider the hydrodynamic systems which are dual to dilatonic  black holes with the near horizon geometries describing 
special semi-local quantum liquid states at low temperature.  The gravitational dual theory 
is the generalized Gubser-Rocha model with linear axion fields \cite{Gubser:2009qt, Zhou:2015qui,Kim:2017dgz}, 
\be\label{eq:actionbg}
S=\int d^4x\sqrt{-g}\,\bigg(R-\frac{1}{4} 
e^{\alpha\phi}\, F^2-\frac{3}{2}(\partial\phi)^2+\frac{6}{L^2}\cosh\phi-\frac{1}{2}\sum_{I=1}^2(\partial\psi_I)^2\bigg)\,,
\ee
where $F_{ab}=\partial_a A_b-\partial_bA_a$ is the field strength for a $U(1)$ gauge field $A_a$. The massless scalar fields $\psi_I=m\,x^i\,\delta_{Ii}$ are known as linear axion fields \cite{Andrade:2013gsa} which explicitly break the spatial translational symmetry in the $x$-$y$ plane while preserving the isotropy.\footnote{The translational symmetry breaking effect can also be realised in the framework of massive gravity \cite{Vegh:2013sk}, see e.g.  \cite{Davison:2013txa}.} 
The dilaton field $\phi$ with a particular choice of potential is used to realize a special quantum critical point at low energy. 
Here we choose a specific form of effective gauge coupling $g_\text{eff}^{2}=
e^{-\alpha\phi}$ which in principle could be promoted to an arbitrary function of the dilaton field $Z[\phi]$. As we shall show in the following, the gauge coupling strength near the horizon is crucial for the properties of breakdown of hydrodynamics. 
When $\alpha=1$, it reduces back to the standard Gubser-Rocha model with linear axion terms.

We shall focus on the neutral zero density systems, i.e. we set  $A=0$.  The gauge field could be viewed as a probe in the black hole background. 
We choose the ansatz for the finite temperature background as 
\bea
\begin{split}
&ds^2=-udt^2+\frac{dr^2}{u}+f(dx^2+dy^2)\,,\\
&\phi=\phi(r)\,,~~~\psi_I=m x_I\,,
\end{split}
\label{eq:bg}
\eea
where $\{t, x, y, r\}$ are spacetime coordinates, and $u, f, \phi$ are functions of the radial coordinate $r$, $x_I={x, y}$ for $I=1, 2$, while $m$ characterizes the strength of momentum relaxation. More details about the equations of motion can be found in appendix \ref{appA}.

For the zero density system, there exist two different solutions. 
One solution has 
a nontrivial dilaton which indicates that in the dual field theory a nontrival source for the scalar operator has been turned on. The other solution is the AdS$_4$-Schwarzschild black hole with linear axions where the dilaton field is trivial, i.e. $\phi=0$.
In this paper we shall focus on the dilatonic black hole.\footnote{Note that these two solutions are two different physical situations, i.e. with or without external scalar source, and it is not proper to study the phase transition between them unless one makes the hairy black hole sourceless.}

The neutral dilatonic black hole solution takes the following analytic form  
\bea\label{eq:neubg}
\begin{split}
u&=\sqrt{r}(r-r_0)\left(r-r_0+\sqrt{2}m\right)\frac{1}{\sqrt{r-r_0+\frac{m}{\sqrt{2}}}}\,,\\
f&= \sqrt{r}\left(r-r_0+\frac{m}{\sqrt{2}}\right)^{3/2}\,,\\
\phi&= \frac{1}{2}\log\left(\frac{r-r_0+\frac{m}{\sqrt{2}}}{r}\right)\,,\\
\psi_I&=m x_I\,.
\end{split}
\eea
Note that the sign of the linear axion does not matter. The system only depends on the $m^2$ and  should be symmetric under $m\to -m$. 
In the following we shall focus on $m>0$. 
This hairy black hole has two horizons with the outer horizon located at $r=r_0$. The AdS boundary is located at $r\to\infty$. 

The effective gauge coupling constant takes the form of $g_\text{eff}^2=e^{-\alpha\phi}$ and it is monotonically decreasing from the UV boundary to the IR horizon when $\alpha>0$ which can be understood in terms of charge screening.  We shall focus our discussion mainly in the regime $\alpha\geq 0$ in most sections of this paper and only comment on the case of $\alpha<0$ in subsection \ref{subs:nalp}. 

The temperature and the entropy density of the dilatonic black hole are 
\bea
\begin{split}
T=\frac{\sqrt{m r_0}}{2^{5/4}\pi}\,,~~~~~~
s=2^{-\frac{11}{4}}m\sqrt{m r_0}\,.
\end{split}
\eea
Obviously this system has a vanishing entropy density at zero temperature and at finite temperature $s\sim T$. This fact was crucial in the proposal of linear resistivity for strange metal in \cite{Davison:2013txa}. The linear resistivity has also been discussed in related model in e.g. \cite{Jeong:2018tua}. 

The charge diffusion constant $D_c$ at zero density can be obtained from Einstein relation
\be\label{eq:cd}
D_c=\frac{\sigma}{\chi}\,,
\ee
where $\sigma$ and $\chi$ are the electrical conductivity and susceptibility. The electrical conductivity takes the form 
\bea
\sigma
=\left(\frac{m}{2\sqrt{2}\pi T}\right)^\alpha\,.~~~~~~
\eea
One particularly interesting observation is that when $\alpha=1$ we have linear resistivity which reminds us of the behavior of the strange metal phase in high T$_c$ systems \cite{Davison:2013txa}.
The susceptibility $\chi$ can be obtained from the retarded Green's function $\langle \rho\rho\rangle|_{\omega=k=0}$. 
At zero frequency and zero momentum, the fluctuation of $A_t$ decouples from other fields, and we have 
$
a''_t+\big(\frac{f'}{f}+\alpha\phi'\big)a'_t=0\,.
$ 
In the background \eqref{eq:neubg}, with the condition $a_t(r_0)=0$ at the horizon, $a_t$ can be solved 
$
a_t
=1-\left(\frac{\frac{Q}{r_0}+1}{\frac{Q}{r}+1}\right)^{\frac{1}{2}\left(1+\alpha\right)}
$ for $\alpha\neq -1$.\footnote{When $\alpha=-1$, we have  $a_t=1-\frac{\text{log}\left(\frac{Q}{r}+1\right)}{\text{log}\left(\frac{Q}{r_0}+1\right)}$.} 
When $r\rightarrow \infty$, the asymptotic expansion of $a_t$ is 
$
a_t=a_1-\frac{a_2}{r}+...\,,
$
and the susceptibility can be obtained from $\chi = \frac{a_2}{a_1}$. The charge diffusion constant can be computed from \eqref{eq:cd}. In appendix \ref{app:weakhydro} we use an alternative approach to calculate $D_c$ following  \cite{Chen:2017dsy,Grozdanov:2018fic} and obtain the same results as above.  

In table \ref{table:diffusion} 
we list the electrical conductivity, susceptibility, and the charge diffusion constant as well as its low temperature behavior. We leave the detail discussions on the IR geometry at low temperature and the IR Green's function to subsection \ref{subsec:ir}.
\begin{table}[H]
\centering
\renewcommand{\arraystretch}{1.5}
\begin{tabular}{|c|c|c|c|c|}
\hline
 & $\sigma$ & $\chi$ & $D_c$ & $T\ll m$ \\ [1ex]
\hline
 $
 \alpha (\neq -1)$ & $\left(\frac{m}{2\sqrt{2}\pi T}\right)^\alpha$ & 
 $\frac{\left(1+\alpha\right) m^\alpha \left(m^2-8\pi^2 T^2\right)}
 {2\sqrt{2}m^{1+\alpha}-2^{3+\frac{3\alpha}{2}}(\pi T)^{1+\alpha}}$ & $\frac{\sigma}{\chi}$
 
 & \tabincell{c}{$D_c\rightarrow \frac{2\sqrt{2}}{\left(1+\alpha\right)m}\left(\frac{m}{2\sqrt{2}\pi T}\right)^\alpha$\,, $\alpha>-1$ \\
 $D_c\rightarrow -\frac{8\pi T}{(1+\alpha)m^2}$\,, $\alpha<-1$} \\ [1ex]
 \hline
  $\alpha=-1 $& $\frac{2\sqrt{2}\pi T}{m}$ & $\frac{m^2-8\pi^2 T^2 }{\sqrt{2}m\,\text{log}\left(\frac{m^2}{8\pi^2 T^2}\right)}$ & $\frac{\sigma}{\chi}$ & $D_c\rightarrow \frac{8\pi T}{m^2}\text{log}\left(\frac{m}{2\sqrt{2}\pi T}\right)$ \\ [1ex]
 \hline
\end{tabular}
\caption{\label{table:diffusion} \small
The electrical conductivity $\sigma$,  susceptibility $\chi$ and charge diffusion constant $D_c$  as a function of $\alpha$.}
\end{table}

The charge current is conserved and the charge density diffuses in the late time near equilibrium regime. 
The diffusive constant $D_c$ is expected to be bounded from below due to the chaotic behavior of operators \cite{Hartnoll:2014lpa, Blake:2016sud}, and from above due to causality and unitarity \cite{Hartman:2017hhp}. 
The butterfly velocity $v_B$ and Lyapunov time $\tau_L$ are fundamental quantities to characterize quantum chaos and they have been studied in \cite{Kim:2017dgz} for this model. More explicitly, at zero density the butterfly velocity and Lyapunov time are  
\be
v_B^2=\frac{16\pi^2 T^2}{24\pi^2 T^2+m^2}\,,
~~~~ 
\tau_L=\frac{1}{2\pi T}\,. \ee
Therefore, in the quantum critical region $T/m\rightarrow 0$ where $v_B\sim T/m$, 
\begin{itemize}
\item when $\alpha>-1$, $D_c\gg v_B^2\tau_L$ and satisfies the lower bound proposal. It is crucial to define an upper bound for the charge diffusion constant;

\item when $\alpha<-1$, $D_c\rightarrow -\frac{1}{1+\alpha}v_B^2\tau_L$, i.e. the charge diffusion constant satisfies a bound defined from quantum chaos; 

\item when $\alpha=-1$, the ratio between $D_c$ and $v_B^2\tau_L$ diverges with logarithmic dependence in $m/T$ instead of a power law dependence.   
\end{itemize}

In the following we shall show that if we use the equilibrium velocity and time $v_{eq}$ and $\tau_{eq}$ from local equilibrium scales following the proposal in \cite{Arean:2020eus}, the upper bound $D_c\leq v_{eq}^2\tau_{eq}$ is always satisfied for the charge diffusion constant. 

\subsection{The near extremal geometry and IR Green's function}
\label{subsec:ir}

At zero temperature, the near horizon geometry of \eqref{eq:neubg} takes a simple form which is conformal to AdS$_2\,\times\,$R$^2$. 
More precisely, we have the IR geometry by taking the limit $r\to 0$ of \eqref{eq:neubg} with $r_0=0$,
\bea
\label{eq:nh1}
\begin{split}
ds^2&=-2^{\frac{3}{4}}m^{\frac{1}{2}} r^{\frac{3}{2}}dt^2+\frac{dr^2}{2^{\frac{3}{4}}m^{\frac{1}{2}} r^{\frac{3}{2}}}+ 2^{-\frac{3}{4}}m^{\frac{3}{2}} r^{\frac{1}{2}}\left(dx^2+dy^2\right)\,,\\
e^\phi&=2^{-\frac{1}{4}} m^{\frac{1}{2}} r^{-\frac{1}{2}}\,.
\end{split}
\eea
Through a coordinate transformation 
$
r=\frac{\sqrt{2}}{m \zeta^2}\,
$ we obtain
\bea\label{eq:nhcads2}
ds^2=\frac{2\sqrt{2}}{m\zeta}
\left(
\frac{1}{\zeta^2}\left(-dt^2+d\zeta^2\right)+
\frac{m^2}{4}\left(dx^2+dy^2\right)
\right)\,.
\eea
Under the scaling transformation $(t, ~\zeta, ~x, ~y)\rightarrow (\lambda t,~ \lambda\zeta,~x,~ y)$, the line element $ds^2\rightarrow \lambda^{-1}ds^2$ which means this geometry is conformal to AdS$_2\,\times\,$R$^2$.  
This geometry is known to describe a semi-local quantum liquid state with finite spatial correlation length while infinite correlation time \cite{Hartnoll:2012wm}. 
Note that our study is different from the holographic systems in \cite{Arean:2020eus, Wu:2021mkk,Jeong:2021zsv} where the extremal IR geometry is AdS$_2\,\times\,$R$^2$ which in general gives a nonzero entropy at zero temperature and might suffer from potential instabilities \cite{Iqbal:2011in}. Note that the important near horizon geometry (\ref{eq:nhcads2}) is completely supported by the translational symmetry breaking parameter $m$. When $m=0$, we do not have this type of near horizon geometry at zero density. This is quite similar to the case in \cite{Andrade:2013gsa} where an AdS$_2\,\times\,$R$^2$ near horizon geometry emerges in the extremal black hole with linear axion fields, while with pure AdS vacuum solution without linear axion fields. 

At extremely low temperature $T\ll m$, i.e. $r_0\ll m$, the geometry in the near horizon regime with $r-r_0\ll m$ is
\bea
\label{eq:nh2}
\begin{split}
ds^2 &=-2^{\frac{3}{4}}(mr)^{\frac{1}{2}}(r-r_0)dt^2+\frac{dr^2}{2^{\frac{3}{4}}(mr)^{\frac{1}{2}}(r-r_0)}+ 2^{-\frac{3}{4}}m^{\frac{3}{2}} r^{\frac{1}{2}}\left(dx^2+dy^2\right)\,,\\
e^\phi&=2^{-\frac{1}{4}} m^{\frac{1}{2}} r^{-\frac{1}{2}}\,.
\end{split}
\eea
Taking the limit $r_0\rightarrow 0$, the above geometry reduces to the IR geometry at zero temperature \eqref{eq:nh1}. Note that now the effective gauge coupling near the horizon takes the form of  $g_\text{eff}^2\sim (T/m)^{\alpha}$ at low temperature.

In the near extremal dilatonic black hole, the IR Green's function $\mathcal{G}_{IR}$ can be computed analytically. As we will show in the next section, the poles of this IR Green's function can play important roles in the whole Green's function in the dual field theory. Note that we will study the charge diffusive hydrodynamics, which are encoded in the equations of motion of fluctuations of gauge fields as discussed in detail in appendix 
\ref{app:a2}.
We first focus on the case $k=0$ for simplicity and comment on nonzero $k$ later. We define $\mathfrak{w}=\frac{\omega}{2\pi T}$ in the following.  We will solve the first equation in \eqref{eq:GI} at low temperature with the near horizon geometry (\ref{eq:nh2}).  
We focus on $r-r_0\ll m$ regime, then the EOM is reduced to a simple form\footnote{Note that here we work in the limit $k\ll \omega$ while not exactly $k=0$.}
\bea\label{eq:fluzerofreq}
\mathfrak{a}''+\left(\frac{1}{r-r_0}+\frac{1-\alpha}{2r}\right)\mathfrak{a}'
+\frac{r_0}{4r}\frac{\mathfrak{w}^2}{\left(r-r_0\right)^2}\mathfrak{a}=0
\eea
where $\mathfrak{a}$ is the gauge invariant quantity of the fluctuations of gauge fields in the diffusive channel as discussed in appendix \ref{app:a2} and also in the beginning of next section. 

The analytic solution is a linear combination of two independent hypergeometric functions
\bea
\begin{split}
\mathfrak{a}=&
\left(r-r_0\right)^{-\frac{i\mathfrak{w}}{2}}
\bigg[c_1~_{2}F_1\left(
\frac{1-\alpha-i\mathfrak{w}}{2},-\frac{i\mathfrak{w}}{2},\frac{1-\alpha}{2},
\frac{r}{r_0}\right)
+\\~~&~~~~~~~~~~~~~~
+c_2\left(\frac{r}{r_0}\right)^{\frac{1+\alpha}{2}}
~_{2}F_1\left(1-\frac{i\mathfrak{w}}{2},\frac{1+\alpha-i\mathfrak{w}}{2},\frac{3+\alpha}{2},\frac{r}{r_0}\right)\bigg]\,.
\end{split}
\eea
Near the horizon $r=r_0$ of (\ref{eq:nh2}), this solution gives rise to the infalling and outgoing solutions with the exponents $(r-r_0)^{-\frac{i\mathfrak{w}}{2}}$ and $(r-r_0)^{+\frac{i\mathfrak{w}}{2}}$, respectively. 
The infalling boundary condition constrains that
\bea
\frac{c_2}{c_1}=
\frac{i\mathfrak{w}}{2}
\frac{
\Gamma\left(\frac{1-\alpha}{2}\right)
\Gamma\left(\frac{1+\alpha-i\mathfrak{w}}{2}\right)
}
{
\Gamma\left(\frac{3+\alpha}{2}\right)
\Gamma\left(\frac{1-\alpha-i\mathfrak{w}}{2}\right)
}
\,.
\eea

Close to the outer boundary of IR geometry (\ref{eq:nh2}), i.e. $\frac{r}{r_0}\rightarrow \infty$, 
\bea
\mathfrak{a}(r)=\mathcal{A}(\omega, T)+\mathcal{B}(\omega, T)r^{\frac{-1+\alpha}{2}}\,.
\eea
When $\alpha<1$, the IR Green's function $\mathcal{G}_{IR}$ is
\bea\label{eq:ir1}
\begin{split}
\mathcal{G}_{IR}(\omega, T)
\propto
\frac{\mathcal{B}(\omega, T)}{\mathcal{A}(\omega, T)}
\propto-i \frac{\mathfrak{w}}{2}
\frac{
\text{Cos}\left(\frac{\pi}{2}\left(\alpha-i\mathfrak{w}\right)\right)
\Gamma\left(\frac{\alpha-1}{2}\right)
\Gamma\left(\frac{1-\alpha+i\mathfrak{w}}{2}\right)
}
{\text{Cos}\left(\frac{\pi}{2}\left(\alpha+i\mathfrak{w}\right)\right)
\Gamma\left(\frac{1-\alpha}{2}\right)
\Gamma\left(\frac{1+\alpha+i\mathfrak{w}}{2}\right)}\Big(\frac{T}{\sqrt{m}}\Big)^{1-\alpha}
\,.
\end{split}
\eea
When $\alpha>1$, the IR Green's function is
\bea\label{eq:ir2}
\begin{split}
\mathcal{G}_{IR}(\omega, T)
\propto
\frac{2i}{\mathfrak{w}} 
\frac
{\text{Cos}\left(\frac{\pi}{2}\left(\alpha+i\mathfrak{w}\right)\right)
\Gamma\left(\frac{1-\alpha}{2}\right)
\Gamma\left(\frac{1+\alpha+\mathfrak{w}}{2}\right)}
{\text{Cos}\left(\frac{\pi}{2}\left(\alpha-i\mathfrak{w}\right)\right)
\Gamma\left(\frac{\alpha-1}{2}\right)
\Gamma\left(\frac{1-\alpha+i\mathfrak{w}}{2}\right)}\Big(\frac{T}{\sqrt{m}}\Big)^{\alpha-1}
\,.
\end{split}
\eea
When $\alpha=1$, close to the boundary of IR geometry, we have 
\bea 
\mathfrak{a}(r)=\mathcal{S}(\omega, T)+\mathcal{R}(\omega, T)\log r\,.
\eea
In this case there is a logarithmic anomaly for the source term. A proper way to get the Green's function is to consider a double trace deformation (e.g. \cite{Faulkner:2012gt}) and then the Green's function depends on the Landau pole of the theory. We will not discuss the poles for this case. As we shall show in the next subsection, close to $\alpha=1$ the poles from UV Green's function slightly mismatch the IR results $\alpha\to 1$ which should be related to this anomaly.  

From the above formula \eqref{eq:ir1} and \eqref{eq:ir2}, we find the poles of the IR Green's function: 
\begin{itemize}

\item When $\alpha<1$, the IR poles are located at $\frac{i\omega}{2\pi T}=2n-1-\alpha$ with positive integer $n= 1, 2, \dots$. 

\item When $\alpha>1$, the IR poles are located at $\frac{i\omega}{2\pi T}=2n-1+\alpha$ with positive integer $n=1, 2, \dots$.

\end{itemize}

Finally we make some comments on the finite $k$ effect on the IR Green's function. There is no general analytical solution from  \eqref{eq:GI} at finite $k$ in the IR regime. Nevertheless, one can get useful information about the location of QNMs when $k\sim \omega$. Note that to solve \eqref{eq:GI}, we should first take the near extremal limit $r_0\ll m$ as well as near horizon $r-r_0\ll m$, then take the limit of the outer boundary of IR via $r/r_0\to\infty$. In the limit $r-r_0\ll m$ we have $u\ll f$ in the IR.  When $k\sim\omega\ll m$, we have $k^2 u\ll \omega^2 f$. At small momentum $k\sim \omega$, the equations \eqref{eq:fluzerofreq} get corrected at order $k^2/m^2$. Therefore it is expected that the location of quasi-normal modes are of almost the same value as the zero frequency result, which is similar to the AdS$_2$ results \cite{Arean:2020eus, Wu:2021mkk, Jeong:2021zsv}. Here we will not consider the case when $k$ becomes even larger for simplicity. We shall call the poles at small frequency $k\leq \omega$ of the above behavior as IR modes. In the next section we will show the comparison between the IR poles discussed above and the corresponding QNMs of UV pole computed from numerics.

\section{Breakdown of hydrodynamics from pole collision}
\label{sec:3}

In this section, we will study the breakdown of hydrodynamic system in the previous section from pole collision in the charge diffusive sector and also examine the related upper bound of the charge diffusion constant $D_c$. We focus on the neutral hydrodynamic system near semi-local quantum liquid state which is dual to the geometry that is conformal to AdS$_2\,\times\,$R$^2$. We also comment on the high temperature effects on convergence radii and diffusion upper bound. 
 
We consider the fluctuations for gauge fields $\delta A_t,~\delta A_x,~\delta A_r$ which decouple from other matter field fluctuations. In momentum space 
$\delta A_{\mu}=a_{\mu}(r)e^{-i\omega t+ik x}\,
$, the dynamical equations are given by \eqref{eq:eoma} in appendix \ref{app:a2}. Using the gauge invariant quantity  $\mathfrak{a}\equiv a_t+\frac{\omega}{k}a_x$, we have the EOM for $\mathfrak{a}$ as
\bea\label{eq:flugia}
\mathfrak{a}''+
\left(\frac{\omega^2f'}{-\omega^2f+k^2u}-\frac{\omega^2fu'}{u(-\omega^2f+k^2u)}+\frac{f'}{f}+\alpha\phi'\right)\mathfrak{a}'+
\left(\frac{\omega^2}{u^2}-\frac{k^2}{uf}\right)\mathfrak{a}=0\,. 
\eea
With the infalling boundary condition near the horizon, we obtain the retarded Green's function. The quasi-normal modes of the dual system \cite{Kovtun:2005ev,Kaminski:2009dh} can be obtained from the Green's function, i.e. the sourceless condition at the boundary. In the remaining part, we will show our numerical results on pole collision and discuss related physics with different IR gauge coupling constant. 

\subsection{Pole collision with a slow mode}
\label{sec:case1}

In systems with a long-lived non-hydrodynamic slow mode, the equilibrium time is considerably longer than Plankian time, i.e. $\tau_{eq}\gg T^{-1}$. This non-hydrodynamic mode slows down the hydrodynamic system's return to equilibrium. The low frequency dependence of the transport typically has a Drude behavior. The slow mode has appeared in previous models, such as systems with momentum conservation weakly broken \cite{Davison:2014lua}, 
holographic quantum critical points with irrelevant deformations \cite{Davison:2018ofp,Davison:2018nxm}, holographic probe branes  \cite{Chen:2017dsy}, and so on. 

In this subsection, we will show that the slow mode also shows up in our model at low temperature when $\alpha>1$. There is a pole collision between the diffusive hydrodynamic mode and the gapped slow mode, after which they translate into two sound-like modes, namely ``diffusion-sound crossover" \cite{Davison:2014lua}. Note that the sound-like modes only exist starting from a finite $k_{eq}$ and they are not exactly hydrodynamic modes. 

In the following, we consider $\alpha=2$ as an example. In this case the effective gauge coupling constant in the IR is $g_\text{eff} \sim T/m$ at low temperature. The electrical conductivity, susceptibility and charge diffusion constant are shown in table \ref{table:diffusion2}. It is interesting to note that we have $\sigma\sim (m/T)^2$ at low temperature which reminds us the resistivity in a Fermi liquid. We have shown in section \ref{sec:2} that at low temperature $D_c\gg v_B^2\tau_L$, with $v_B^2\to \frac{16\pi^2 T^2}{m^2}, \tau_L=\frac{1}{2\pi T}$. We will discuss the upper bound on charge diffusion using the local equilibration scales from pole collision. 
\begin{table}[H]
\centering
\renewcommand{\arraystretch}{1.5}
\begin{tabular}{|c|c|c|c|c|}
\hline
 & $\sigma$ & $\chi$ & $D_c$ & $T\ll m$ \\ [1ex]
 \hline
$\alpha=2$ & ~~$\frac{m^2}{8\pi^2 T^2}$~~ & ~~
$\frac{3m^2(m^2-8\pi^2 T^2)}{2\sqrt{2}m^3-64\pi^3 T^3}$ ~~
& ~~$\frac{\sqrt{2}m^3-32\pi^3 T^3}{12m^2\pi^2T^2-96\pi^4 T^4}$ ~~
&~~ $D_c\rightarrow \frac{m}{6\sqrt{2}\pi^2 T^2}$ ~~\\ [1ex]
\hline
\end{tabular}
\caption{\label{table:diffusion2} 
\small The electrical conductivity $\sigma$,  susceptibility $\chi$ and charge diffusion constant $D_c$ for $\alpha=2$.}
\end{table}

\begin{figure}[h!]
  \centering
\includegraphics[width=0.45\textwidth]{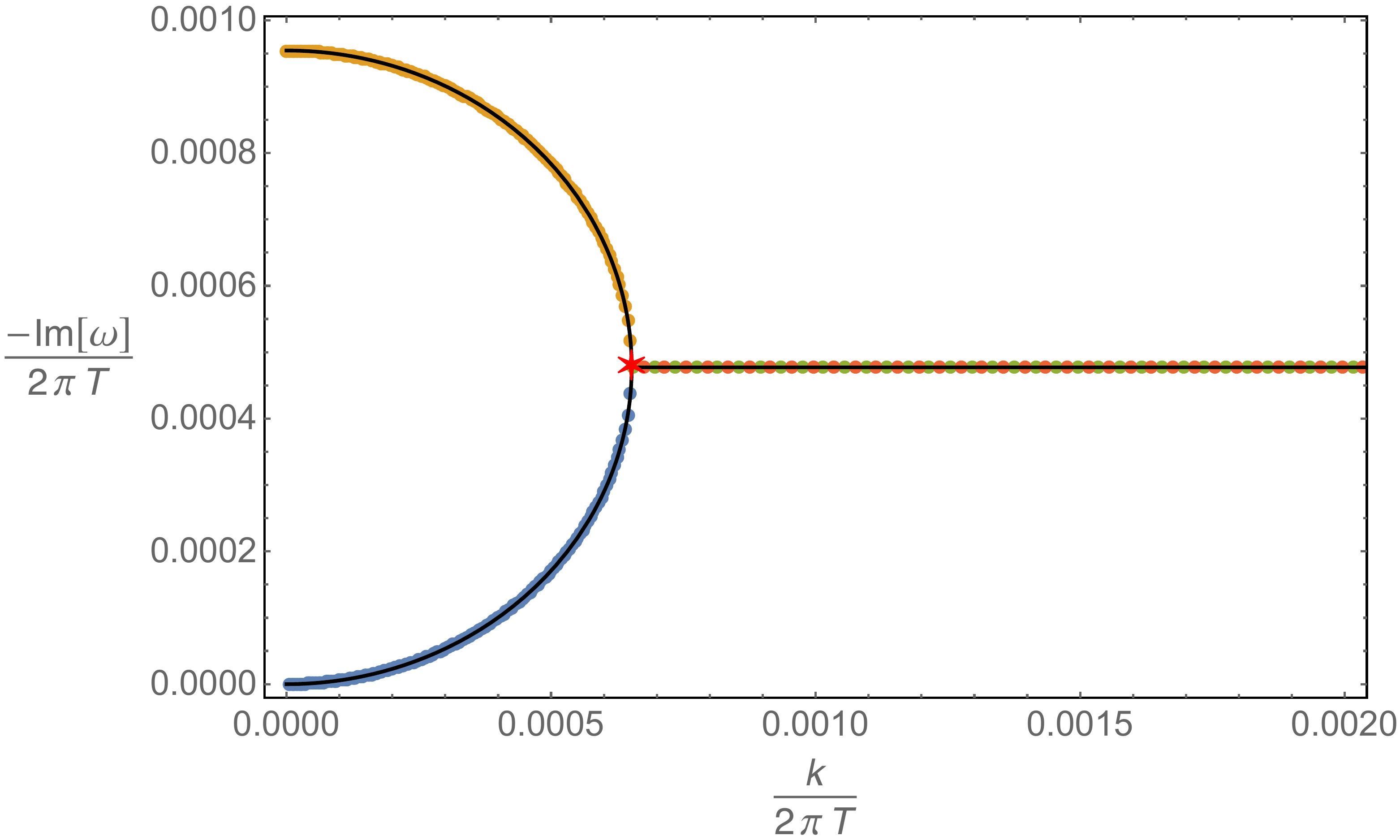}
\hspace{10mm}
\includegraphics[width=0.44\textwidth]{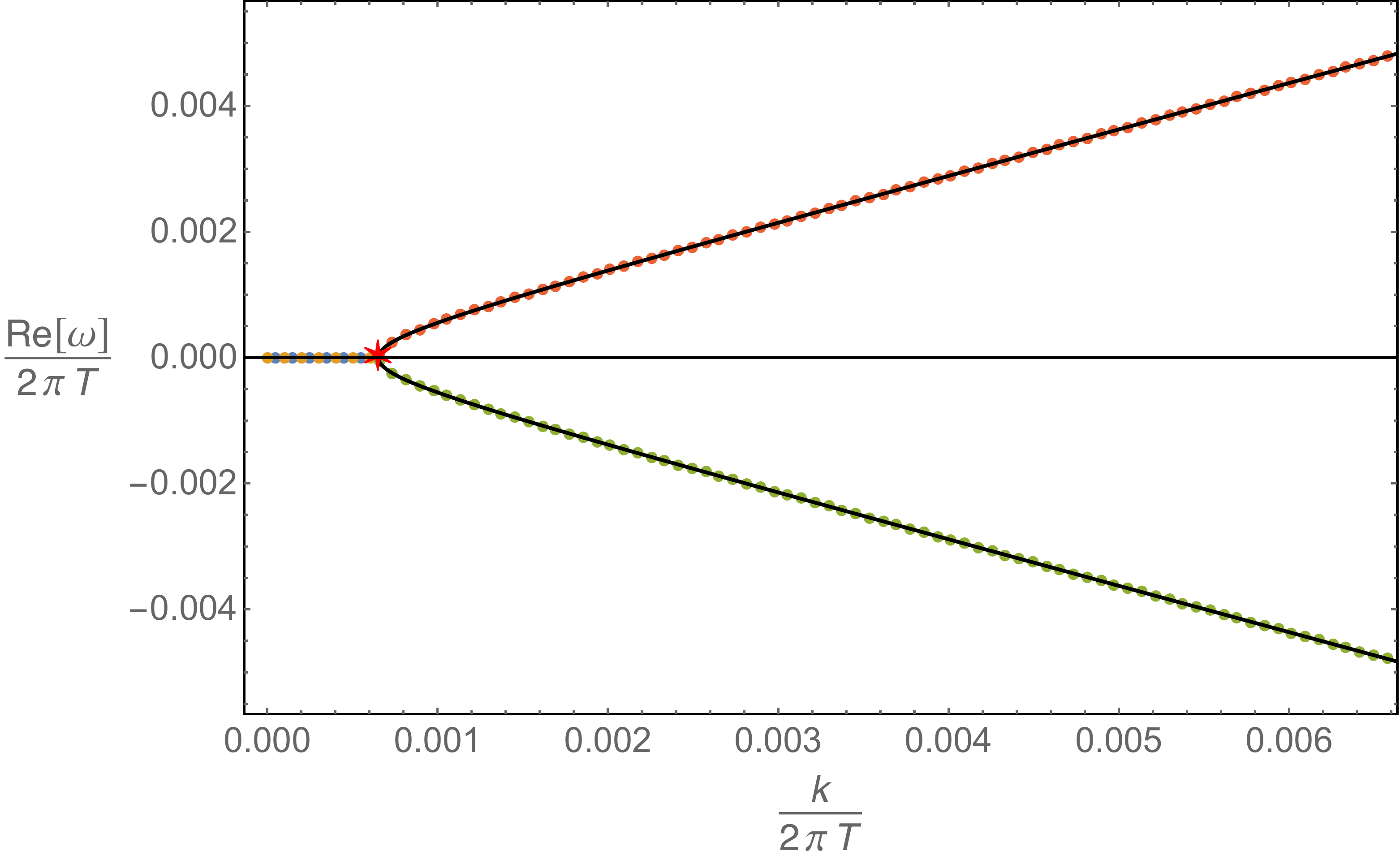}
  \caption{\small Pole collision between the hydrodynamic mode (blue dots) and the non-hydrodynamic slow mode (orange dots) for $\alpha=2$ and $T/m \simeq 1.34\times10^{-4}$. 
  They collide at $(k_{eq}, -i\omega_{eq})$ (red star) and translate into two sound-like modes with opposite real parts, in green and red dots. 
  The black lines are analytic dispersion relations obtained from telegrapher equation. 
  }
  \label{fig:PCalpha2}
\end{figure}

Both the hydrodynamic and the first non-hydrodynamic modes are  quasi-normal modes of the system, which 
are shown in Fig. \ref{fig:PCalpha2}, with $\alpha=2$ and $T/m\simeq 1.34\times10^{-4}$. The left and right plots are for the imaginary and real parts of frequencies of  the lowest two quasi-normal modes as a function of real $k$. The two modes collide at a real $k$ and a pure imaginary $\omega$. 
For small momentum, these two quasi-normal modes are a hydrodynamic mode and the first non-hydrodynamic mode (i.e. a slow mode) respectively. 
Moreover, the hydrodynamic mode is diffusive, i.e.  $\omega=-iD_c k^2$ while the non-hydrodynamic mode behaves as
$\omega=-i\Gamma+ i D_c k^2$. 
In the full finite $k\leq k_{eq}$ regime, 
both the hydrodynamic mode and the first non-hydrodynamic modes are pure imaginary. These two modes 
obey the ``semicircle law" and display pole collision at $(k_{eq}, -i \omega_{eq})$ where they merge into two sound-like  modes with opposite real parts. 
In the large $k\gg k_{eq}$ limit, the real parts are linear in $k$, i.e. $\text{Re}(\omega)=\pm v_s k$ where $v_s$ is the velocity of the sound-like waves.  

\vspace{.3cm}

For $\alpha=2$, when $\omega\ll T$, the quasi-normal modes can be derived analytically from the matching method as discussed in appendix \ref{app:weakhydro}. It turns out that the dynamics of both the hydrodynamic mode and the first non-hydrodynamic slow mode are governed by a simple telegrapher equation
\bea\label{eq:teeq}
\omega^2+\frac{i}{\tau}\omega-\frac{D_c}{\tau}k^2=0\,,
\eea
from which we can obtain the dispersion relations
\bea
\label{eq:dispersion}
\omega_{\pm}=-\frac{i}{2\tau}\left(1\pm \sqrt{1-4D_c\tau k^2}\right)\,,
\eea
where $D_c$ and $\tau$ are defined in \eqref{eq:Dtau}. 
The dispersion relations \eqref{eq:dispersion} are shown in black lines in Fig. \ref{fig:PCalpha2}, from which we find that the telegrapher equation fits the first two quasi-normal modes perfectly well when $|\omega|\ll T$. We can get a lot of information from the telegrapher equation:\footnote{Here we focus on the real momentum behavior. In appendix \ref{app:pccomplex} the quasi-normal modes with respect to complex momentum near equilibrium momentum are discussed.}
\begin{itemize}
\item The pole collision between the hydrodynamic diffusion mode and the slow mode occurs at $k_c=\frac{1}{\sqrt{4D_c\tau}}\,,~\omega_c= -\frac{i}{2\tau}$. Their absolute values define equilibrium momentum and equilibrium frequency as 
\bea
\label{eq:collisionqh}
(k_{eq}\,,~ \omega_{eq})=\left(\frac{1}{\sqrt{4D_c\tau}}\,,~ \frac{1}{2\tau}\right)\,,
\eea
which is labeled as a red star in Fig. \ref{fig:PCalpha2}.
\item 
When $k\leq k_{eq}$, $\omega_{\pm}$ are imaginary and represented by the black curves in the left subdiagram. 
\item
When $k>k_{eq}$, $\text{Im}(\omega_{\pm})=-\frac{i}{2\tau}$ remain a constant and $\text{Re}(\omega_{\pm})\to \pm v_sk$ with $v_s=\sqrt{\frac{D_c}{\tau}}$ when $D_c\tau k^2\gg 1$, indicating that these are two sound-like modes. 
\end{itemize}
The convergence radii of hydrodynamic expansions are set by  
the scales defined from the pole collision point, i.e. $(k_{eq},  \omega_{eq})=\left(\frac{1}{\sqrt{4D_c\tau}}\,,~
\frac{1}{2\tau}\right)$, from which we have the equilibrium time and the equilibrium velocity \cite{Arean:2020eus}
\bea\label{eq:tauveq}
\begin{split}
\tau_{eq}=\frac{1}{\omega_{eq}}\,,~~~v_{eq}=\frac{\omega_{eq}}{k_{eq}}\,.
\end{split}
\eea
It is interesting to note that we have $v_{eq}=v_s$. At low temperature, for the case $\alpha=2$ that we considered,
\be
v_{eq}^2\sim 1\,,~~~\tau_{eq}\sim \frac{m}{T^2}\,.
\ee
We see that the equilibration velocity $v_{eq}$ is a constant and is much larger than the butterfly velocity $v_B$. Meanwhile, the equilibration time is much longer than the Planckian time \cite{jan-plankian,Hartnoll:2021ydi}  or the Lyapunov time of the system. 

For the charge diffusion constant at low temperature, on the one hand, it is bounded from below $D_c\gg v_B^2\tau_L$, on the other hand, from \eqref{eq:collisionqh} and \eqref{eq:tauveq} we see that the diffusion constant is bounded from above as 
\bea
D_c\lesssim \frac{1}{2}\,v_{eq}^2\,\tau_{eq}\,.
\eea
The symbol ``$\simeq$" here indicates that $D_c$ saturates this bound in the quantum critical region. This is a typical feature of diffusion upper bound with a slow mode as the first non-hydrodynamic mode in the system. We will discuss the related universality in subsection \ref{sec:general}.

\subsection{Pole collision with an IR mode}
\label{sec:case2}
In the previous subsection, we have shown that the breakdown of the hydrodynamics is due to the presence of a slow mode which collides with the hydrodynamic diffusion mode. By tuning the bulk gauge coupling constant in the IR, i.e. the parameter $\alpha$, we show there are situations that the first non-hydrodynamic mode is an IR pole in $\mathcal{G}_{IR}$ of the strongly coupled semi-local quantum liquid which is described by the geometry conformal to AdS$_2\,\times\,$R$^2$ . For the case that the quantum critical states are described by the IR geometry of AdS$_2\,\times\,$R$^2$, the first non-hydrodynamic mode has been shown to be always an IR pole. Here we observe that this is true only for a special regime of the IR effective gauge coupling. 

\begin{figure}[H]
  \centering
\subfigure
{
\begin{minipage}{0.48\linewidth}
\includegraphics[width=1\linewidth]{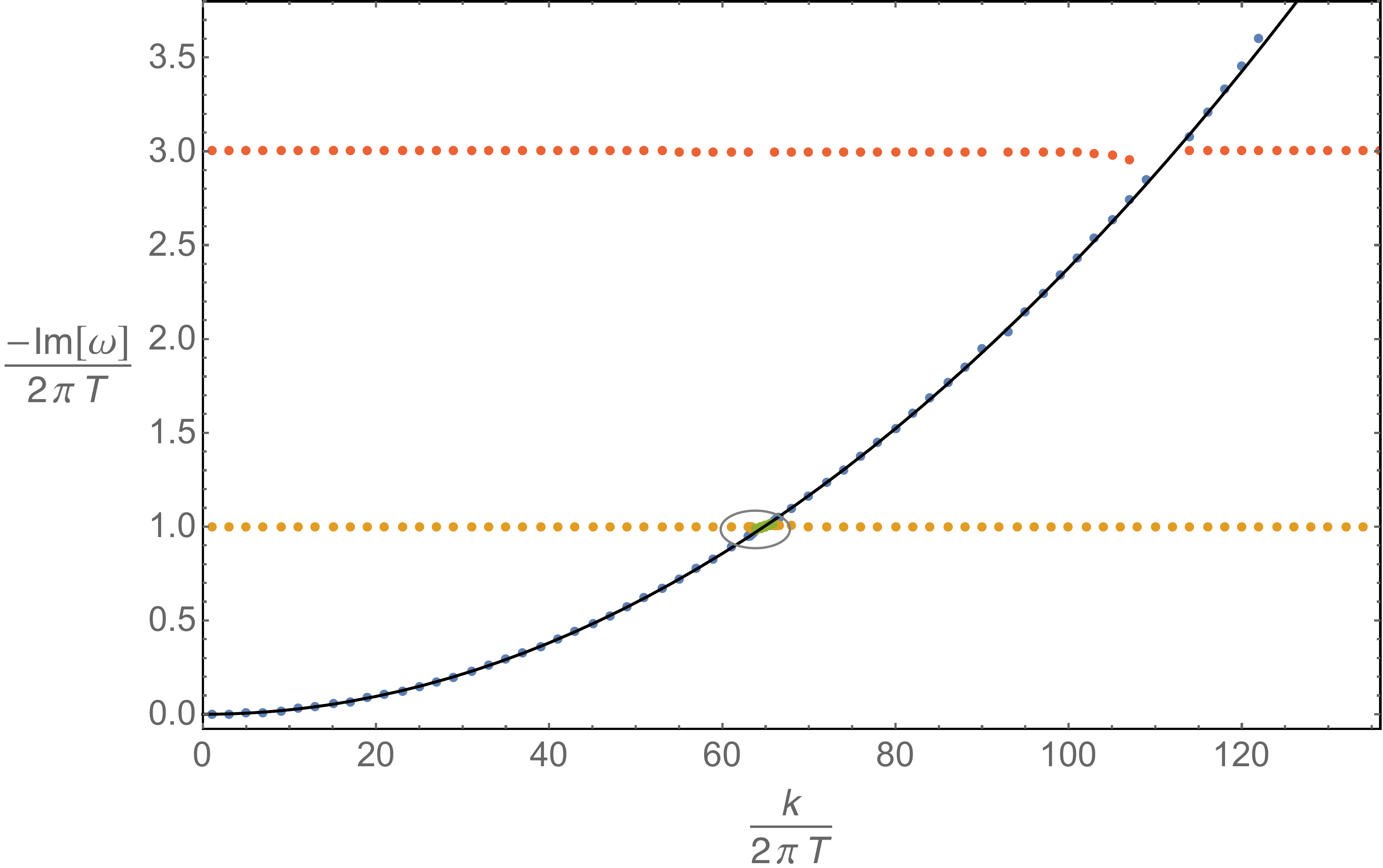}
\end{minipage}
}
\hspace{2mm}
\subfigure{
\begin{minipage}{0.45\linewidth}
\includegraphics[width=0.9\linewidth]{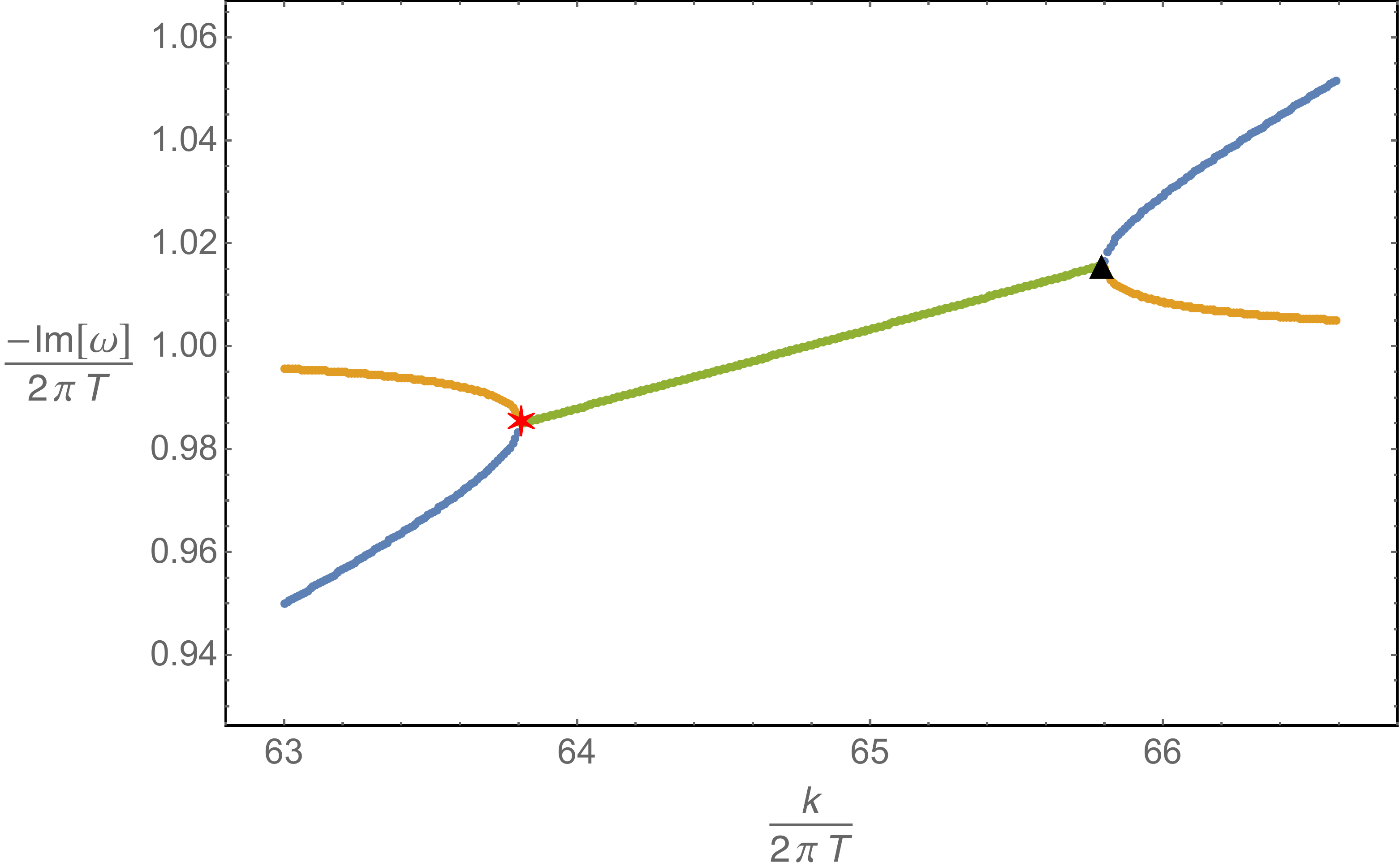}
\vspace{2mm}
\\
\includegraphics[width=0.9\linewidth]{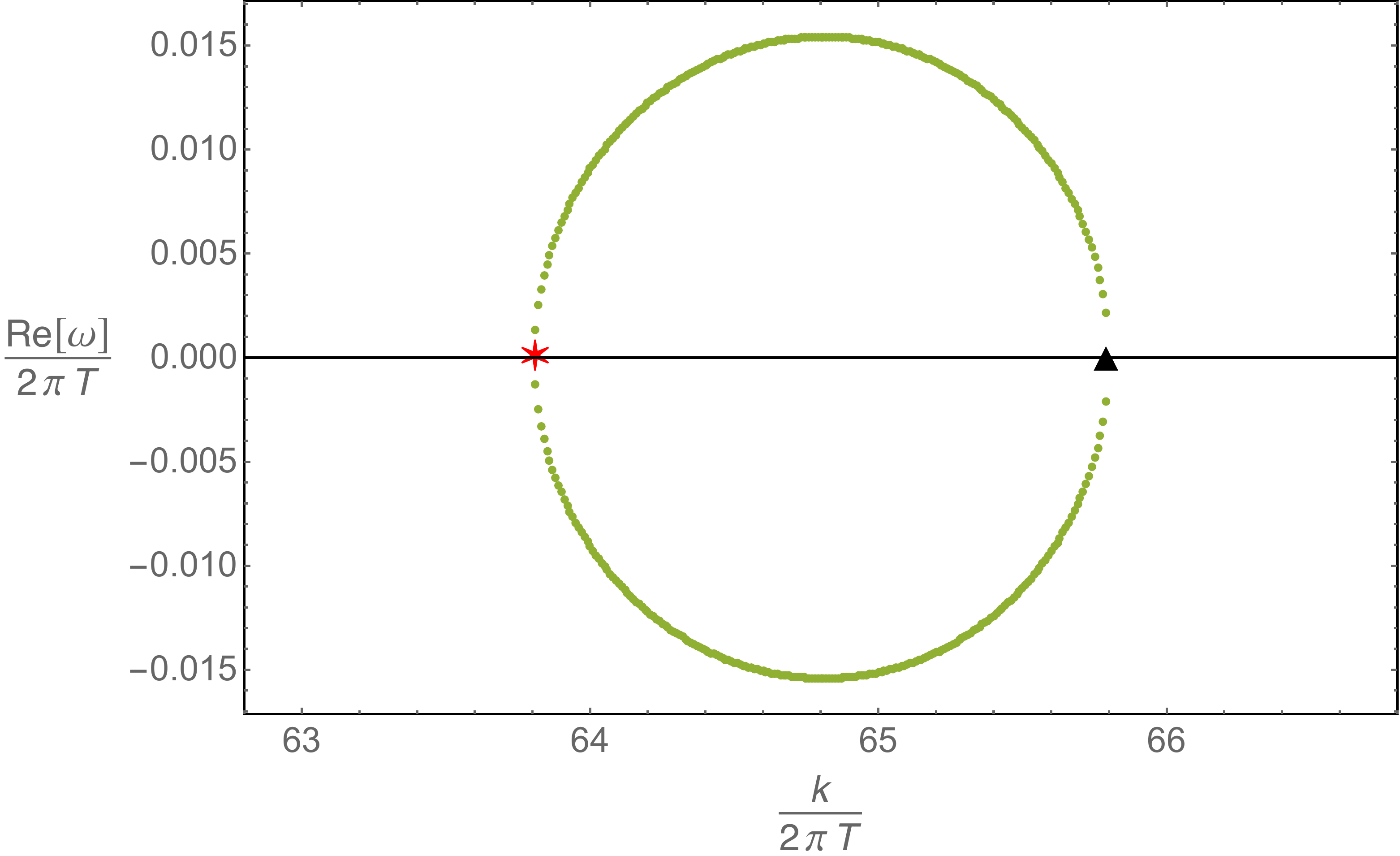}
\end{minipage}
}
  \caption{\small Pole collision between the hydrodynamic  diffusive mode and the first non-hydrodynamic mode (an IR mode) when $T/m\simeq 1.34\times 10^{-5}$ and $\alpha=0$. The right plots are the real and imaginary parts of frequencies of these modes as a function of $k$ close to the collision points.}
  \label{fig:diff-IR}
\end{figure}

In this subsection we focus on the case $\alpha=0$ in a parallel description with subsection \ref{sec:case1}. Now the effective gauge coupling in the IR is $g_\text{eff}\sim 1$. The electrical conductivity $\sigma$,  susceptibility $\chi$ and charge diffusion constant $D_c$ are shown in table \ref{table:diffusion3}. We also have $D_c\gg v_B^2\tau_L$ at low temperature in this case. 
\begin{table}[H]
\centering
\renewcommand{\arraystretch}{1.5}
\begin{tabular}{|c|c|c|c|c|}
\hline
 & $\sigma$ & $\chi$ & $D_c$ & $T\ll m$ \\ [1ex]
 \hline
$\alpha=0$ & ~~~~$1$~~~~& ~~$\frac{m}{2\sqrt{2}}+\pi T$~~&~~ $(\frac{m}{2\sqrt{2}}+\pi T)^{-1}$ ~~&~~ $D_c\rightarrow \frac{2\sqrt{2}}{m}$~~ \\ [1ex]
\hline
\end{tabular}
\caption{\label{table:diffusion3}\small The electrical conductivity $\sigma$,  susceptibility $\chi$ and charge diffusion constant $D_c$ for $\alpha=0$.}
\end{table}
The quasi-normal modes are shown in Fig. \ref{fig:diff-IR}.  The hydrodynamic diffusive mode (in blue dots) exist for a large regime of momentum. It obeys the dispersion relation (in black line) $\omega=-iD_c k^2$ with \bea
D_c\simeq \frac{2\sqrt{2}}{m}\,.
\eea
The non-hydrodynamic modes are a tower of pure imaginary IR modes located at $\text{Im}[\omega]=-n\cdot 2\pi T$, with $n=1,3,5...$. 
As shown in section \ref{subsec:ir}, the poles of IR Green's function is almost independent of $k$ 
for small momentum $k\leq \omega$. Therefore, these non-hydrodynamic modes of the dual system have a clear origin from 
the semi-local quantum liquid, due to the deep IR geometry that is conformal to AdS$_2\,\times\,$R$^2$.  This statement can be generalized to a series of $\alpha$ as illustrated in Fig. \ref{fig:IRmode} in the next subsection,
which show that the non-hydrodynamic excitations come from the dynamics of IR quantum critical physics. 

As shown in Fig. \ref{fig:diff-IR}, the hydrodynamic diffusive mode collides with the first non-hydrodynamic mode with an IR origin at $(k_{eq}, -i\omega_{eq})$, which is indicated by a red star as we zoom in the region bounded by a gray circle. The collision occurs at real $k$ where the diffusive mode and first IR mode merge into two complex modes with the same imaginary part and opposite real parts (Fig. \ref{fig:diff-IR}).  
They split back into the diffusive mode and the IR mode. These behaviors are different from the observations in \cite{Arean:2020eus, Wu:2021mkk,Jeong:2021zsv} where the collision occurs at complex momentum and frequency. The behaviors of quasi-normal modes for the complex momentum close to the equilibrium momentum are shown in appendix \ref{app:pccomplex}. 

\begin{figure}[H]
  \centering
\includegraphics[width=0.5\textwidth]{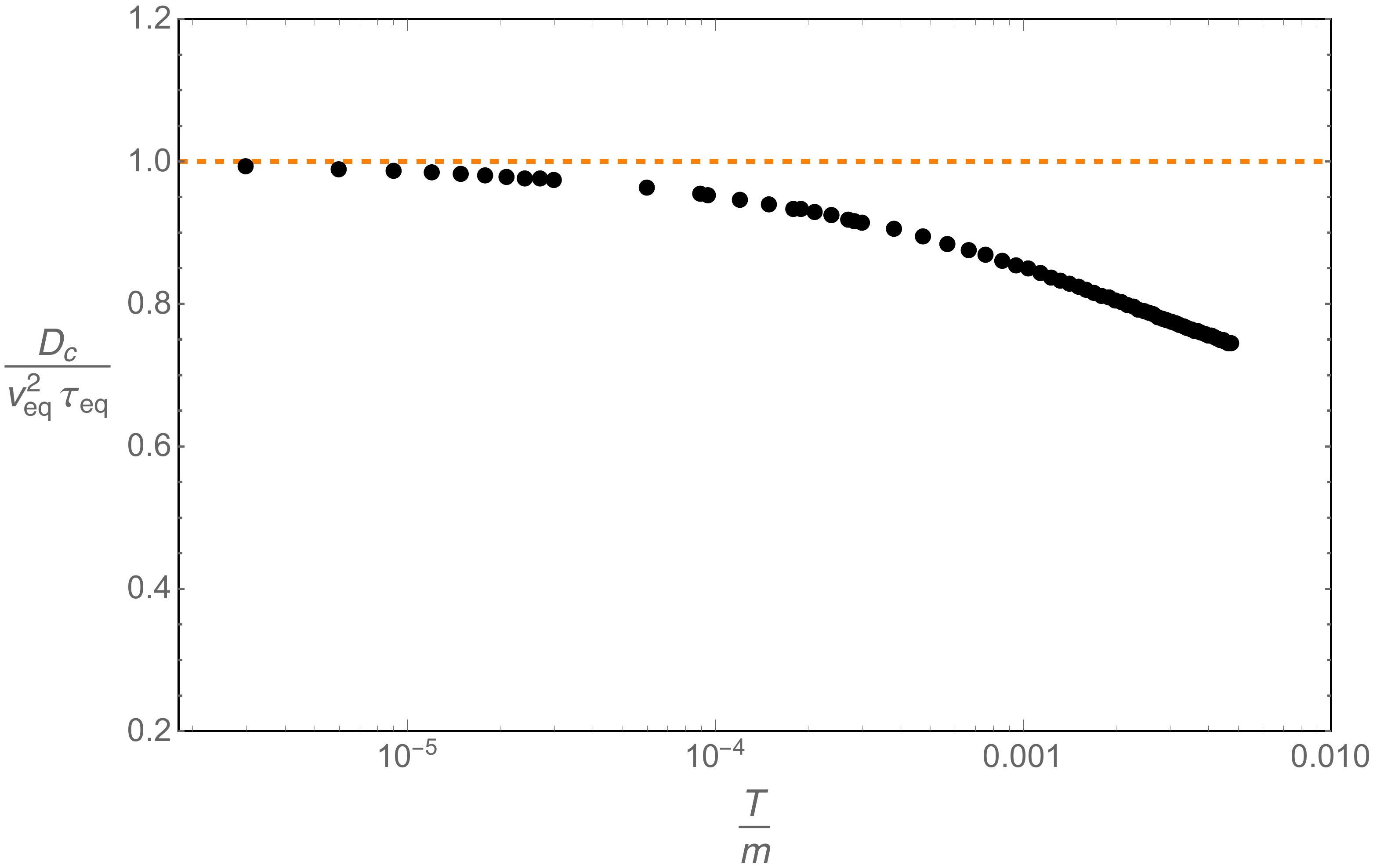}
  \caption{\small The charge diffusion constant $D_c$ is bounded from above by the equilibrium time and equilibrium velocity defined from pole collision, as $T/m \rightarrow 0$. }
  \label{fig:upbound}
\end{figure}

The most interesting character is displayed in the scaling behaviors of the equilibration scales.  
As $T/m \rightarrow 0$, we have numerically confirmed that
\be \omega_{eq}\sim T\,,~~~~k_{eq}^2\sim mT\ee 
and therefore 
\be v_{eq}^2\sim \frac{T}{m}\,,~~~~~\tau_{eq}\sim \frac{1}{T}\,.\ee
One can immediately see that the equilibration velocity is much larger than the butterfly velocity, while the equilibrium time is of the same order as the Lyapunov time or Planckian time.
As a result, we have $D_c \sim v_{eq}^2\tau_{eq}$. 
We numerically compute the ratio $D_c/(v_{eq}^2\tau_{eq})$ as a function of $T/m$, as shown in Fig. \ref{fig:upbound}. It is apparent that $D_c\lesssim v_{eq}^2 \tau_{eq}$ always hold and 
\be D_c\rightarrow v_{eq}^2 \tau_{eq}~~~~\text{as}~ T/m\rightarrow 0\,.\ee 

Similar bounds have been observed in heat diffusion and crystal diffusion \cite{Arean:2020eus, Wu:2021mkk,Jeong:2021zsv} with the IR geometry of AdS$_2\,\times\,$R$^2$ . Here we emphasize that, in all these examples the diffusion constants are independent of $T$ when such pole collision occurs. 

\subsection{More on pole collision for general $\alpha$ at low temperature}
\label{sec:general}

We have shown two kinds of pole collisions in the previous two subsections for two different choices of $\alpha$ in \eqref{eq:actionbg}.\footnote{We focus on the cases $\alpha\geq 0$ in this subsection and comment on the case $\alpha<0$ in subsection \ref{subs:nalp}.} The hydrodynamic diffusive mode collides with a slow mode with a long lifetime ($\tau T\gg 1$) when $\alpha=2$ (or $g_\text{eff}\sim T/m$) where the dispersion relation obeys the semi-circle law, while collides with an IR mode ($\alpha=0$, or  $g_\text{eff}\sim 1$) which is the lowest pole of the retarded Green's function of the conformal to  AdS$_2\,\times\,$R$^2$ geometry. 
These two different kinds of pole collisions indicate two different origins of breakdown of hydrodynamics at low temperature. In this subsection we study the pole collision via tuning the IR effective gauge coupling constant (i.e. varying $\alpha$) and discuss the related diffusion upper bound. 

\begin{figure}[H]
  \centering
\includegraphics[width=0.45\textwidth]{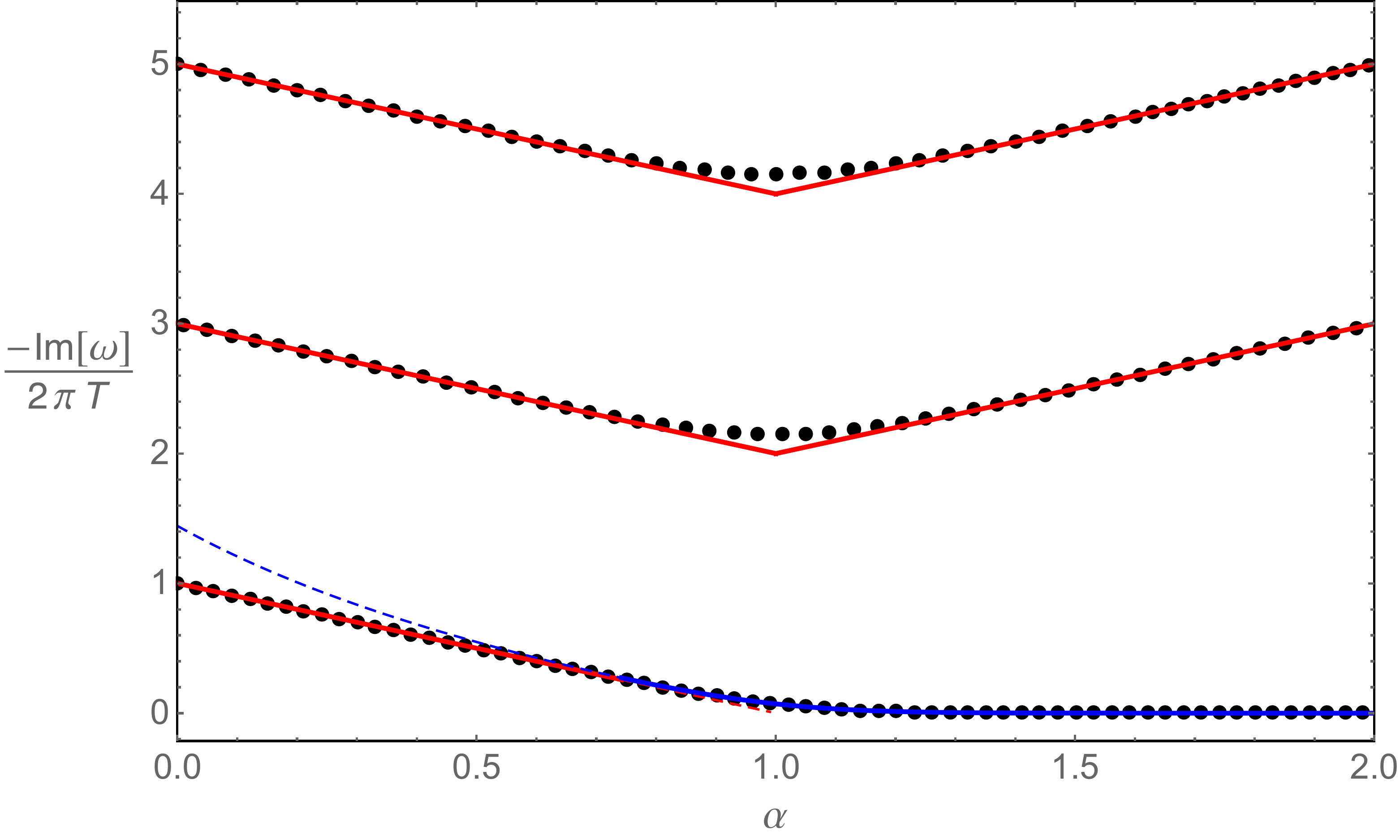}~~
\includegraphics[width=0.45\textwidth]{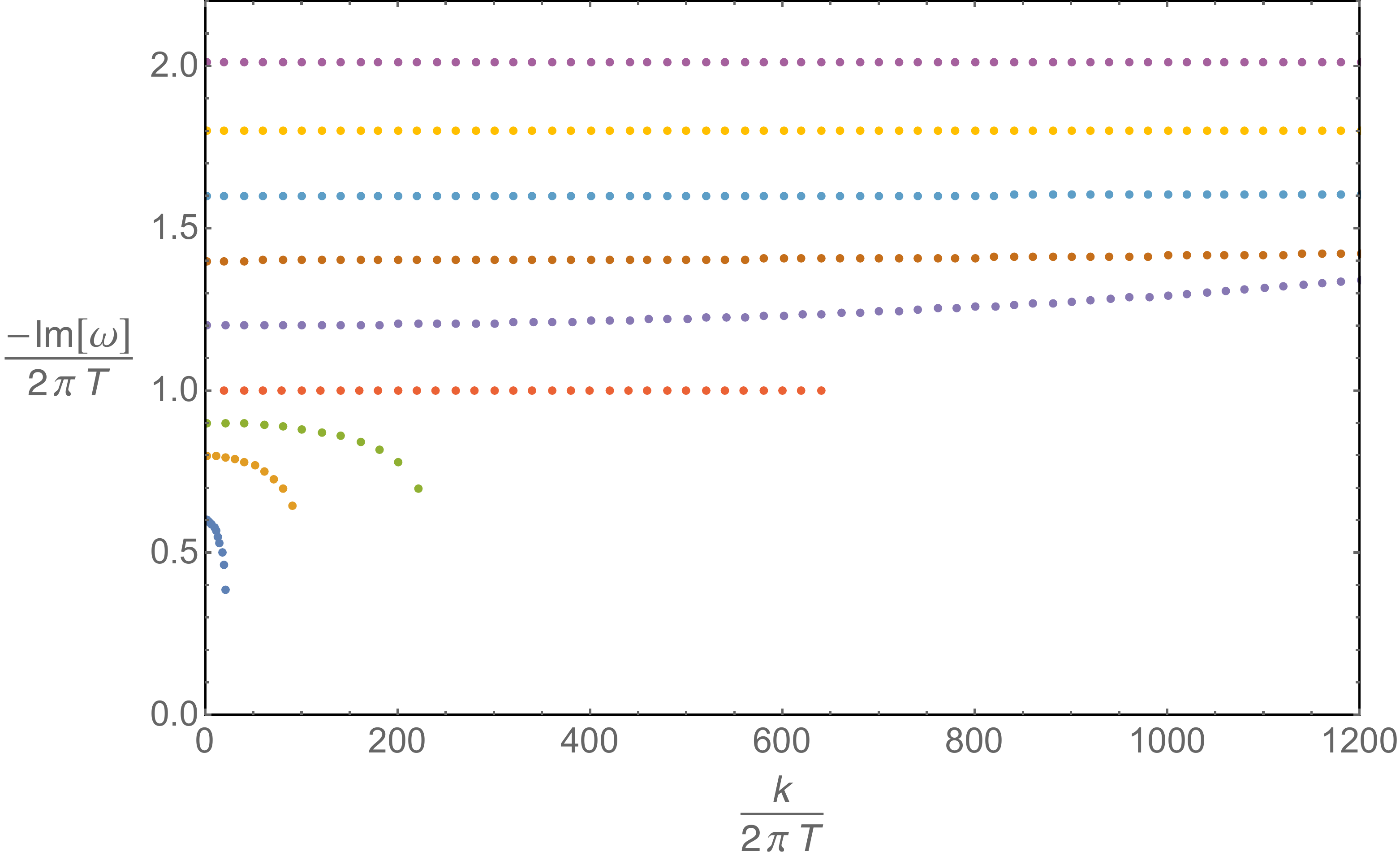}
  \caption{\small {\em Left:} The first three lowest quasi-normal modes at zero momentum (black and blue dots) as a function of $\alpha$ at low temperature $T/m\simeq 1.34\times 10^{-7}$. The red lines are the IR poles from the IR Green's function while the blue line is the pole obtained from a matching method shown in appendix \ref{app:weakhydro}. {\em Right:} The frequency of the first non-hydrodynamic quasi-normal mode as a function of momentum for $\alpha=-1,-0.8,-0.6,-0.4,-0.2, 0, 0.1, 0.2, 0.4$  from top to down at the same low temperature as the left plot.
  }
  \label{fig:IRmode}
\end{figure}

Both the hydrodynamic and the first non-hydrodynamic modes are the poles of the retarded Green's function, i.e. the quasi-normal modes. In the left plot of Fig. \ref{fig:IRmode}, we show the behavior of the first three quasi-normal modes at zero momentum as a function of $\alpha$. The black  dotted points are the quasi-normal modes obtained numerically from the retarded Green's function, while the red lines are the poles obtained analytically from the IR Green's function  $\mathcal{G}_{IR}(\omega)$ in the conformal to AdS$_2\,\times\,$R$^2$  geometry, and the dashed blue line is the quasi-normal modes from the near-far matching calculation as shown in appendix \ref{app:weakhydro}. 

More explicitly, the IR Green's function $\mathcal{G}_{IR}(\omega)$ at zero momentum  in the conformal to  AdS$_2\,\times\,$R$^2$  geometry has been calculated in section \ref{subsec:ir}, and we find a tower of IR poles at pure imaginary values $\frac{i\omega}{2\pi T}=2n-1-\alpha$ for $\alpha<1$, $\frac{i\omega}{2\pi T}=2n-1+\alpha$ for $\alpha>1$ with $n=1,2...$. 
The lowest three (for $\alpha<1$) or two (for $\alpha>1$) modes of these infinite IR poles are represented in red lines in the left plot of Fig. \ref{fig:IRmode}, which coincides with the UV poles from numerical results represented in black dots.\footnote{Note that when $\alpha=1$, there seems a derivation between numerical result and analytical result. As we have discussed in section \ref{subsec:ir}, when $\alpha=1$, there is a logarithmic term to define the dual Green's function in IR and the analytical results should be modified due to this logarithmic term. }
Note that for $\alpha>1$ the lowest IR mode starts from $\frac{i\omega}{2\pi T}=1+\alpha$, which is not the lowest non-hydrodynamic mode. 
This is due to the appearance of a slow mode when $\alpha$ increases, whose lifetime is much larger than $1/T$. 
The slow mode can be calculated from a near-far matching method as shown in appendix \ref{app:weakhydro}. The analytical result on $\omega_*$ from \eqref{eq:Dtau} is presented in the blue line in the left plot of Fig. \ref{fig:IRmode}, which matches very well with the black dots obtained numerically for $\alpha>1$. 

In the right plot of Fig. \ref{fig:IRmode}, we show
the frequency of the first non-hydrodynamic mode as a function of momentum. Note that for $0<\alpha<1$ we only plot the curves up to special $k$'s because they collide with the hydrodynamic pole almost at the locations we stopped. Different from the AdS$_2\,\times\,$R$^2$ case in  \cite{Arean:2020eus, Wu:2021mkk,Jeong:2021zsv}, now the non-hydrodynamic pole descending from the IR mode discussed in subsection \ref{subsec:ir} has nontrivial dependence on $k$, especially for $0<\alpha<1$. Although we do not have the IR Green's function for general $k$ since it is not possible to get the analytical solution and it is also numerically difficult because the separation of scales is highly nontrivial, nevertheless we shall name these modes as IR modes as they coincide with the analytical results on IR modes in regime $k<\omega$ and they are naturally inherited from the IR modes for general $k$. 

\begin{figure}[h!]
  \centering
\includegraphics[width=0.45\textwidth]{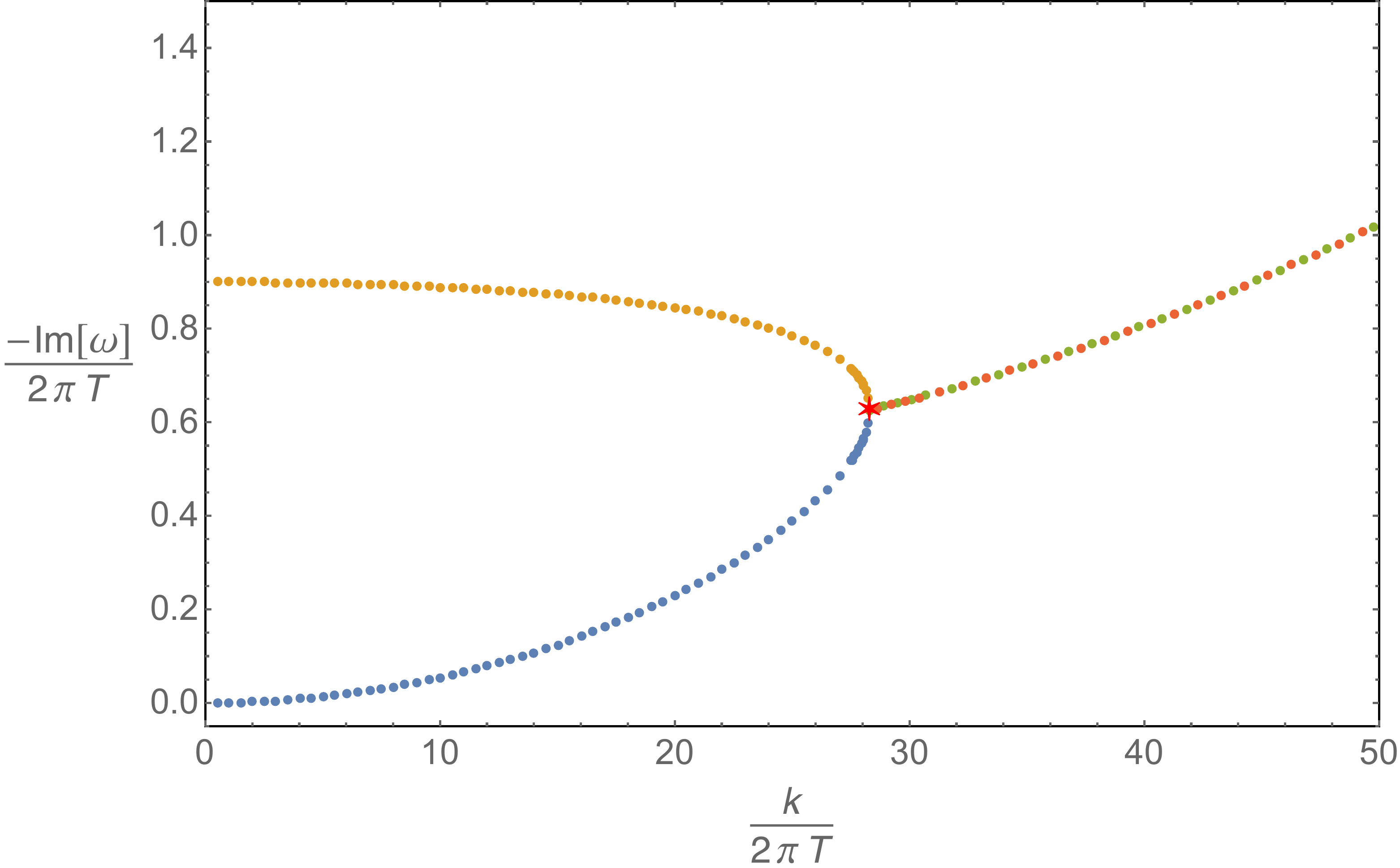}
\hspace{10mm}
\includegraphics[width=0.44\textwidth]{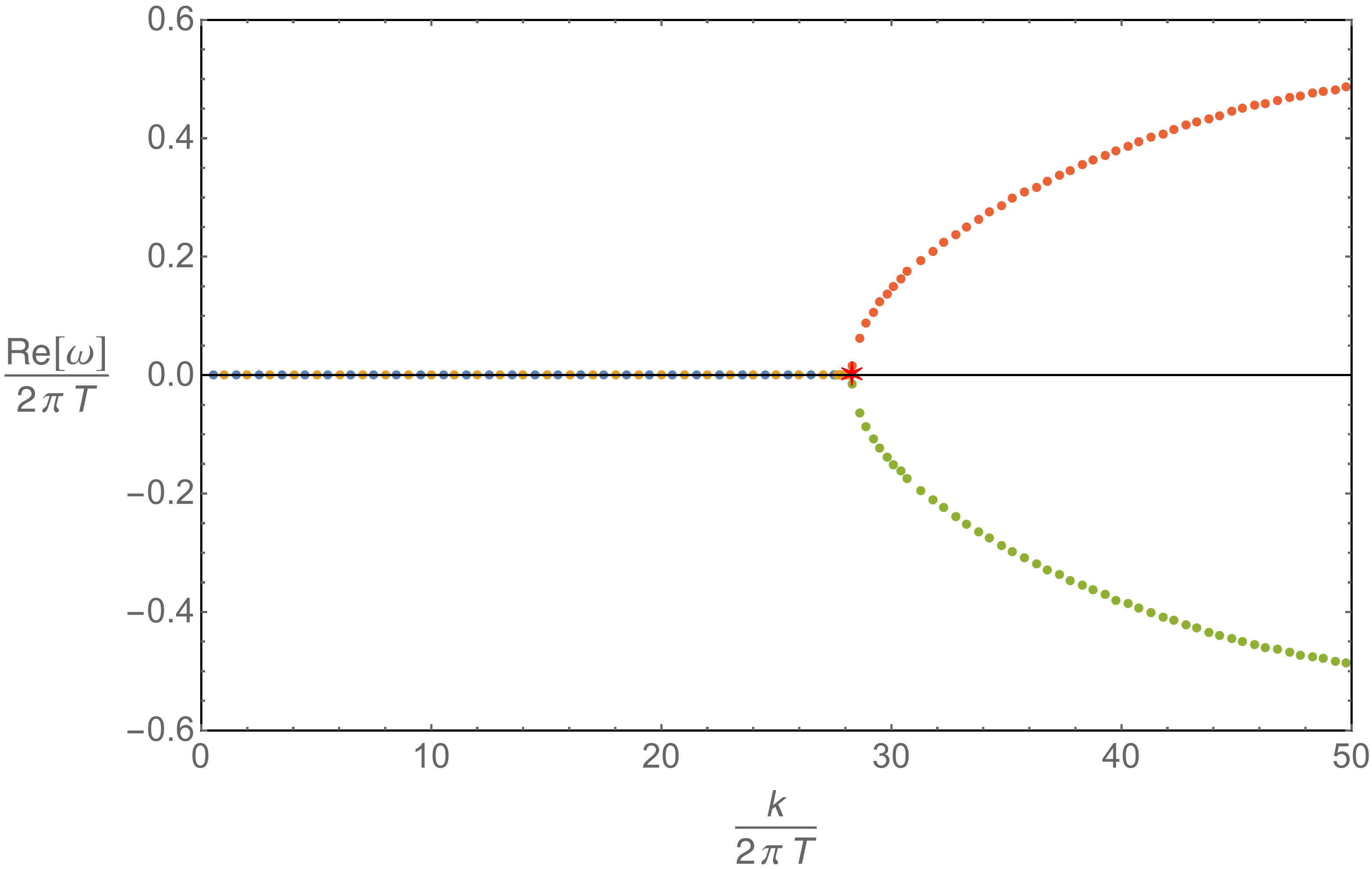}
  \caption{\small The imaginary and real parts of frequencies of the hydrodynamic and IR modes at low temperature $T/m=1.34\times 10^{-5}$ for $\alpha=1/10$. 
  }
  \label{fig:cross}
\end{figure}

We consider an example with $\alpha=1/10$ at low temperature $T/m=1.34\times 10^{-5}$. In this case the first non-hydrodynamic mode is an IR mode. The behavior of the hydrodynamic and IR modes are shown in Fig. \ref{fig:cross}. Similar to the $\alpha=0$ case, the collision also occurs at real momentum and pure imaginary frequency. The difference is that before the collision, for $\alpha=1/10$ the frequency of the IR mode has a nontrivial dependence on $k$, while for  $\alpha=0$ the frequency is almost independent of $k$.  Nevertheless we still have similar scaling behavior of local equilibrium scales and diffusion upper bound as we will discuss in the following. When we further increase $k$, we will see the second collision with the second IR pole. 

\vspace{.3cm}

The collision between the hydrodynamic mode and non-hydrodynamic mode indicates the scales $(\omega_{eq}, k_{eq})$ where hydrodynamics breaks down. Similar to the discussion in previous subsections, we could define the equilibrium time $\tau_{eq}$ and equilibrium velocity $v_{eq}$ for each value of $\alpha$. In the following we discuss the scaling behaviors of $(v_{eq}, \tau_{eq})$ near the quantum critical state with respect to temperature. 
\begin{itemize}
\item The hydrodynamic pole collides with a slow mode. This occurs when $T\tau\gg 1$\footnote{Note that this happens at  $\alpha> 1$. However, for the crossover regime $\alpha\sim 1$ the IR pole and the slow mode are of the same order, and the telegrapher equation seems also apply. Therefore we also discuss the scaling behavior for $\alpha<1$ in this case.}, the trajectory of the hydrodynamic mode and slow mode obey the telegrapher equation (\ref{eq:teleq}, \ref{eq:Dtau}). From the pole collision we have  
\bea
(k_{eq}, ~\omega_{eq})=\left(\frac{1}{\sqrt{4D_c\tau}}\,,~ \frac{1}{2\tau}\right)\,.
\eea
From the formula for $D_c$ and $\tau$ in (\ref{eq:Dtau}), we obtain the following scaling behavior at low temperature 
\be\label{eq:dep1}
\begin{cases}
\,~\omega_{eq}\sim 
k_{eq}   \sim m\left(\frac{T}{m}\right)^{\alpha} \, \quad &\text{for} \;\;  \alpha> 1\\
\,~ \omega_{eq}\sim T\,,~~~~k_{eq}\sim m\left(\frac{T}{m}\right)^{\frac{1+\alpha}{2}}\quad &\text{for} \;\;  \alpha\leq 1~~\text{while}~ T\tau>1\\
	\end{cases}\,,~~
\ee
from which we have 
\be\label{eq:dep2}
\begin{cases}
\,~v_{eq}\sim T^0\,,~~~~\tau_{eq}   \sim \frac{m^{\alpha-1}}{T^\alpha} \, \quad &\text{for} \;\;  \alpha> 1\\
\,~ v_{eq}\sim \left(\frac{T}{m}\right)^{\frac{1-\alpha}{2}}\,,~~~\tau_{eq}\sim \frac{1}{T}\quad &\text{for} \;\;  \alpha\leq 1~~{\text{while}}~ T\tau>1\\
	\end{cases}\,;~~
\ee

\item The hydrodynamic pole collides with an IR mode. This occurs when $\alpha< 1$, 
and the locations of the pole collision 
can only be obtained numerically. We have checked numerically and found 
that
\be\label{eq:dep3}
\omega_{eq}\sim T\,,~~~~~~ k_{eq}\sim m \left(\frac{T}{m}\right)^{\frac{1+\alpha}{2}}\ee
from which we have
\be
\label{eq:dep4}
v_{eq}\sim \left(\frac{T}{m}\right)^{\frac{1-\alpha}{2}}\,,~~~~~~\tau_{eq}\sim \frac{1}{T}\,.
\ee
\end{itemize}
The combinations of equilibrium velocity and equilibrium time give the scaling of $v_{eq}^2\tau_{eq}\sim m^{\alpha-1}/T^\alpha$ which is of exactly the same order as the charge diffusion constant $D_c$ at low temperature as shown in table \ref{table:diffusion}. 
There are two special cases with $\alpha=0$ as the first one, and in such a system the scaling $\omega_{eq}\sim k_{eq}^2\sim T$ shares similarities with the observations in \cite{Arean:2020eus, Wu:2021mkk,Jeong:2021zsv}. Another special example is when $\alpha=1$, the scaling $\omega_{eq}\sim k_{eq}\sim T$ is similar to the hydrodynamic system dual to the Schwartzschild black hole at high temperature \cite{Grozdanov:2019kge}. Here we also have different scalings of the equilibrium frequency and equilibrium momentum and these scalings depend crucially on the parameter $\alpha$ which characterize the IR gauge coupling constant $g_\text{eff}^2\sim (T/m)^\alpha$.

\begin{figure}[h!]
  \centering
\includegraphics[width=0.46\textwidth]{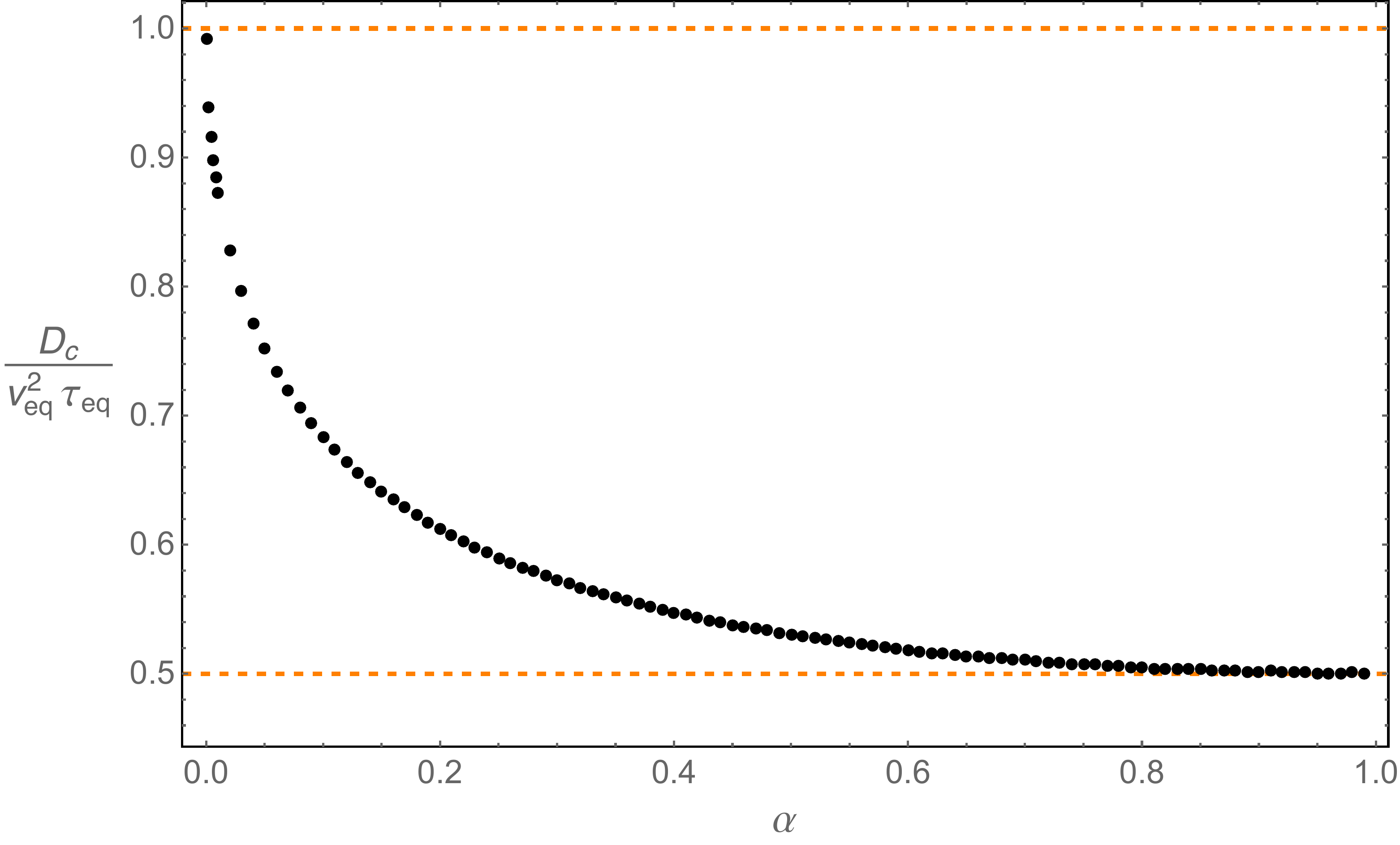}~~~
\includegraphics[width=0.44\textwidth]{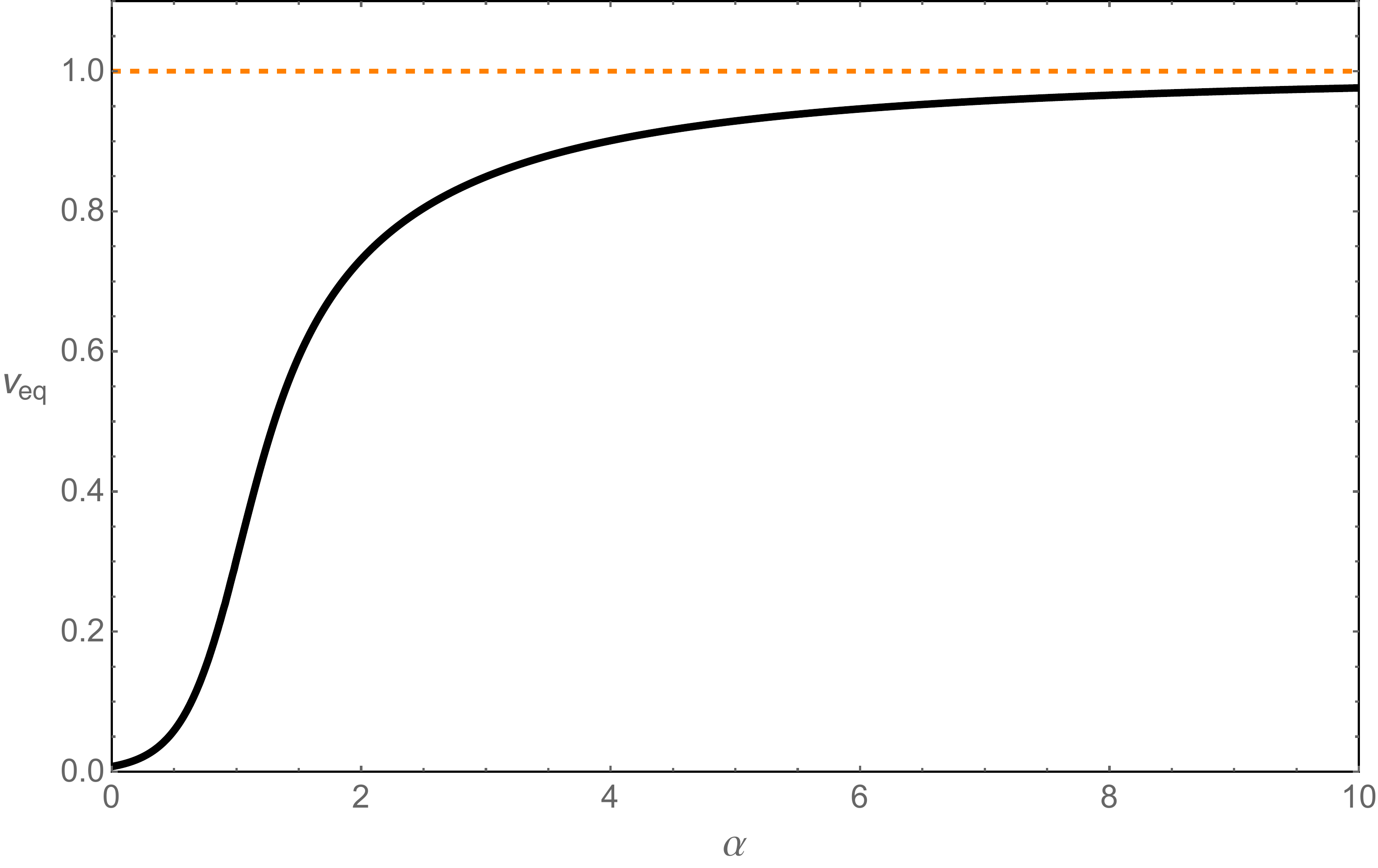}
  \caption{\small Two quantities $D_c/(v_{eq}^2 \tau_{eq})$ ({\em left}) and  $v_{eq}$ ({\em right}) as a function of $\alpha$ at low temperature $T/m\simeq 3\times 10^{-6}$. The diffusion upper bound is always satisfied for $\alpha\geq 0$. 
  }
  \label{fig:c(alpha)}
\end{figure}

In the low temperature limit, for arbitrary $\alpha\geq 0$ 
the behavior of $D_c/(v_{eq}^2 \tau_{eq})$ is shown  
in the left plot of Fig. \ref{fig:c(alpha)}. 
We found that in the case of pole collision with an IR mode $D_c/(v_{eq}^2 \tau_{eq})$ decrease sharply from $1$ when $\alpha=0$, and gradually reaches $1/2$ as a feature of the slow mode phase via a crossover. When $\alpha>1$, the case of pole collision with a slow mode, $D_c/(v_{eq}^2 \tau_{eq})$ equals to $1/2$. The right plot in  Fig. \ref{fig:c(alpha)} shows the equilibrium velocity is always smaller than the speed of light. 

\begin{figure}[h!]
 \centering
  \includegraphics[width=0.47\textwidth]{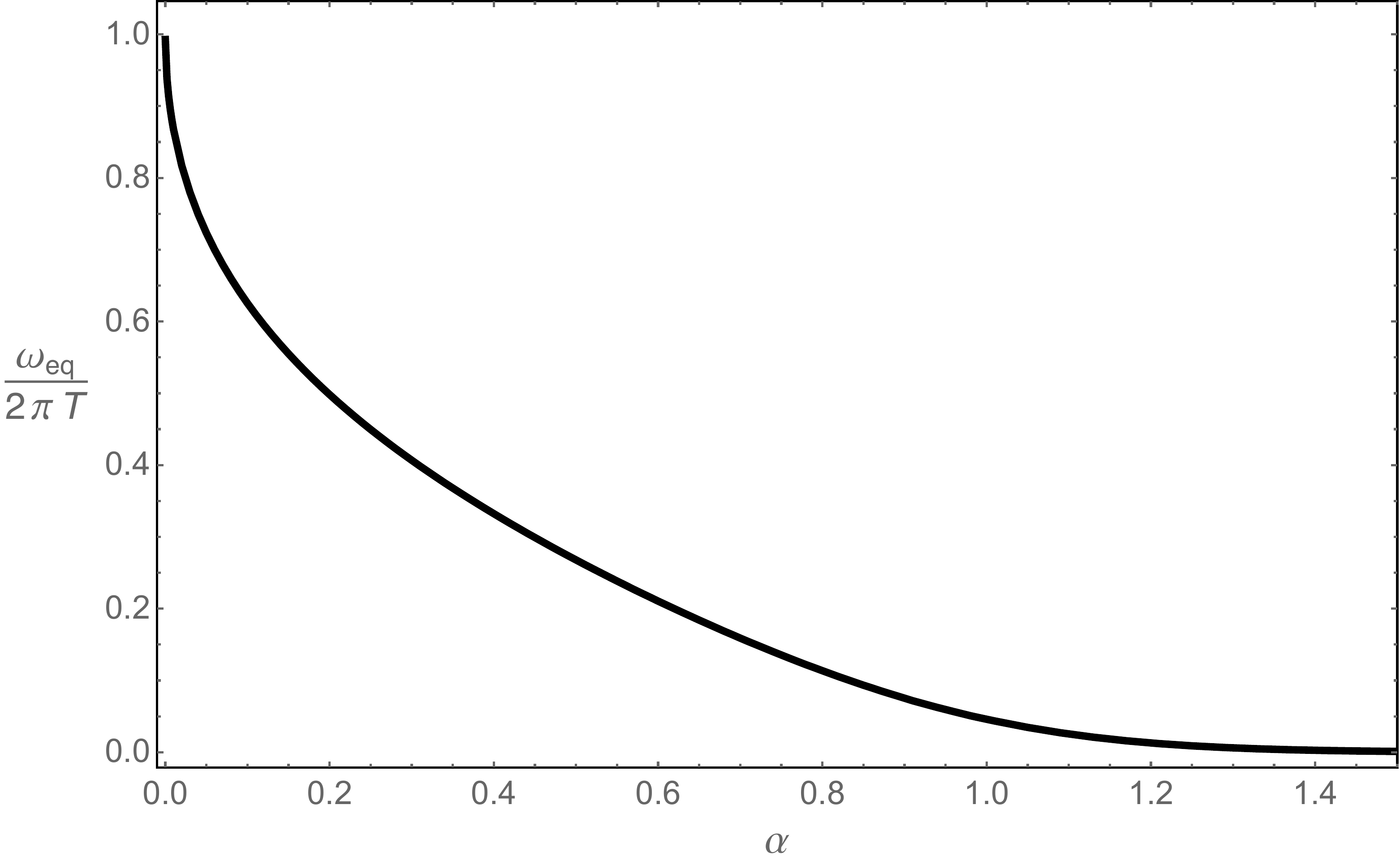}
  \includegraphics[width=0.46\textwidth]{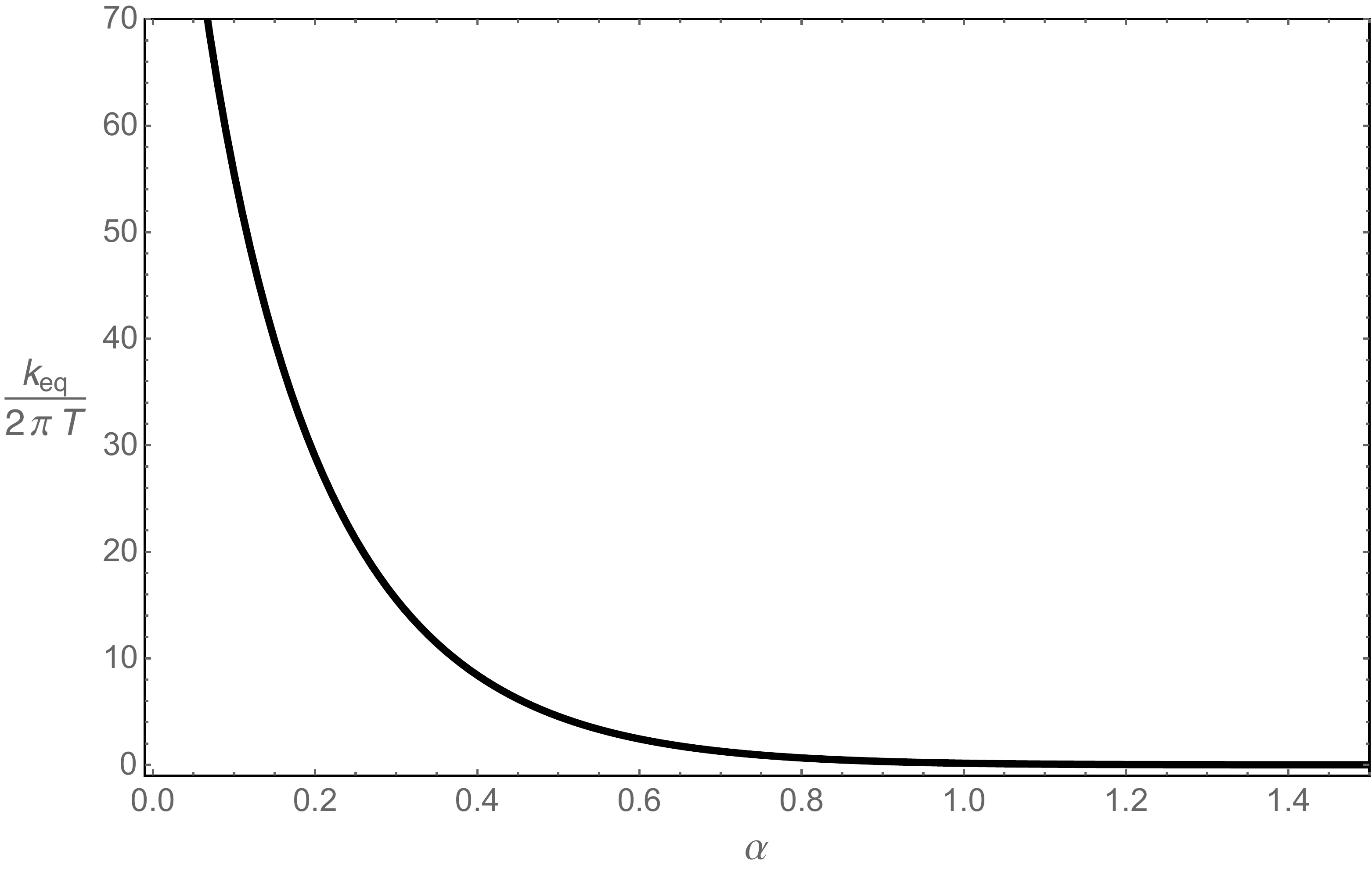}
  \caption{\small The convergence radii $\omega_{eq}/(2\pi T)$  
  and $k_{eq}/(2\pi T)$  
  as a function of $\alpha$ (i.e. the IR effective gauge coupling) at low temperature $T/m\simeq 3\times 10^{-6}$.  
  }
  \label{fig:Rwalp}
\end{figure}

We have found that effects of $\alpha$, i.e. the IR effective gauge coupling constant $g_\text{eff}^2\sim (T/m)^\alpha$, plays a prominent role in the origin of the first non-hydrodynamic mode. The equilibrium frequency and momentum as function of $\alpha$ is shown in Fig. \ref{fig:Rwalp}. Note that for $\alpha<1$, although the first non-hydrodynamic mode is an IR pole,  $\omega_{eq}/(2\pi T)$ depends on $\alpha$ in a non-trivial way. This is due to the fact that now the non-hydrodynamic pole depends on $k$ as shown in Fig. \ref{fig:IRmode} and is different from the cases of critical states dual to AdS$_2$ in \cite{Arean:2020eus, Wu:2021mkk,Jeong:2021zsv} where $\omega_{eq}/(2\pi T)$ only depends on the IR conformal dimension of the density operator. Here at finite $k$ we do not have a clear notion of conformal dimension in IR at low temperature due to the lack of analytical solution as discussed in Sec. \ref{subsec:ir}. Moreover, the convergence radius of the diffusion dispersion relation, which is given by $k_{eq}/(2\pi T)$, is monotonically decreasing when we increase $\alpha$ or equivalently decrease the IR effective gauge coupling  $g_\text{eff}$. This behavior is consistent with the intuition that hydrodynamics works better for a strongly coupled quantum many body system. Moreover this behavior is independent of the nature of the first non-hydrodynamic mode at low temperature. Note that monotonic behavior for the convergence radius as functions of coupling has been found in \cite{Baggioli:2020loj} using the experimental data for fluids. There are also studies in \cite{Choi:2020tdj, Grozdanov:2021gzh} showing the non-monotonic dependence of the coupling strength in the theory. It is  interesting to understand the dependence of hydrodynamic convergence on the coupling constant better. 

\subsection{Comments on pole collision at general temperature}

The above discussions are mainly for low temperature where the extremal near horizon geometry is conformal to AdS$_2\,\times\,$R$^2$. In this subsection we briefly comment on the behavior of pole collisions beyond the low temperature regime. 

\begin{figure}[h!]
  \centering
  \includegraphics[width=0.45\textwidth]{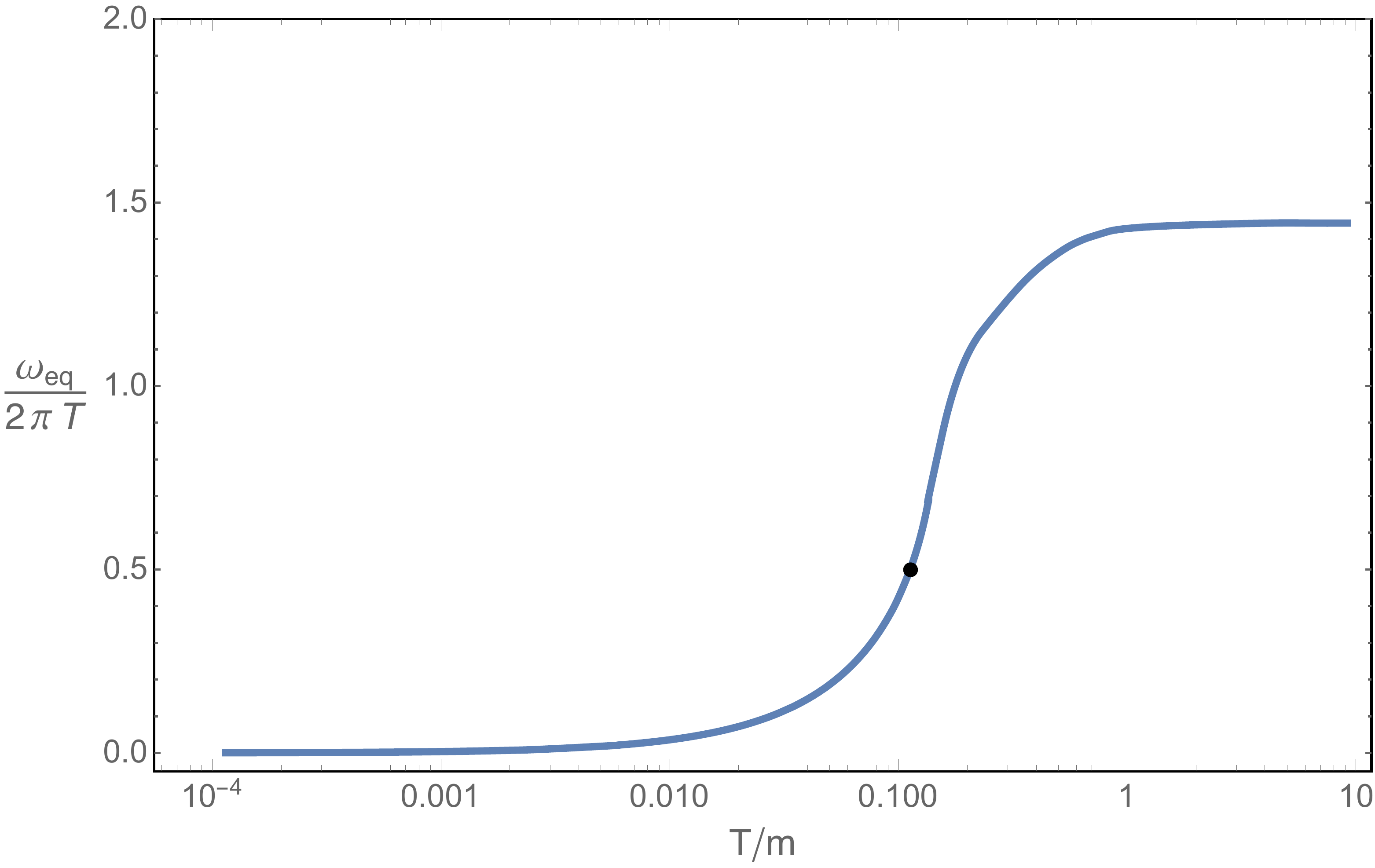}~~
\includegraphics[width=0.45\textwidth]{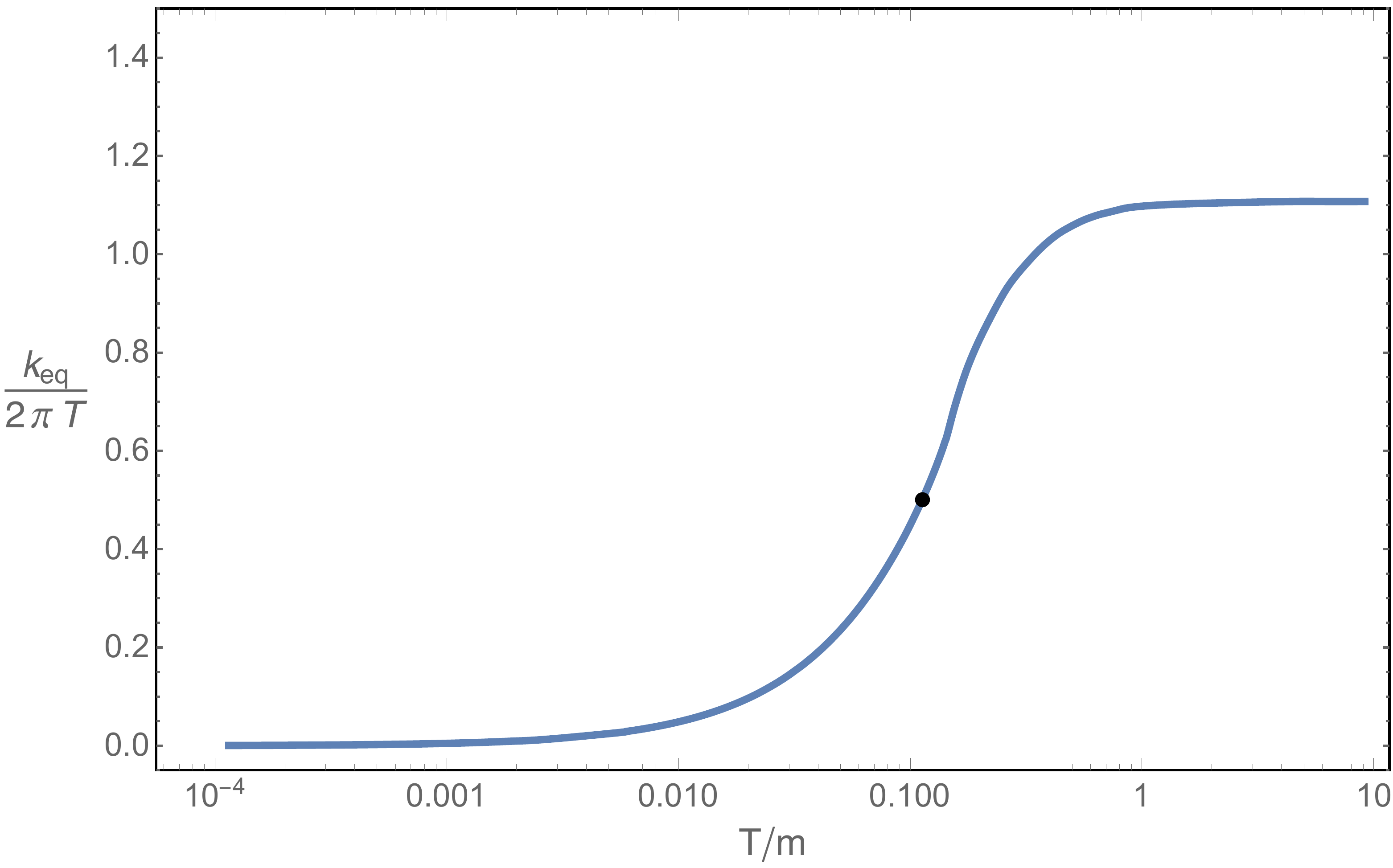}\\
  \includegraphics[width=0.45\textwidth]{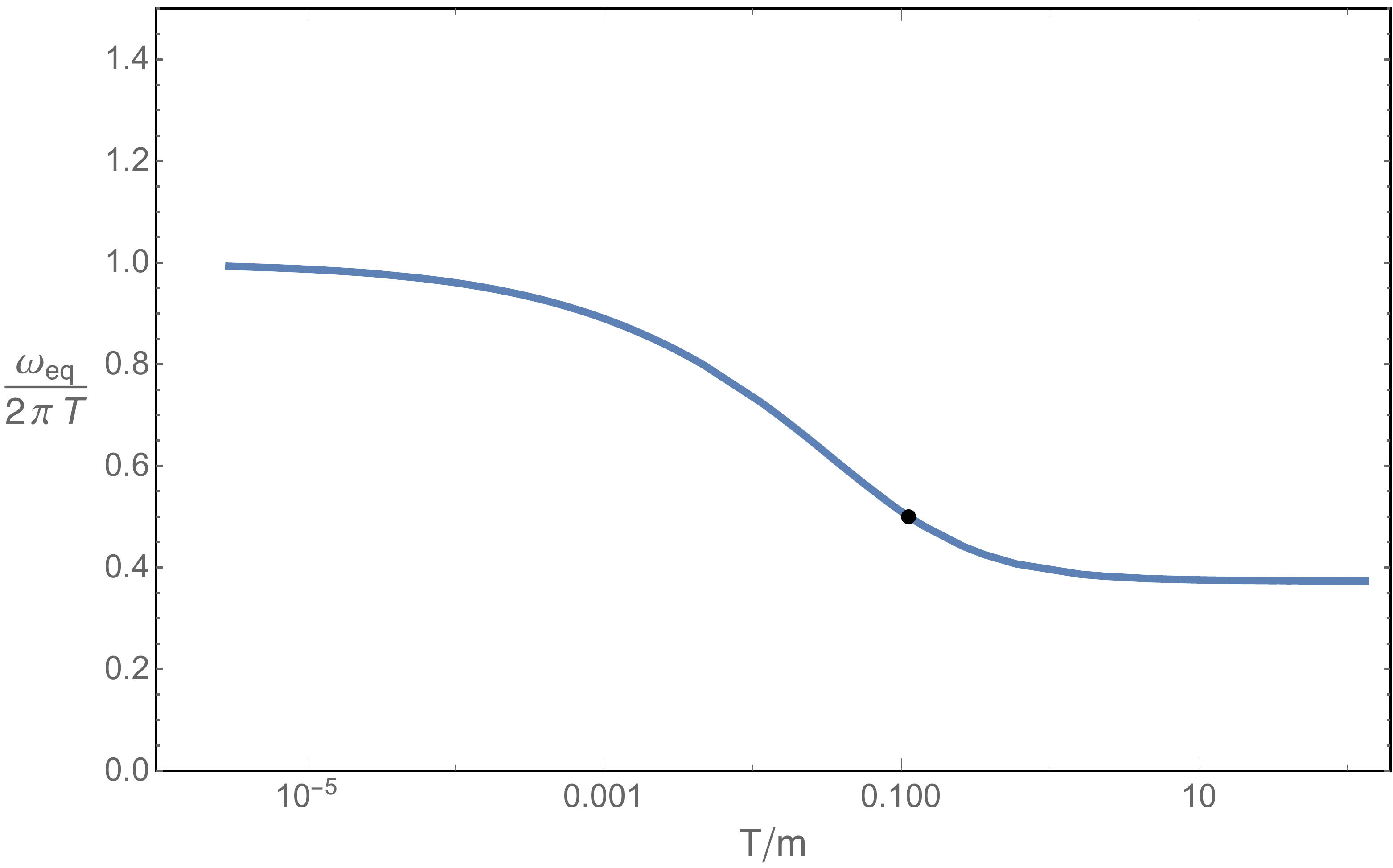}~~
\includegraphics[width=0.45\textwidth]{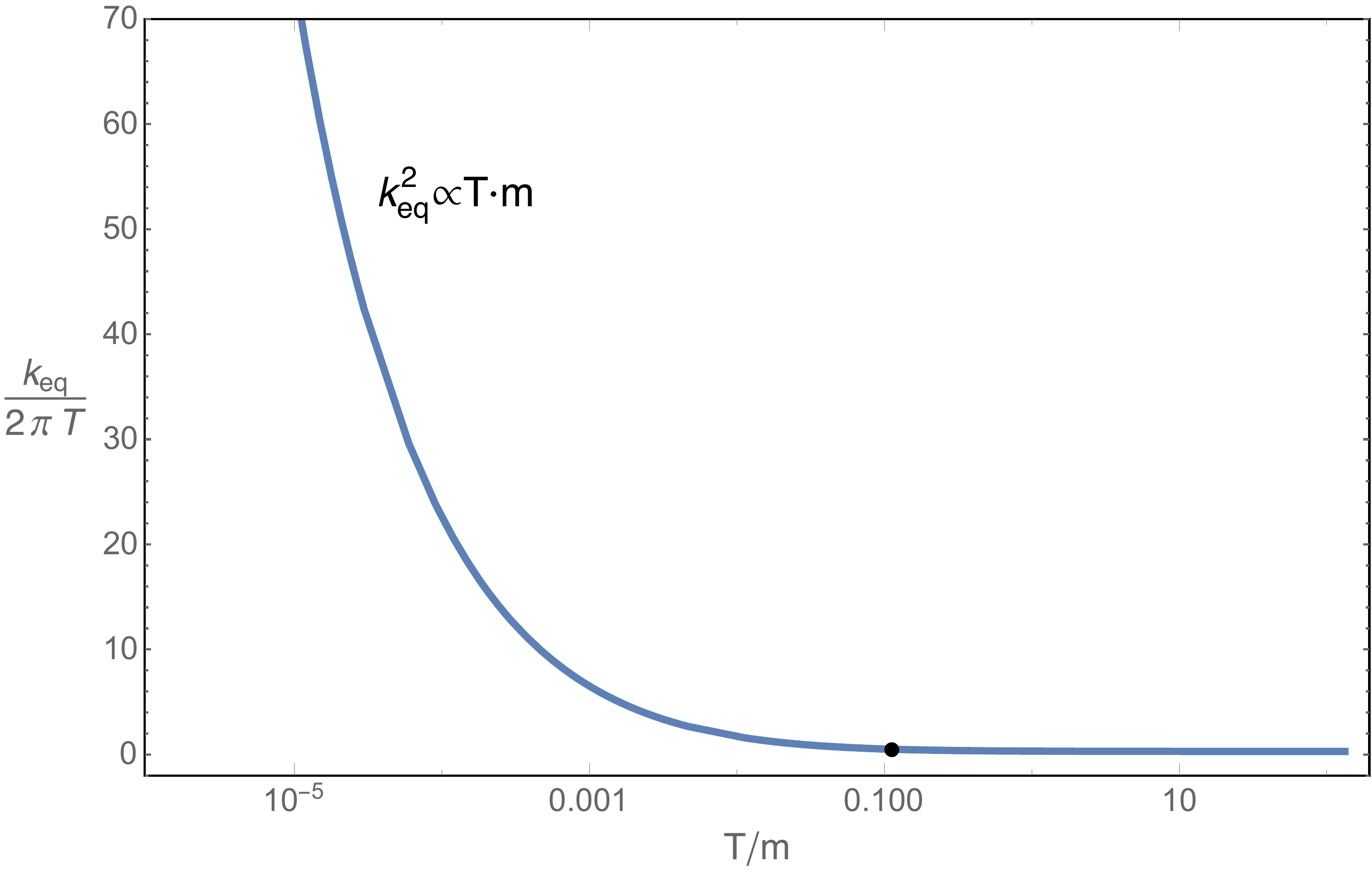}
  \caption{\small The equilibrium momentum and equilibrium frequency as a function of $T$ for $\alpha=2$ ({\em top two}) 
  and $\alpha=0$ ({\em down two}). The black dot is for the temperature at which the dilaton becomes trivial, i.e. the Schwartzschild black hole. Note that at high temperature, when $\alpha=2$, $k_{eq}/(2\pi T)\rightarrow 1.107$  while when $\alpha=0$, $k_{eq}/(2\pi T)\rightarrow 0.148$.
  }
\label{fig:tuneT0}
\end{figure}

We focus on two special cases, i.e. $\alpha=2$ and $\alpha=0$. Their low temperature behaviors have been studied in detail in sections \ref{sec:case1} and \ref{sec:case2}. In the following we tune the temperature to study the breakdown of hydrodynamics and the upper bound for the diffusion constant. In Fig. \ref{fig:tuneT0}, we show the behavior of the equilibrium frequency and equilibrium momentum as function of $T/m$. We find that when it is a slow mode (i.e. top plots) as the first non-hydrodynamic mode at low temperature, the convergence radius $k_{eq}/(2\pi T)$ increases when we increase $T/m$, while when it is an IR mode (i.e. down plots) at low temperature, the convergence radius decreases when we increase $T/m$. Note that in the plots the black dot is the result for the Schwartzschild black hole since at $T/m=1/(2^{3/2}\pi)$ the dilatonic black hole reduces to a  Schwartzschild black hole. For $\alpha=2$, below $T/m\simeq 0.15$, the collision occurs at real momentum while above this value the collision occurs at complex momentum.\footnote{This is consistent with the following observation on the slow mode at zero momentum. When we increase the temperature up to a certain temperature, the pure imaginary slow mode splits into a pair of complex QNMs with opposite real parts while same imaginary part. 
Then the poles moves up when we further increase the temperature. When $T/m\geq 1$, the QNMs will be almost fixed and not sensitive to the temperature any more.} We have $\omega_{eq}\sim k_{eq}\sim T^2/m$ at small $T$ while we have $\omega_{eq}\sim k_{eq}\sim T$ at large $T$. For $\alpha=0$, the collision always occurs at real momentum. The equilibrium momentum behaves as $k_{eq}\propto \sqrt{T m}$ at low temperature while $k_{eq}\propto T$ at high temperature. 

\begin{figure}[h!]
  \centering
  \includegraphics[width=0.46\textwidth]{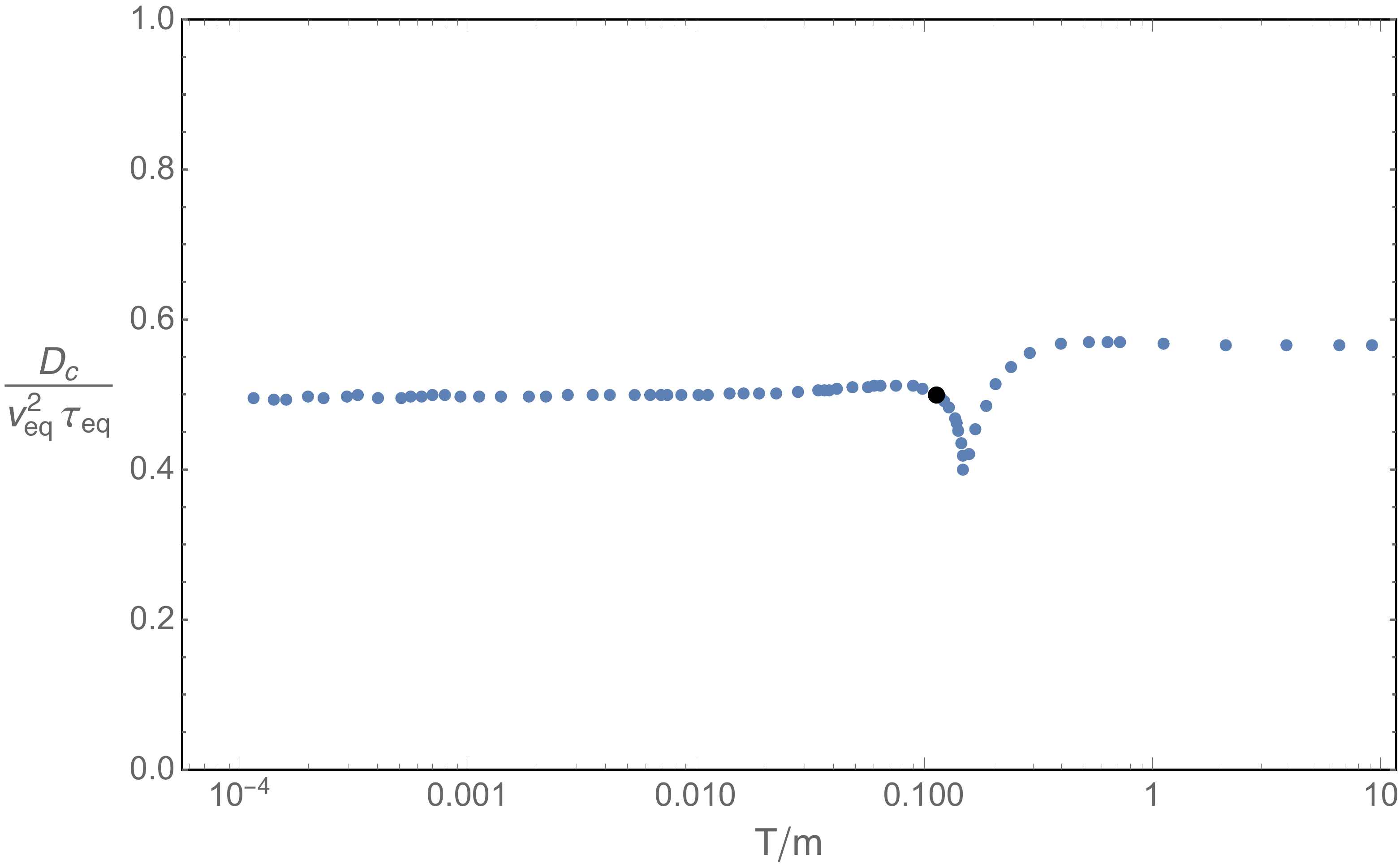}~~
\includegraphics[width=0.44\textwidth]{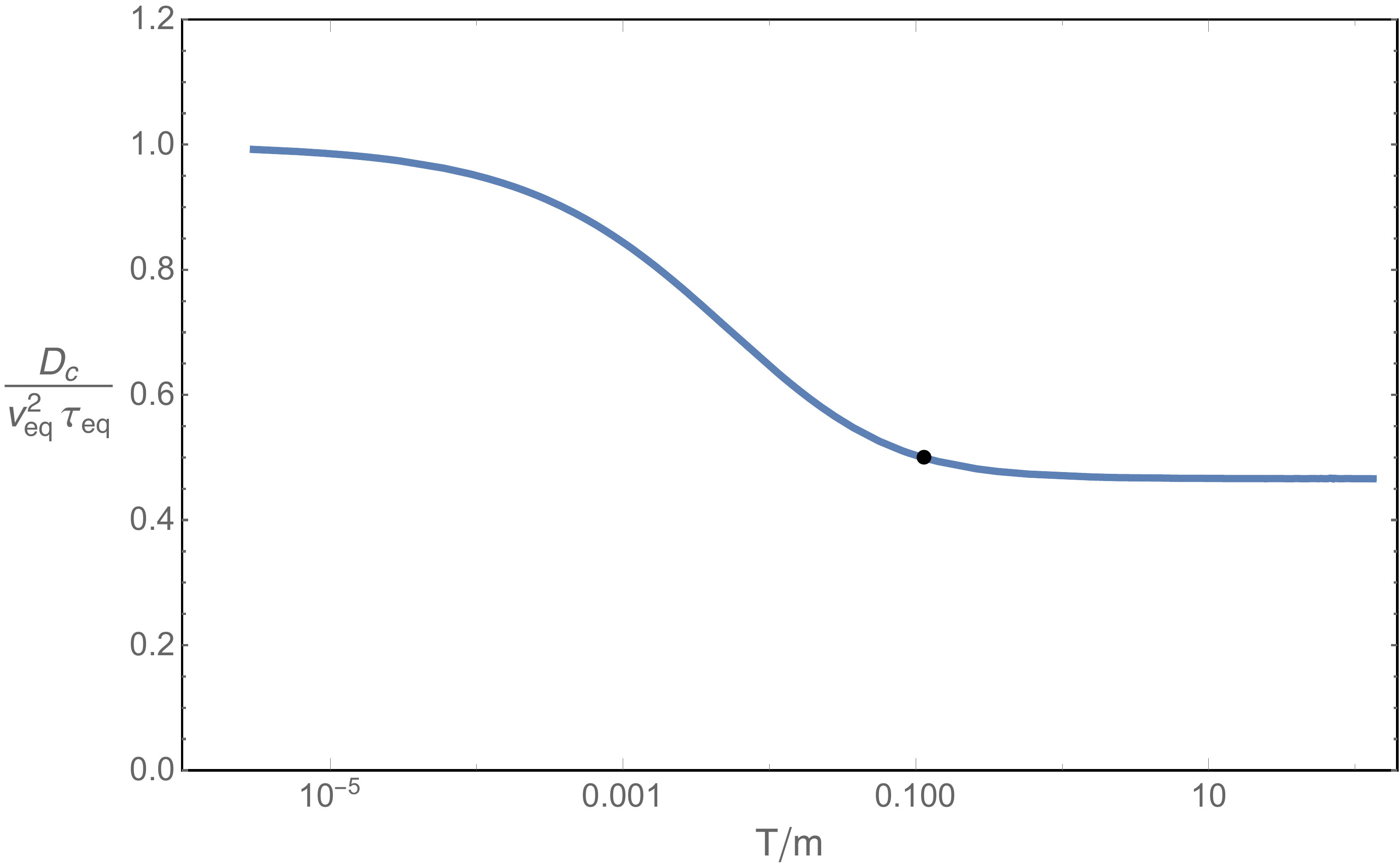}
  \caption{\small  $D_c/(v_{eq}^2 \tau_{eq})$ as a function of $T$ for $\alpha=2$ ({\em left}) 
  and $\alpha=0$ ({\em right}). For $\alpha=2$, interesting ``non-smooth" behavior of $D_c/v_{eq}^2\tau_{eq}$ was observed when the location of collision changes from real momentum to complex momentum. 
  }
  \label{fig:tuneT2}
\end{figure}

The upper bound for the diffusion constant can be examined for general temperature. In Fig. \ref{fig:tuneT2} we show the ratio $D_c/(v_{eq}^2\tau_{eq})$ as a function of $T/m$ for these two choices of values of $\alpha$ and find that the diffusion upper bound is always satisfied. For the case $\alpha=2$, there is an interesting ``non-smooth" behavior for $D_c/(v_{eq}^2\tau_{eq})$ close to $T/m\simeq 0.15$ at which the location of pole collision changes from 
real momentum to complex momentum. It is interesting to check if this behavior is typical when the location of collision changes from real momentum to complex momentum. 

\subsection{Comments on pole collision for negative $\alpha$}
\label{subs:nalp}

Finally we briefly comment on the behavior of the pole collision for negative $\alpha$ at low temperature. Note that the effective gauge coupling in the IR at low temperature is of the form $g_\text{eff}^2\sim (T/m)^\alpha$. When $\alpha>0$ it correctly captures the charge screening effect while for $\alpha<0$ it departs from the intuition of the charge screening effect. Nevertheless we consider one simple example of negative $\alpha=-10^{-4}$ to analyze the pole collision at low temperature for complementary.  

\begin{figure}[h!]
  \centering
 \includegraphics[width=0.44\textwidth]{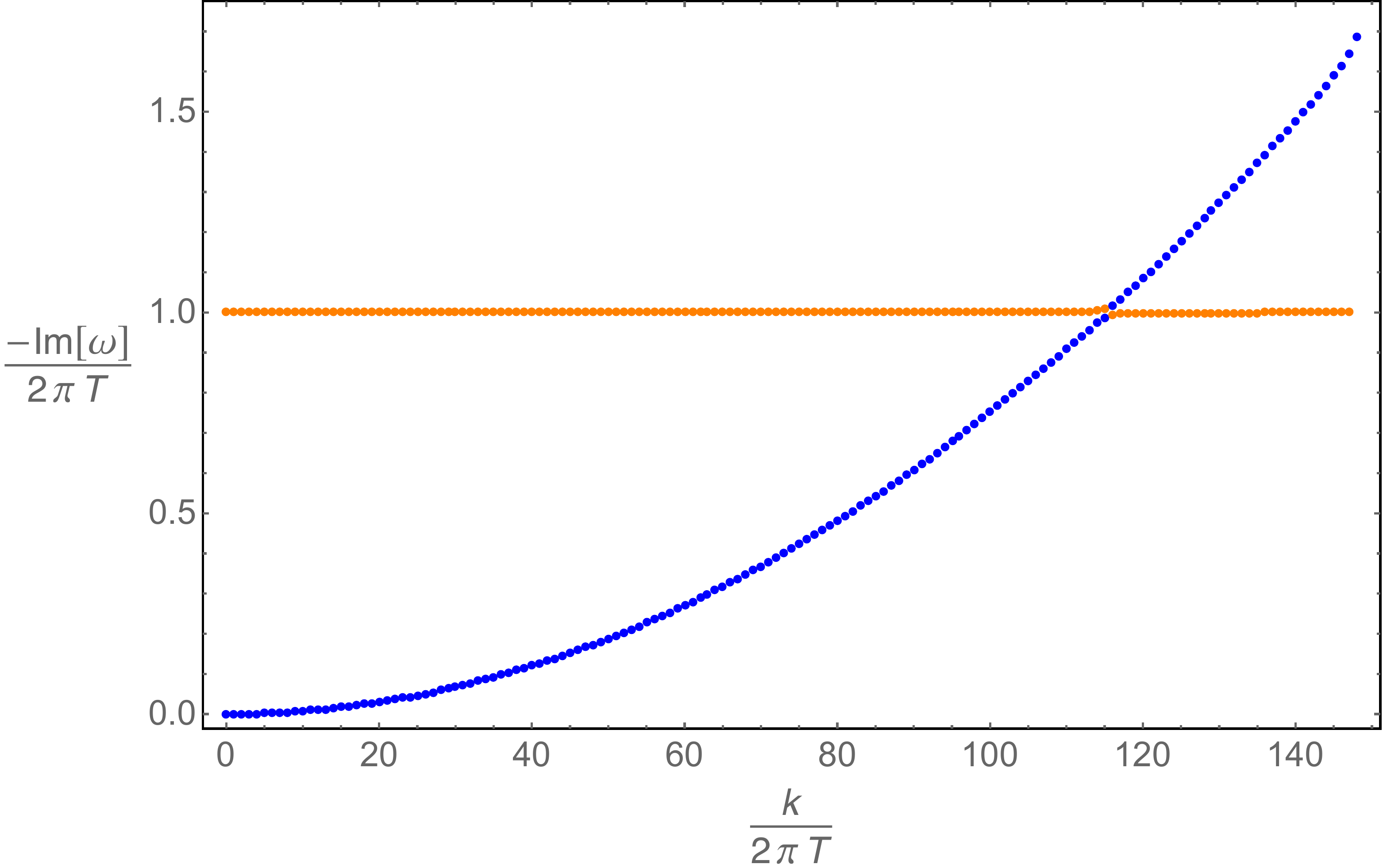}~~~
  \includegraphics[width=0.44\textwidth]{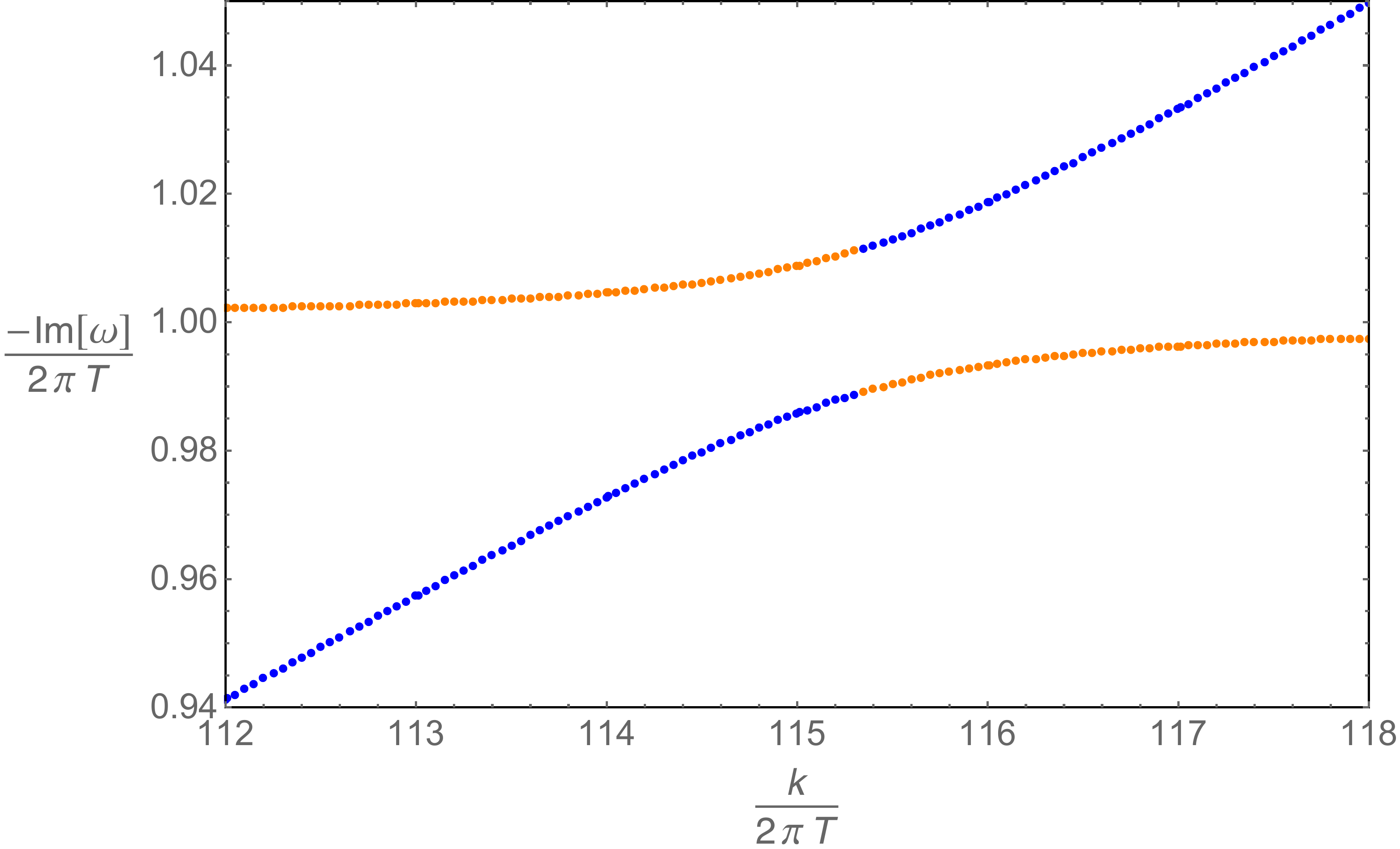}
   \caption{\small Frequencies of the hydrodynamic mode (blue) and the first IR mode (orange) at low temperature $T/m=4.23\times 10^{-6}$ for $\alpha=-10^{-4}$. The right plot is an enlarged version of the left one close to $k=k_{eq}$ which shows that there is no pole collision for real $k$.  }
  \label{fig:negalp0}
\end{figure}

Similar to the discussions for $\alpha\in [0,1)$, we now have the first non-hydrodynamic mode inherited from the IR modes. 
 In Fig. \ref{fig:negalp0}, we show the frequencies of the hydrodynamic and the first non-hydrodynamic modes  
 as a function of real $k$ at low temperature $T/m=4.23\times 10^{-6}$.
 We find that different from the cases of $\alpha\geq 0$, now we do not have pole collision at real momentum. 
 
\begin{figure}[h!]
  \centering
\includegraphics[width=0.3\textwidth]{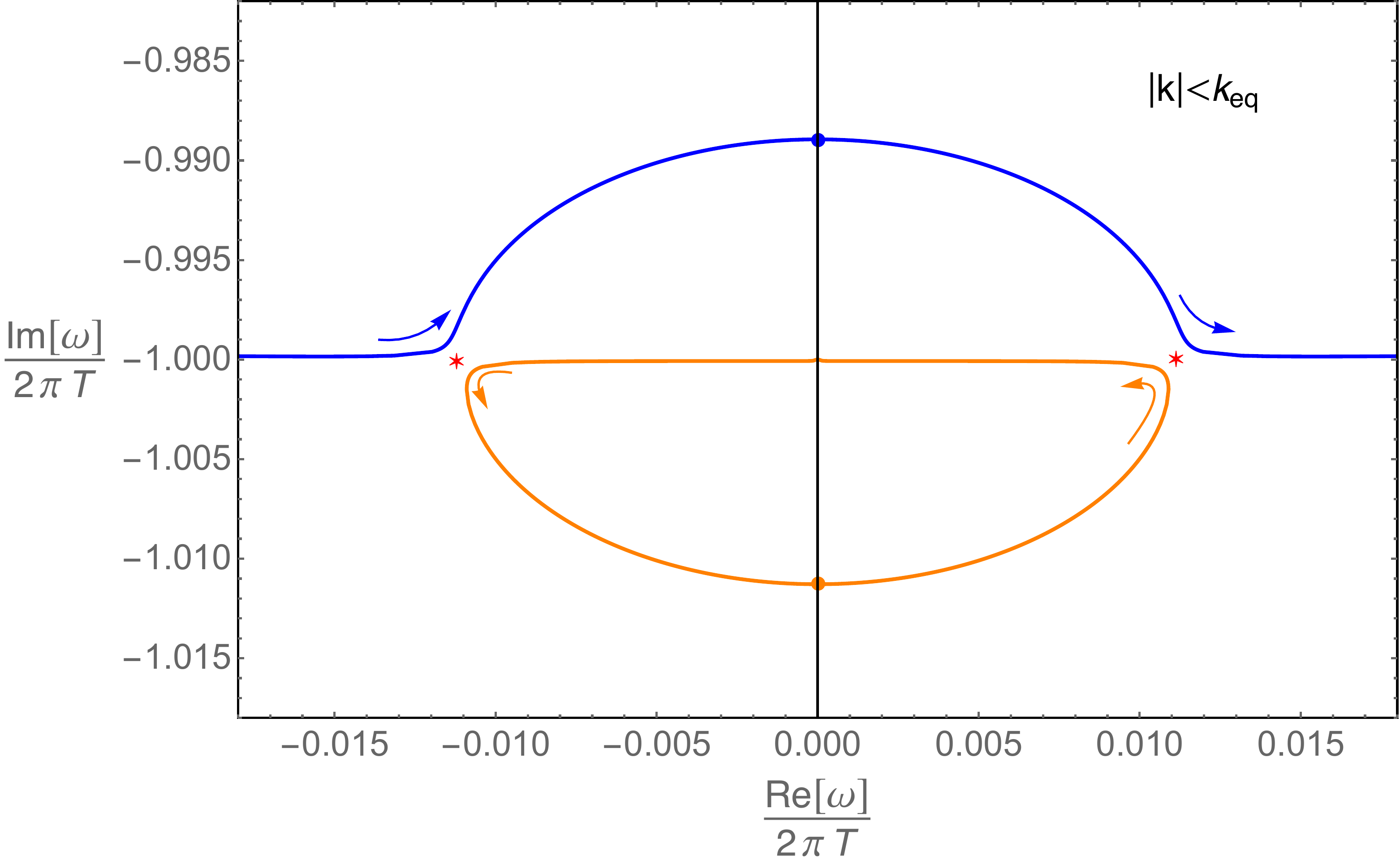}
~
\includegraphics[width=0.3\textwidth]{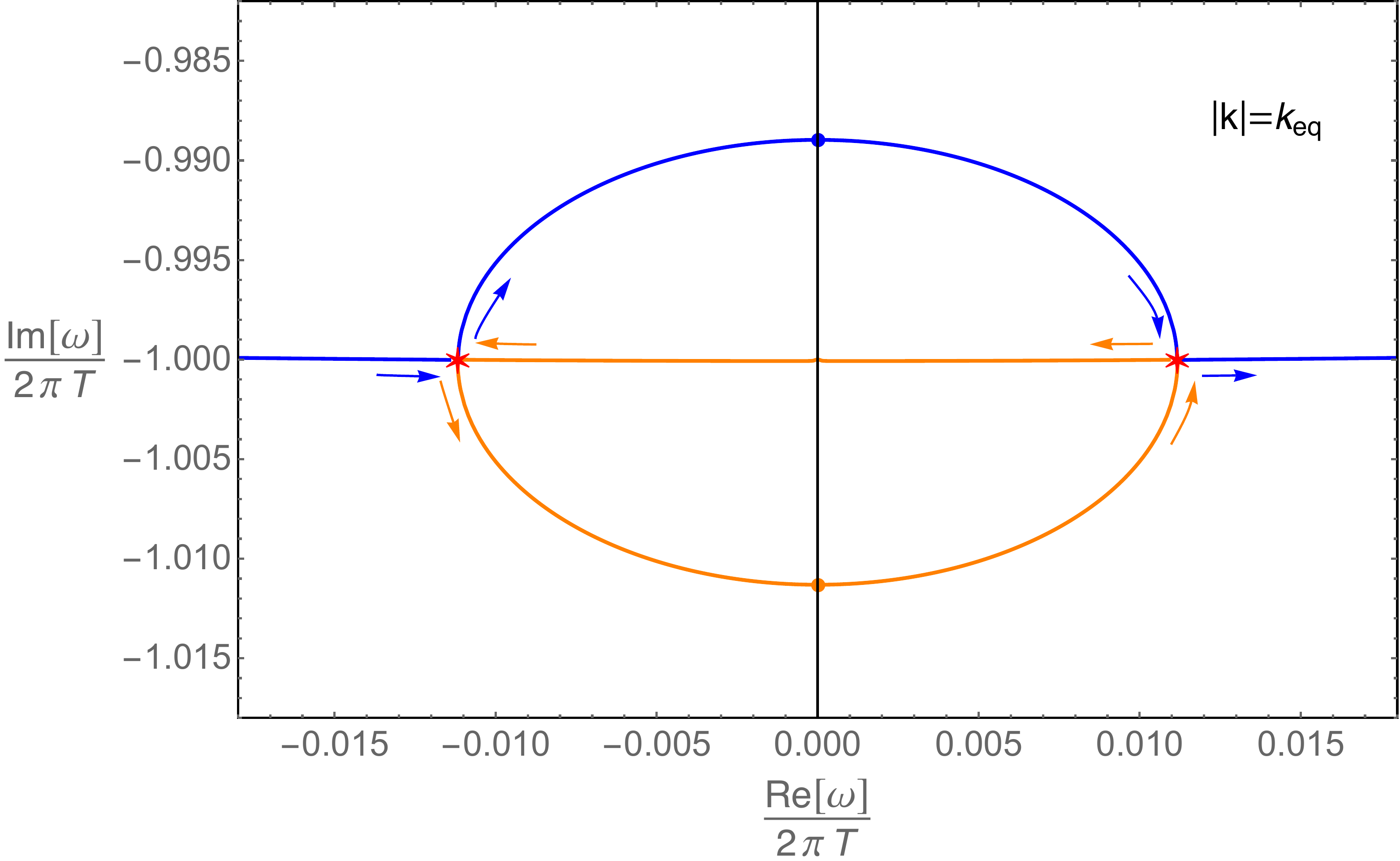}
~
\includegraphics[width=0.3\textwidth]{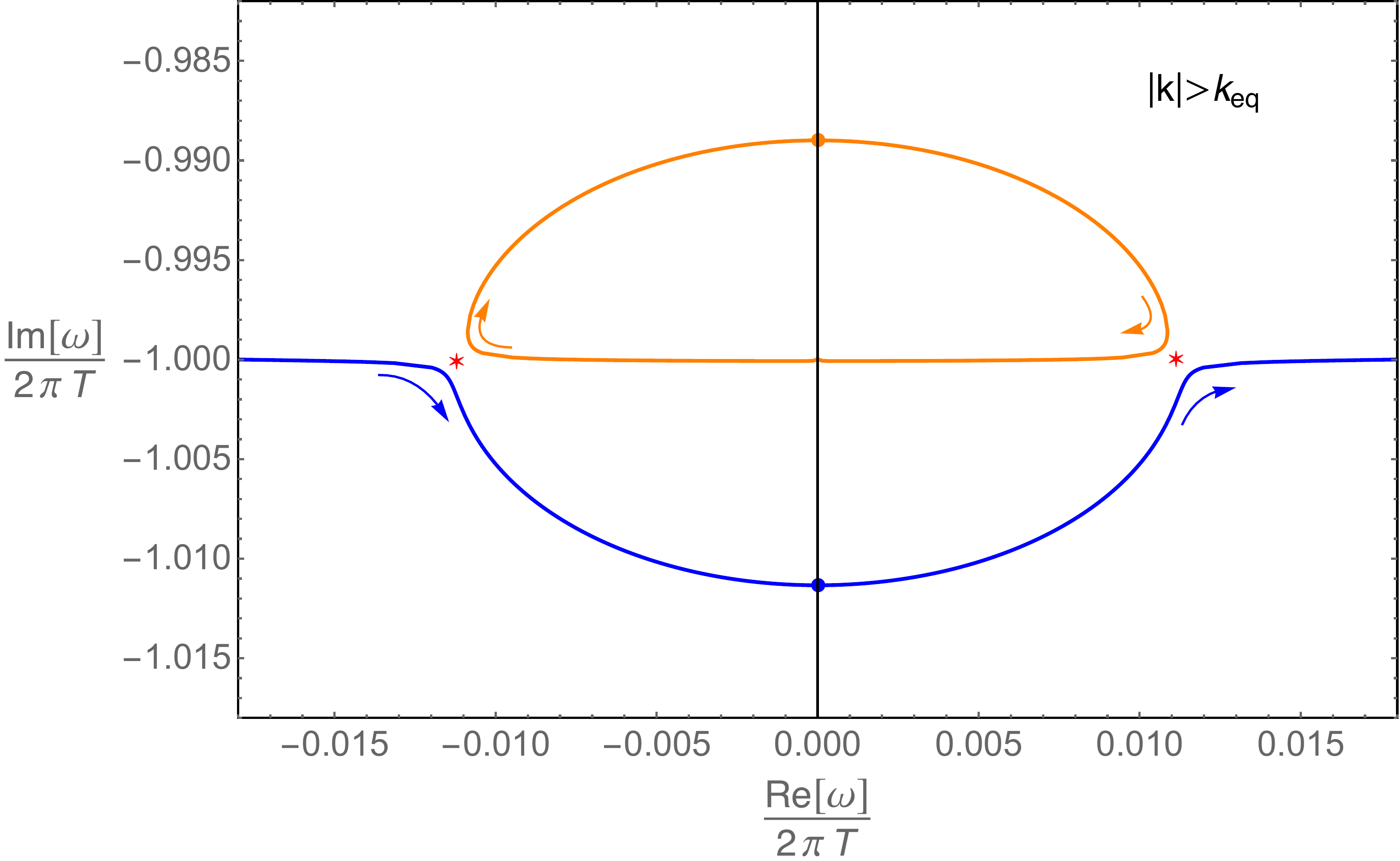}
\caption{\small The complex frequencies of the hydrodynamic mode (blue) and the lowest IR mode (orange) by tuning the phase of the complex value of $k$ at low temperature $T/m=4.23\times 10^{-6}$ for $\alpha=-10^{-4}$. The arrow indicates the value of the phase from $0$ to $\pi$. The collisions occur at $(k=k_{eq}e^{i\varphi_k}, \omega=\omega_{eq}e^{i\varphi_{\omega}})$ with $k_{eq}/(2\pi T)\simeq 115.3266$ and $\omega_{eq}/(2\pi T)\simeq 1$, while  $\varphi_k\simeq 0.0036\pi$ and $0.9964\pi$, $\varphi_\omega\simeq \mp0.4964\pi$. 
Note that $\varphi_\omega \simeq \varphi_k-\pi/2$ and $D_c/(v_{eq}^2
\tau_{eq})=0.9994$. From left to right, the modulus of the complex momentum are fixed in each plot to be $|k|/(2\pi T)= 115.3236 ~(\text{\em left}),~115.3266 ~(\text{\em middle}),~115.3296 ~(\text{\em right}).$
   }
  \label{fig:negalp}
\end{figure}

Now the pole collision occurs at complex value of momentum. Close to the collision point $k_{eq}$, in Fig. \ref{fig:negalp} we plot the frequencies as a function of the phase of the momentum with fixing modulus $|k|$ with $|k|< k_{eq}, |k|= k_{eq}, |k|> k_{eq}$ respectively. The orange curve represents the IR mode while the blue curve is the hydrodynamic mode at complex momentum which form a big closed curve and we only show a part of it closed to the location of pole collision. The arrow is the evolution when we increase the phase of the complex momentum from $0$ to $\pi$.\footnote{Note that different from \cite{Grozdanov:2019kge}, we use the phase $\varphi$ in  $k=|k|e^{i\varphi}$ instead of $k^2$.}  We find that the behavior of frequencies are slightly different from the case shown in appendix \ref{app:pccomplex}, there is no obvious topological change. However, there is an interesting reconnection of quasi-normal modes crossing $k_{eq}.$ The hydrodynamic mode and non-hydrodynamic mode exchange their positions after the collision and this is the reason that in Fig. \ref{fig:negalp0} we use different colors for a single curve. Note that similar behavior has been observed before in  \cite{Choi:2020tdj, Jeong:2021zsv}. At the collision points, the phases of the complex frequency $\varphi_k$ and complex momentum $\varphi_\omega$ satisfy $\varphi_\omega\simeq\varphi_k-\pi/2$. We have checked that the diffusion upper bound is satisfied. 
 
 The above behavior is expected to be quite general for any negative value of $\alpha$ since we have checked that for real $k$ the behaviors of the hydrodynamic mode and non-hydrodynamic mode are quite similar to the plot shown in Fig. \ref{fig:negalp0}. For any negative $\alpha$, the first non-hydrodynamic mode is expected to be an IR mode and its pole collision with the hydrodynamic mode occurs at complex momentum.
 One might also expect that the diffusion upper bound is always saturated for any negative $\alpha$ similar to the example we have checked in this subsection, then from the left plot in 
 Fig. \ref{fig:c(alpha)}, across $\alpha=0$ one finds similar ``non-smooth" behavior of $D_c/(v_{eq}^2\tau_{eq})$ as the left plot in Fig. \ref{fig:tuneT2}. It would be extremely interesting to study the physical conditions under which that the pole collision occurs at real or complex momentum.

\section{Conclusion and discussion}
\label{sec:4}

We have studied the breakdown of diffusive hydrodynamics in  holographic neutral states described by the generalized  Gubser-Rocha model with linear axion fields. The holographic systems have near horizon geometries conformal to   AdS$_2\,\times\,$R$^2$ in the extremal limit which are known as  special semi-local quantum critical states \cite{Hartnoll:2012wm}. We introduced a general gauge coupling which is characterized by a parameter $\alpha$ via  $g_\text{eff}^2=e^{-\alpha\phi}$ where $\phi$ is the dilaton field. The convergence radius of hydrodynamic expansion is determined by the pole collision between the hydrodynamic and lowest non-hydrodynamic modes. We focused on the low temperature physics where the effective gauge coupling near the horizon 
$g_\text{eff}^2\sim (T/m)^
\alpha$. We found that when $\alpha>1$, the first non-hydrodynamic mode which collides with the charge diffusive mode is a slow mode. When $\alpha<1$, the first non-hydrodynamic mode is the lowest IR pole. This observation indicates that the origin of universality for the breakdown of charge diffusive hydrodynamics close to a quantum critical state crucially depends on the effective gauge coupling strength. In other words, the breakdown of hydrodynamics in a quantum critical phase only exhibits partially universality being inherited from the quantum critical groundstate.  Moreover, we found that at low temperature the pole collision occurs at real momentum for $\alpha\geq 0$ while complex momentum for negative $\alpha$. At high  temperature, we checked two typical examples with different origins of lowest non-hydrodynamic mode at low temperature and found that the pole collision occurs at real momentum for $\alpha=0$ and complex momentum for $\alpha=2$. 

The different origins of the non-hydrodynamic modes at low temperature result in totally different scaling behaviors of the $(\omega_{eq}, k_{eq})$ which characterize the convergence radius of the dispersion relation of hydrodynamics in momentum space, i.e. \eqref{eq:dep1} and \eqref{eq:dep3}. Following the proposals in \cite{Arean:2020eus}, we define the equilibrium  velocity $v_{eq}=\omega_{eq}/k_{eq}$ and the equilibrium time $\tau_{eq}=1/\omega_{eq}$ from the local equilibration scales which are set by pole collision. At low temperature, when the non-hydrodynamic mode is a slow mode, we found that the local equilibrium time is always larger than the Planckian time or Lyapunov time, and the equilibrium velocity is larger than the butterfly velocity, and the diffusion upper bound is automatically saturated as $D_c = \frac{1}{2}v_{eq}^2\tau_{eq}$. When the non-hydrodynamic mode is an IR mode, we found that the  local equilibrium time is of the same order as the Planckian time or Lyapunov time, while the equilibrium velocity is larger than the butterfly velocity, and now the upper bound for the diffusion constant becomes  $D_c\lesssim c(\alpha)v_{eq}^2\tau_{eq}$ with $c(\alpha)$ an monotonically decreasing function starting from $c(0)=1$ to $c(1)=1/2$. 

We also studied the effect of the IR gauge coupling constant at low temperature and the temperature effect for fixed $\alpha$ on the convergence radius of hydrodynamic diffusion dispersion relations. We found that at low temperature, when we decrease the IR gauge coupling constant, the convergence radius $k_{eq}/(2\pi T)$ monotonically decreases. For fixed $\alpha$, the behavior of convergence radius $k_{eq}/(2\pi T)$ with respect to $T/m$ depends on the origin of the first non-hydrodynamic mode at low temperature. When the first non-hydrodynamic mode is a slow mode, $k_{eq}/(2\pi T)$ increases when we increase the temperature, while when the first non-hydrodynamic mode is an IR mode, $k_{eq}/(2\pi T)$ decreases when we increase the temperature. In both cases, $k_{eq}/(2\pi T)$ approaches a constant at high temperature. Moreover, the upper bound for the charge diffusion constant is always satisfied at any temperature. 

For the case that the hydrodynamic pole collides with an IR mode, we have the following interesting observations compared with the results from hydrodynamics near AdS$_2\,\times\,$R$^2$ quantum critical points \cite{Arean:2020eus}. Firstly, in our model the extremal IR geometry is conformal to AdS$_2\,\times\,$R$^2$ and this results in zero entropy at zero temperature while still has semi-local quantum critical behavior. However, except the case of $\alpha=0$ now in general the first non-hydrodynamic mode has nontrivial dependence on $k/T$ for larger $k$ as shown in the right plot of Fig. \ref{fig:IRmode}, this is different from the AdS$_2\,\times\,$R$^2$ case. Secondly, the equilibrium frequency is always proportional to $T$ while the dependence of equiblirium momentum on $T$ depends on $\alpha$. Interestingly when $\alpha=0$ we have the diffusion constant independent of $T/m$ at low temperature and $\omega_{eq}\sim k_{eq}^2\sim T$ which is quite similar to the results in \cite{Arean:2020eus}.
Thirdly, in our model at low temperature the pole collision could occur for both complex momentum if $\alpha<0$ and real momentum if $0\leq\alpha<1$, while the pole collision only occurs for complex momentum in \cite{Arean:2020eus}. This might be related to the fact that we focused on the charge diffusive hydrodynamics. It would be interesting to study the breakdown of other diffusion dispersion relations in the hydrodynamic systems we considered. Finally, the upper bound for the diffusion constant is always satisfied in both these two models. 

The holographic effective field theory for diffusive hydrodynamics has been studied in \cite{Glorioso:2018mmw, deBoer:2018qqm}.  
The convergence of the diffusive hydrodynamics should put a special cutoff scale for the effective field theory which might be able to manifest in the action. Therefore, it would be interesting to derive the effective field theory for our model since it incorporated two different origins of non-hydrodynamic modes with a tuning parameter $\alpha$ which results in different convergence radius. Moreover, for the case of pole collision with a slow mode, there exists a quasi-hydrodynamic picture  \cite{Davison:2014lua, Grozdanov:2018fic} as well as an effective field theory description \cite{Baggioli:2020whu}. It would be interesting to construct an analogous description to include the IR mode into the effective theory.

The holographic charge diffusive hydrodynamics has been understood through a semi-holographic description in terms of the IR degrees of freedom together with a Goldstone boson which arises from the spontaneous breaking of $U(1)\times U(1)$ down to the diagonal $U(1)$ \cite{Nickel:2010pr}. Our study indicates that at low temperature the IR effective gauge coupling  $g_\text{eff}^2\sim (T/m)^\alpha$ seems to control the  
coupling strength between IR gauge fields and the Goldstone boson. When $g_\text{eff}^2> T/m$ the IR gauge fields play an important role and the first non-hydrodynamic mode is an IR mode, while when $g_\text{eff}^2< T/m$ the first non-hydrodynamic mode depends on all the energy scales since \eqref{eq:Dtau} is an integration along the whole radial direction. It would be very interesting to construct explicitly such a semi-holographic description.  

\subsection*{Acknowledgments}
We would like to thank Matteo Baggioli, Hyun-Sik Jeong, Keun-Young Kim, and Ya-Wen Sun  
for useful discussions. 
This work is supported in part by the National Natural Science Foundation of China grant No.11875083.  

\newpage
\appendix
\section{Equations of motion}
\label{appA}
In this appendix, we show the equations of motion for the background and the fluctuations. 

\subsection{Equations of background}

For the action (\ref{eq:actionbg}) discussed in section \ref{sec:2}, we have the following equations of motion
\bea
\begin{split}
 &R_{\mu\nu}-\frac{1}{2}g_{\mu\nu}\left[R-\frac{1}{4}e^{\alpha\phi} F^2-\frac{3}{2}(\partial\phi)^2+6\cosh \phi-\frac{1}{2}\sum_{I=1}^{2}(\partial \psi_I)^2\right]\\
 &~~~~~=\frac{1}{2}e^{\alpha\phi} F_{\mu \rho}F_\nu^\rho+\frac{3}{2}\partial_\mu\phi\partial_\nu\phi+\frac{1}{2}\sum_{I=1}^2(\partial_\mu\psi_I\partial_\nu\psi_I)\,,\\
 &\nabla_\mu(e^{\alpha\phi} F^{\mu\nu})=0\,,\\
 &\nabla^2\phi-\frac{\alpha}{12}e^{\alpha\phi} F^2+2\sinh\phi=0\,,\\
 &\nabla^2\psi_I=0\,.
\end{split}
\eea
For the ansatz \eqref{eq:bg} of the background, we have the equations
\bea
\begin{split}
u''+\left(-\frac{f'^2}{2f^2}+\frac{3}{2}\phi'^2\right)u-\frac{m^2}{f}&=0\,,\\
f''+\frac{3}{2}f\phi'^2-\frac{f'^2}{2f}&=0\,,\\
\frac{uf'^2}{f^2}+\frac{2f'u'}{f}-3u\phi'^2+\frac{2m^2}{f}-6(e^\phi+e^{-\phi})&=0\,,\\
\phi''+(\frac{u'}{u}+\frac{f'}{f})\phi'+\frac{1}{u}(e^\phi-e^{-\phi})&=0\,.
\end{split}
\eea
The first three equations are from the Einstein equation while the last equation is from the equation of motion for dilaton field. There are four equations for three fields, among which only three are independent. One can check that, for example, the first equation can be obtained from linear combination of the other equations and the derivative of the first order equation from Einstein equation. 

Note that in this work, we focus on vanishing chemical potential cases, i.e. $A_t=0$. For a nonzero $A_t$, i.e. finite chemical potential, there is an analytical solution when $\alpha=1$ \cite{Zhou:2015qui, Kim:2017dgz}, 
\bea
\begin{split}
u(r)&=r^2g(r)h(r),~~~f(r)=r^2g(r)\,,\\
h(r)&=1-\frac{m^2}{2(Q+r)^2}-\frac{(Q+r_0)^3}{(Q+r)^3}\left(1-\frac{m^2}{2(Q+r_0)^2}\right),~~~g(r)=\left(1+\frac{Q}{r}\right)^{\frac{3}{2}}\,,\\
A_t(r)&=\sqrt{3Q(Q+r_0)\left(1-\frac{m^2}{2(Q+r_0)^2}\right)}\left(1-\frac{Q+r_0}{Q+r}\right)\,,\\
\phi(r)&=\frac{1}{3}\log\left(g(r)\right)\,,
\end{split}
\label{eq:chargebg}
\eea
while for general $\alpha$ one needs to solve the system numerically.

The nontrivial dilatonic neutral black hole solution \eqref{eq:neubg} is the case $Q=-r_0+\frac{m}{\sqrt{2}}$ of \eqref{eq:chargebg}. There exists another neutral black hole solution, i.e. the well-known AdS$_4$ Schwartzschild solution with axion charge, which has trivial dilaton and can be viewed as the case $Q=0$ of \eqref{eq:chargebg}. It takes the following form, 
\be
\label{eq:schbg}
u=r^2\left(1-\frac{r_0^3}{r^3}\right)-\frac{m^2}{2}\left(1-\frac{r_0}{r}\right),~~~f=r^2,~~~\phi=0\,.
\ee

For the dual theory of the background \eqref{eq:schbg}, the electrical conductivity, susceptibility and diffusion constant are as follows \cite{Kim:2017dgz}, 
\bea
\sigma=1\,,~~~\chi=\frac{1}{6}\left(4\pi T+\sqrt{16\pi^2T ^2+6m^2}\right)\,,~~~D_c=\frac{6}{4\pi T+\sqrt{16\pi^2T ^2+6m^2}}\,.
\eea
Note that different from the quantities for the background of \eqref{eq:neubg}, here they do not depend on the parameter $\alpha$ due to the fact that in \eqref{eq:schbg} we have trivial dilaton field.  
The butterfly velocity and Lyapunov time are 
\bea
v_B^2=\frac{6\pi T}{4\pi T+\sqrt{16\pi^2T^2+6m^2}}\,,~~~~\tau_L=\frac{1}{2\pi T}\,.
\eea
Therefore, we have
\bea
\frac{D_c}{v_B^2 \tau_L}=2\,.
\eea
This is quite different from the dual theory of the dilatonic black hole introduced in Sec. \ref{sec:2}.

\subsection{Equations of fluctuations}
\label{app:a2}

At zero density, the fluctuations of gauge field and metric field decouple from each other. We focus on the charge diffusive hydrodynamics and  will consider the fluctuations of gauge field in momentum space. Without loss of generality, we make the Fourier transformation  
\bea
\begin{split}
\delta A_{\mu}&=a_{\mu}(r)e^{-i\omega t+ik x}\,,\\
\end{split}
\eea
and obtain the dynamical equations for the fluctuations. 
The fluctuations $\{a_t, a_x, a_r\}$ and $a_y$ are decoupled due to they have different parity when $y\to -y$. The first three fluctuations are parity even and contribute to the diffusive process in the hydrodynamic limit, while the last fluctuation is parity odd and do not consist hydrodynamic mode.

\subsubsection{Diffusive channel}

The equations of motion for the fluctuations $a_t, a_x, a_r$ are
\bea
\label{eq:eoma}
\begin{split}
a_t''+
\left(\frac{f'}{f}+\alpha\phi'\right)a_t'+i\omega a_r'-
\frac{k^2}{uf}a_t-\frac{k\omega}{uf}a_x+i\omega\left(\frac{f'}{f}+\alpha\phi'\right)a_r
&=0
\,,\\
a_x''+
\left(\frac{u'}{u}+\alpha\phi'\right)a_x'-i k a_r'+
\frac{\omega^2}{u^2}a_x+\frac{k\omega}{u^2}a_t-
i k \left(\frac{u'}{u}+\alpha\phi'\right)a_r&=0\,,\\
\frac{i\omega}{u}a_t'
+\frac{ik}{f}a_x'
+\left(\frac{k^2}{f}-\frac{\omega^2}{u}\right)a_r&=0\,.
\end{split}
\eea
Note that these equations are invariant under $U(1)$ gauge transformation 
\be
\delta A_\mu \rightarrow \delta A_\mu-\partial_\mu\Lambda\,,~~~
\Lambda=e^{-i\omega t+ikx}\lambda(r)\,.
\ee
One can always choose the radial gauge $a_r=0$ to do the calculation.\footnote{The radial gauge will be used in appendix \ref{app:weakhydro}. There is a residual gauge transformation which is useful in the calculation of the retarded Green's function \cite{Amado:2009ts}. 
}  In this gauge, the above equations reduce to 
\bea
\label{eq:radialgauge}
\begin{split}
\partial_r\left(f e^{\alpha\phi} a_t'\right)-\frac{ e^{\alpha\phi}}{u} \,k\left(\omega  a_x+k a_t\right)&=0\,,\\
\partial_r\left(u e^{\alpha\phi}  a_x'\right)+\frac{ e^{\alpha\phi}}{u}\omega \left(\omega  a_x+k a_t\right)&=0\,,\\
\frac{\omega }{u}a_t' +\frac{k }{f}a_x'&=0\,.
\end{split}
\eea

Another way to solve the equations \eqref{eq:eoma} is to use the $U(1)$ gauge invariant variables \cite{Kovtun:2005ev} which are defined as
\bea
\begin{split}
\mathfrak{a}=a_t+\frac{\omega}{k}a_x\,,~~~
\mathfrak{b}= a_r+\frac{i a_x'}{k}\,,
\end{split}
\eea
we have the decoupled equations for these variables
\bea
\begin{split}
\mathfrak{a}''+
\left(\frac{\omega^2f'}{-\omega^2f+k^2u}-\frac{\omega^2fu'}{u(-\omega^2f+k^2u)}+\frac{f'}{f}+\alpha\phi'\right)\mathfrak{a}'+
\left(\frac{\omega^2}{u^2}-\frac{k^2}{uf}\right)\mathfrak{a}&=0\,,\\
\mathfrak{b}''+
\left(\frac{3u'}{u}+\alpha\phi'\right)\mathfrak{b}'
+\left(\frac{u''}{u}+\alpha \phi''+
\frac{u'^{2}}{u^2}+\frac{2\alpha u'\phi'}{u}+\frac{\omega^2}{u^2}-\frac{k^2}{uf}
\right)\mathfrak{b}&=0\,.
\end{split}
\label{eq:GI}
\eea

The above two different methods to calculate the Green's function and quasi-normal modes are equivalent \cite{Amado:2009ts}. 
In this paper, we will use both of them.  
To study the pole collisions in the charge diffusive process, we solve the quasi-normal modes of the first equation in \eqref{eq:GI}. To calculate the telegrapher's equation and related parameters, we will work in the radial gauge and use equations \eqref{eq:radialgauge}.

\subsubsection{Transverse channel}
The equation of motion for $a_y$ is
\bea
a_y''+\left(\frac{u'}{u}+\alpha\phi'\right)a_y'+\left(\frac{\omega^2}{u^2}-\frac{k^2}{uf}\right)a_y=0\,.
\eea
The equation of $a_y$, together with $h_{ty}, h_{xy}$ in general finite density case, is related to the parity-odd channel. 
We do not consider this sector in this work. 

\section{$\tau$ from the matching method}
\label{app:weakhydro}

As shown in section \ref{sec:3}, when $\alpha>1$ the first non-hydrodynamic mode has the feature of $\omega\ll T$ as $T\rightarrow 0$. The hydrodynamic mode and the first non-hydrodynamic mode are well fitted by the telegrapher equation. In this appendix, we shall semi-analytically solve the equations \eqref{eq:eoma} to show the first two quasi-normal modes satisfying the telegrapher equations \cite{Davison:2018ofp, Davison:2018nxm, Chen:2017dsy, Grozdanov:2018fic}.

We work in the radial gauge, i.e $a_r=0$. Our strategy to solve the equations \eqref{eq:radialgauge} is 
as follows. We first divide the radial direction outside the horizon into inner regime $r-r_0\ll T$ and outer regime $r-r_0\gg \omega$, then we solve \eqref{eq:eoma} separately in these regimes and match the solutions in the matching regime $\omega \ll r-r_0\ll T$ to get the solution in the full spacetime. 
\begin{itemize}
    \item In the outer regime $r-r_0\gg \omega, k$, we can solve the system order by order in $\omega$ and $k$. The leading order solution satisfies equations 
\bea
\begin{split}
\partial_r\left(f e^{\alpha\phi} a_t'\right)=0\,,~~~~~~
\partial_r\left(u e^{\alpha\phi}  a_x'\right)=0\,.
\end{split}
\eea
Therefore, the solutions take the following form\footnote{Variables with a tilde are defined on boundary 
in momentum space. The $x^\mu$ dependence of the fields can be realized by Fourier transformation and replacing $-i\omega\to \partial_t,~ ik\to \partial_x$.}
\bea
\label{eq:outsol}
\begin{split}
 a_t(r)&=\tilde{a}_t+\tilde{j}_t R_1(r)+\mathcal{O}(\omega, k)\,,\\
 a_x(r)&=\tilde{a}_x+\tilde{j}_x R_2(r)+\mathcal{O}(\omega, k)\,,
\end{split}
\eea
with
\bea
\begin{split}
R_1(r)=\int_{r}^{\infty}\frac{ds}{f e^{\alpha\phi}}\,,
~~~~~~R_2(r)=\int_{r}^{\infty}\frac{ds}{u e^{\alpha\phi}}\,.
\end{split}
\eea
Note that $\tilde{j}_t$ and $\tilde{j}_x$ in \eqref{eq:outsol} are the dual charge density and current  
from the holographic dictionary. Plugging \eqref{eq:outsol} into the constraint equation in \eqref{eq:radialgauge}, we obtain the dual conservation equation (via replacing $-i\omega\to \partial_t,~ ik\to \partial_x$)
\be\label{eq:Ward}
\partial_t \tilde{j}^t+\partial_x \tilde{j}^x=0\,,
\ee
where we have used $\tilde{j}^t=-\tilde{j}_t, \tilde{j}^x=\tilde{j}_x$ since we work in the most plus convention for the dual field theory. 
In the hydrodynamic limit, the outer regime can be extended to $r\to r_0$,  where $R_1(r)$ is regular, while there is a logarithmic divergence in $R_2(r)$ as $u(r)\rightarrow 4\pi T(r-r_0)$. One can rewrite $R_2(r)$ in terms of the sum between the regular part $r_2(r)$ and the logarithmic divergent part
\bea
R_2(r)=r_2(r)-\frac{1}{4\pi T e^{\alpha\phi_0}}\text{log}\left(r-r_0\right)\,,
\eea
where
\bea
r_2(r)=\int_{r}^\infty ds\left(\frac{1}{u e^{\alpha\phi}}-\frac{1}{4\pi T e^{\alpha\phi_0}}\,\frac{1}{s-r_0}\right)\,,~~~
\phi_0=\frac{1}{2}\log\frac{m}{\sqrt{2}r_0}
\,.
\eea

\item In the inner regime $r-r_0\ll T$, the solutions for \eqref{eq:radialgauge} can be written as 
\bea
\begin{split}
\label{eq:aIR}
a_t&=a_t^{(1)}(r)+a_t^{(2)}(r)\left(r-r_0\right)^{-\frac{i\omega}{4\pi T}}\,,\\
a_x&=a_x^{(1)}(r)+a_x^{(2)}(r)\left(r-r_0\right)^{-\frac{i\omega}{4\pi T}}\,,
\end{split}
\eea
where $a_t^{(1,2)}, a_x^{(1,2)}$ are regular functions when $r\to r_0$, with frequency and momentum dependence. The second terms are from the standard infalling boundary condition while the first terms are from the gauge transformation of the fields which are of pure gauge \cite{Amado:2009ts}. 
The functions here are constrained by regular conditions as (via replacing $-i\omega\to \partial_t,~ ik\to \partial_x$)
\bea
\label{eq:relation}
a_t^{(2)}(r_0)=0\,,~~~\partial_ta_x^{(1)}(r_0)-\partial_x a_t^{(1)}(r_0)=0\,.
\eea
The second equation above indicates that $a_t^{(1)}, a_x^{(1)}$ are pure gauge.

\item In the overlap regime, i.e. $\omega, k\ll r-r_0\ll T$, we match the solutions \eqref{eq:outsol} and \eqref{eq:aIR}. 
We split the inner solution \eqref{eq:aIR} into regular part and logarithmic divergence
\bea
\label{eq:insol}
\begin{split}
a_t=a_t^{(1)}\,,~~~
a_x=a_x^{(1)}+a_x^{(2)}+\frac{\partial_t a_x^{(2)}}{4\pi T}\text{log}\left(r-r_0\right)\,.
\end{split}
\eea
Comparing the above solution in the overlap regime with the outer solution \eqref{eq:outsol}, we have
\be\label{eq:nfeq}
a_t^{(1)}=\tilde{a}_t+\tilde{j}_t R_1(r_0)\,,~~~a_x^{(1)}+a_x^{(2)}=\tilde{a}_x+\tilde{j}_x r_2(r_0)\,,~~~\partial_t a_x^{(2)}=-e^{-\alpha\phi_0}\tilde{j}_x\,.
\ee
\end{itemize}

From \eqref{eq:relation} and \eqref{eq:nfeq}, we have the relation 
\be
\label{eq:main}
r_2(r_0)\partial_t \tilde{j}_x-R_1(r_0)\partial_x \tilde{j}_t=-(d\tilde{a})_{tx}-e^{-\alpha\phi_0}\tilde{j}_x
\ee
where $-(d\tilde{a})_{tx}$ is the external electric field along $x$-direction in the dual field theory that can be switched off. Therefore we have the full set of equations for quasi-hydrodynamics \eqref{eq:Ward} and \eqref{eq:main}. To get the above form, it is crucial to work in the limit $\omega\ll T$. 
Note that in \eqref{eq:main}, comparing to the standard constitutive equation (i.e. Fick's law of diffusion), we have an additional $\partial_t j_x$ term which is contributed from the non-hydrodynamic mode. 
From \eqref{eq:Ward} and \eqref{eq:main}, we obtain the dispersion relation of the operator $\tilde{j}_t$ or $\tilde{j}_x$ as
\bea
\omega^2+\frac{i}{r_2(r_0)e^{\alpha\phi_0}}\omega-\frac{R_1(r_0)}{r_2(r_0)}k^2=0\,.
\eea

It takes the same form as the telegrapher equation
\bea\label{eq:teleq}
\omega^2+\frac{i}{\tau}\omega-\frac{D_c}{\tau}k^2=0\,,
\eea
where 
\bea
\label{eq:Dtau}
\tau=\int_{r_0}^\infty dr\left(\frac{e^{\alpha\phi_0}}{u e^{\alpha\phi}}-\frac{1}{4\pi T}\,\frac{1}{r-r_0}\right)\,,~~~~~~
D_c=\int_{r_0}^{\infty}dr\,\frac{e^{\alpha\phi_0}}{f e^{\alpha\phi}}\,.
\eea

Note that the above two integration can be calculated analytically, and we have checked that $D_c$ is the same as that in table  \ref{table:diffusion}.
The pole collision location can be obtained from the telegrapher equation \eqref{eq:teleq} and we have \be
\omega_{eq}=\frac{1}{2\tau}\,,~~~~ k_{eq}=\sqrt{\frac{1}{4D_c\tau}}\,.\ee  
Therefore we have the equilibrium time $\tau_{eq}=2\tau$.

Furthermore, from the equation \eqref{eq:Dtau}, 
we can 
analyze the scaling behaviors of $\tau$. 
It is useful to start from the following equation 
\bea
e^{\alpha\phi_0}r_2(r)=I_2(r)+C(\alpha, r_0,m)\,,
\eea
where $C(\alpha, r_0,m)$ is the integration constant to satisfy $r_2(\infty)=0$. 
When $\alpha>1$, 
$C(\alpha, r_0,m)\propto m^{-1}(T/m)^{-\alpha}$ and $I_2(r_0)\propto T^{-1}$ at low temperature $T/m \rightarrow 0$. Therefore in this case we have 
\be 
\tau m\propto \left(\frac{T}{m}\right)^{-\alpha}\,,~~~\alpha>1,
\ee
i.e. $\tau\propto T^{-\alpha}$ when $m$ fixed. While for $\alpha\leq 1$, $\tau \propto T^{-1}$ for fixed $m$.

Note that in the above derivation, we have assumed that $\omega\ll T$ (i.e. $\alpha>1$). However, the final results \eqref{eq:Dtau} on $D_c$ applies for any $\alpha$, and the results on $\tau$ applies for the case $\omega <T$. The numerical integration of $\tau$ as a function of $\alpha$ could be found in the dashed blue line in Fig. \ref{fig:IRmode} and it matches the exact value of the quasi-normal modes quite well in the regime $\omega <T$.

\section{Pole collision in complex momentum space}
\label{app:pccomplex}
We have shown that for cases $\alpha\geq 0$ at low temperature the pole collision between hydrodynamic mode and the first non-hydrodynamic mode occurs at real momentum, while  
for negative $\alpha$ the pole collision occurs at complex momentum. In this appendix we show more details on pole collisions in subsections \ref{sec:case1} and \ref{sec:case2} when we promote the momentum to be complex, i.e., $k_x=|k|e^{i\varphi}, \varphi\in [0, 2\pi)$ and study the behavior of complex quasi-normal modes when we change the phase while fixing the module of $k_x$ close to the collision momentum $k_{eq}$. 

\begin{figure}[H]
  \centering
\subfigure{
\begin{minipage}{0.35\linewidth}
\includegraphics[width=1\linewidth]{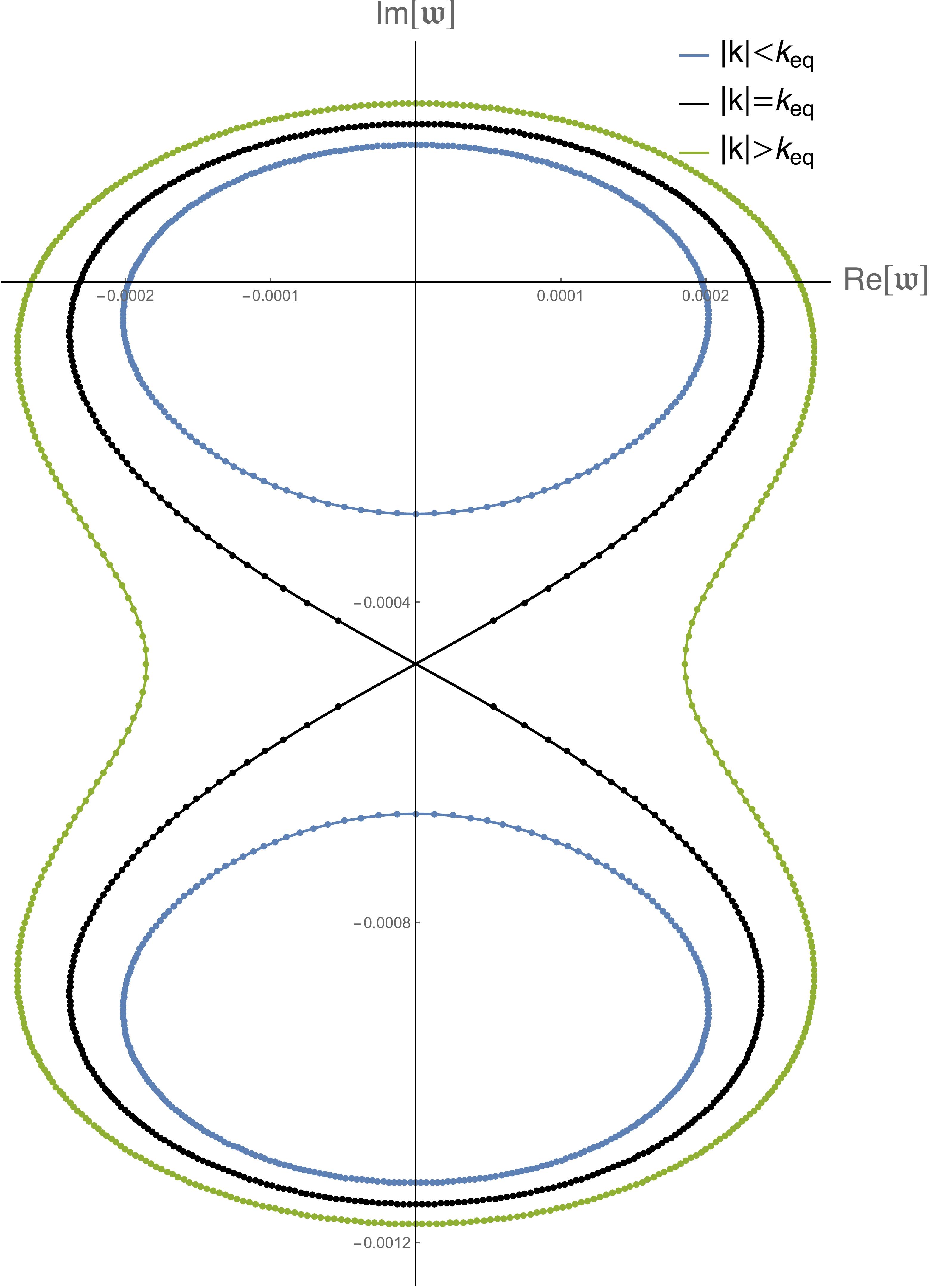}
\end{minipage}
}
\hspace{2mm}
\subfigure{
\begin{minipage}{0.28\linewidth}
\includegraphics[width=1\linewidth]{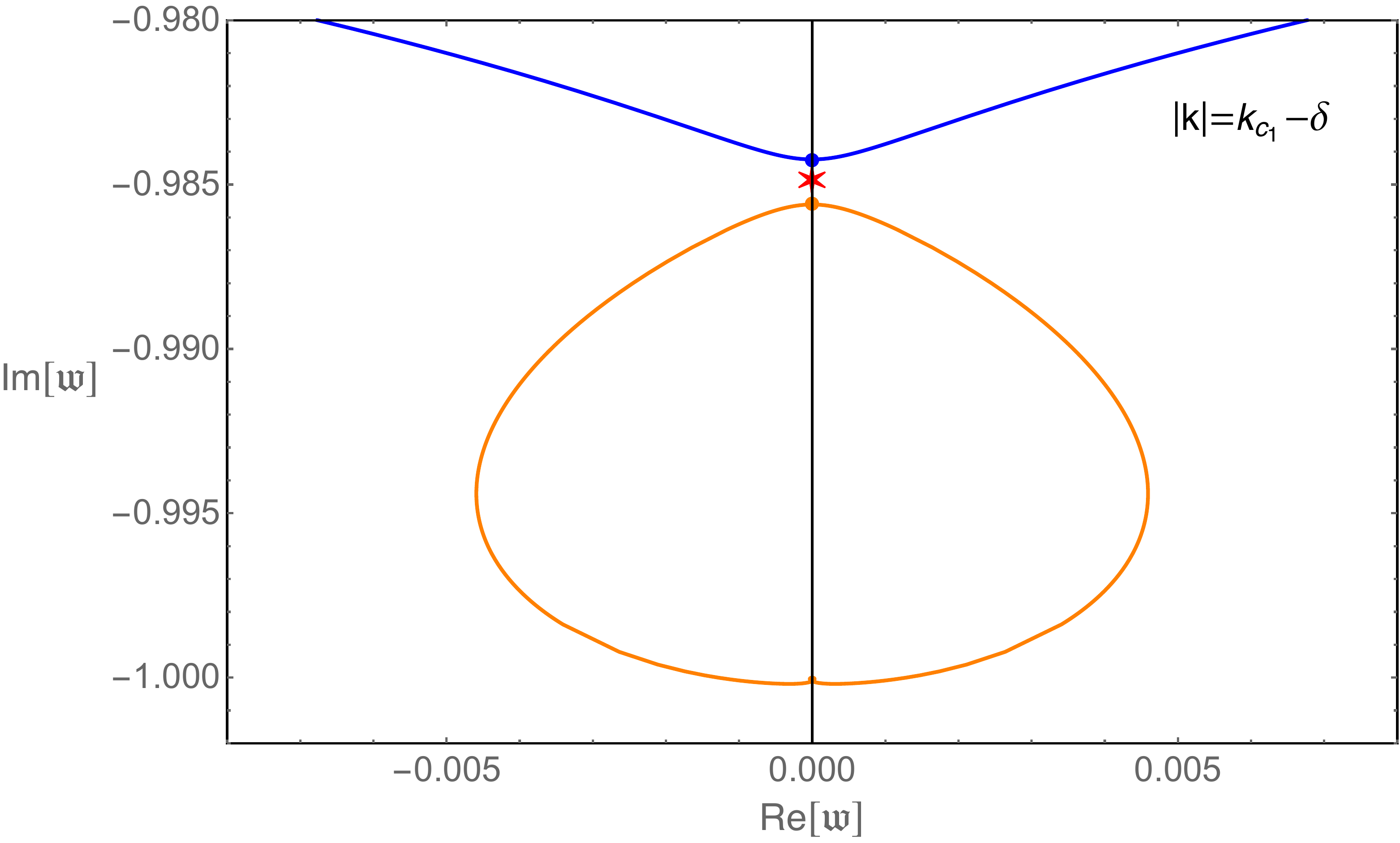}
\vspace{2mm}
\\
\includegraphics[width=1\linewidth]{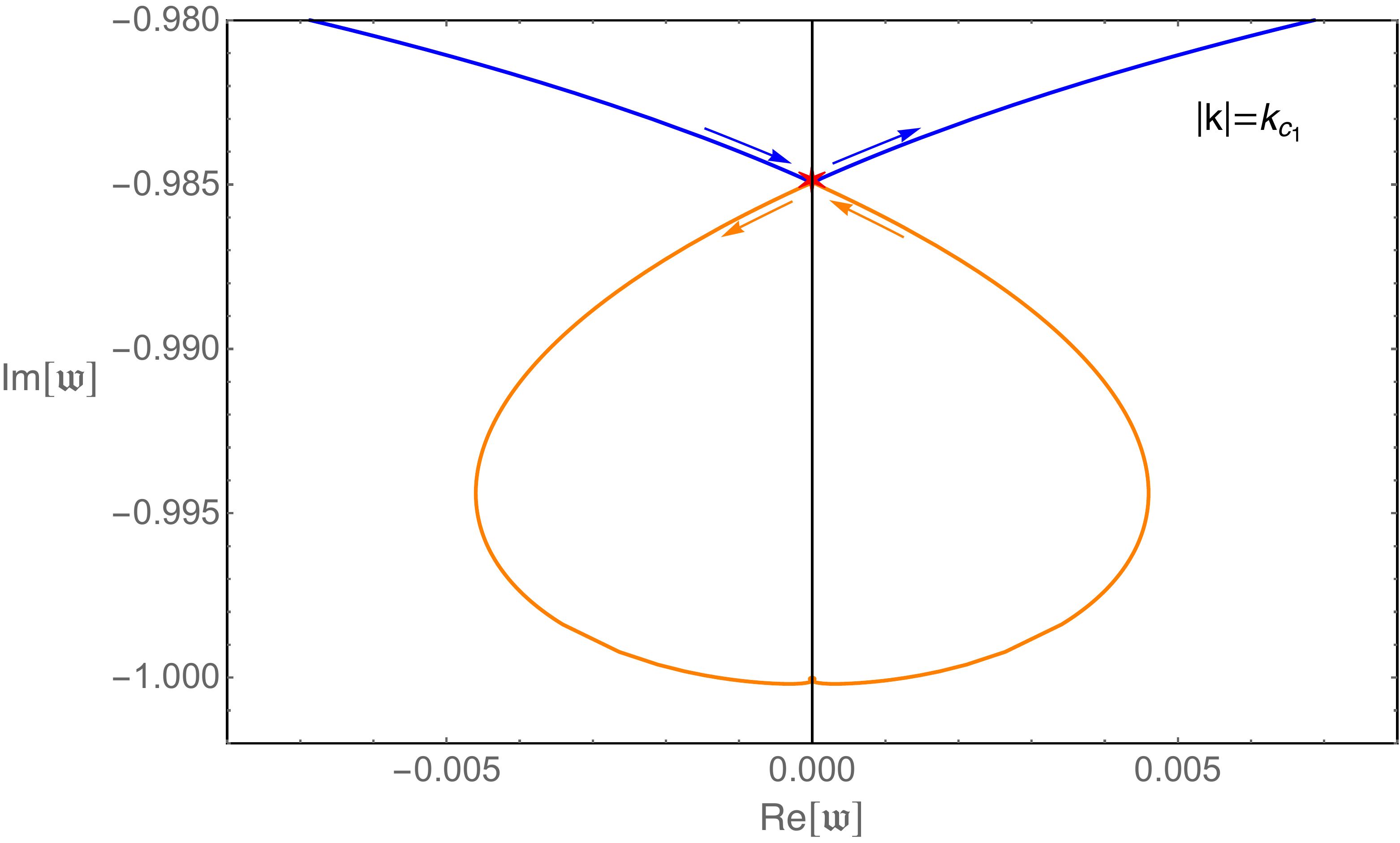}
\\
\vspace{2mm}
\includegraphics[width=1\linewidth]{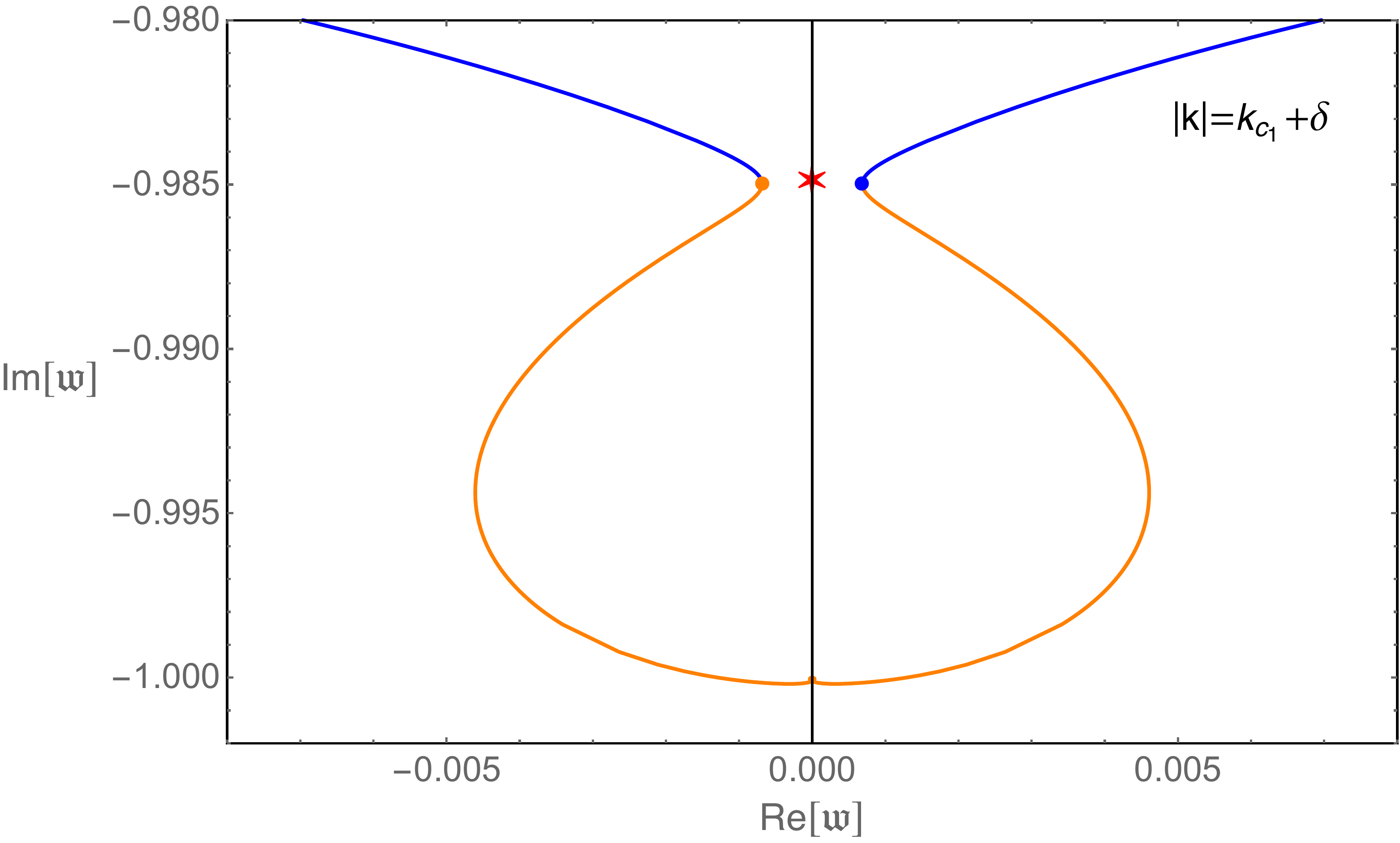}
\end{minipage}
}
\subfigure{\begin{minipage}{0.28\linewidth}
\includegraphics[width=1\linewidth]{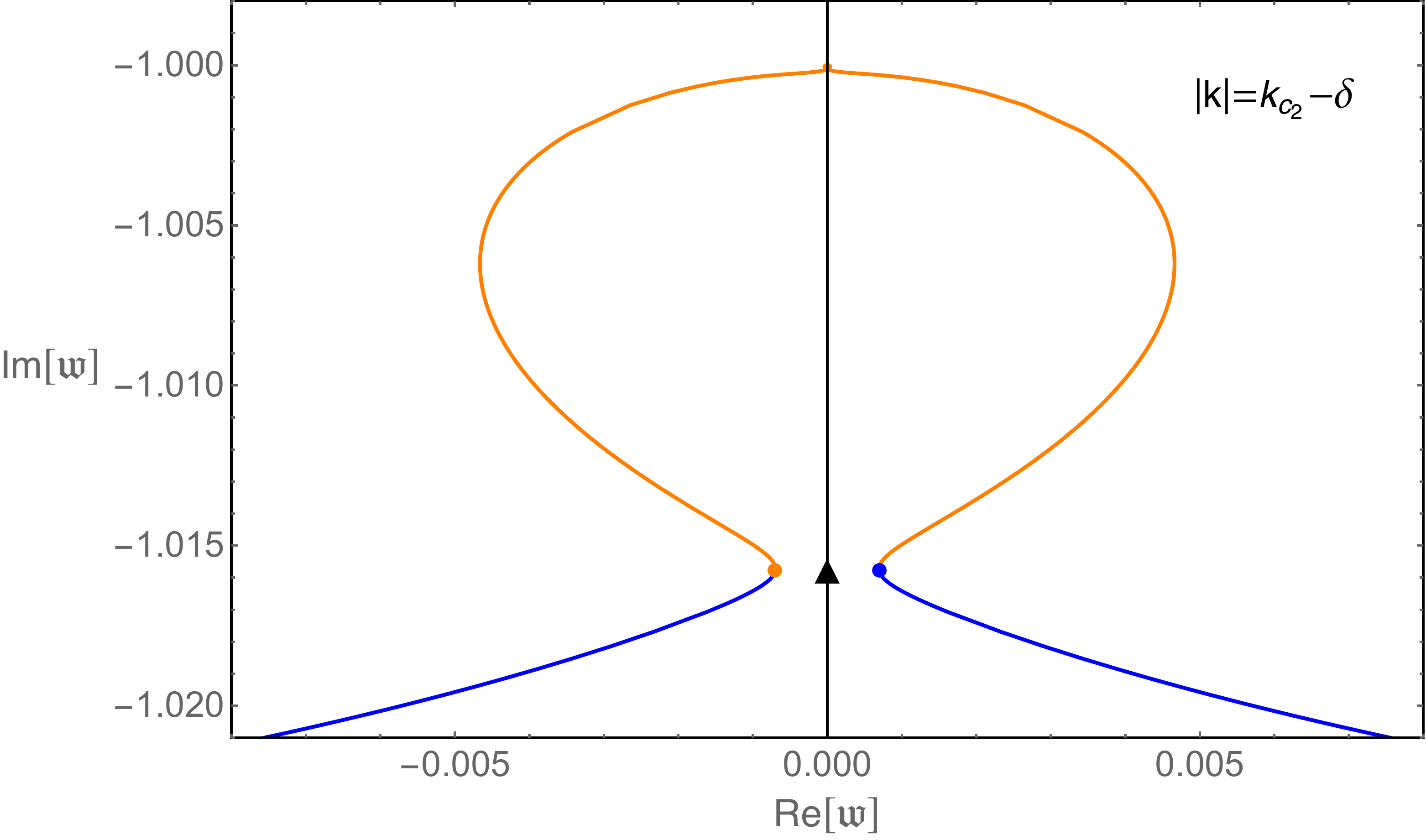}
\vspace{2mm}
\\
\includegraphics[width=1\linewidth]{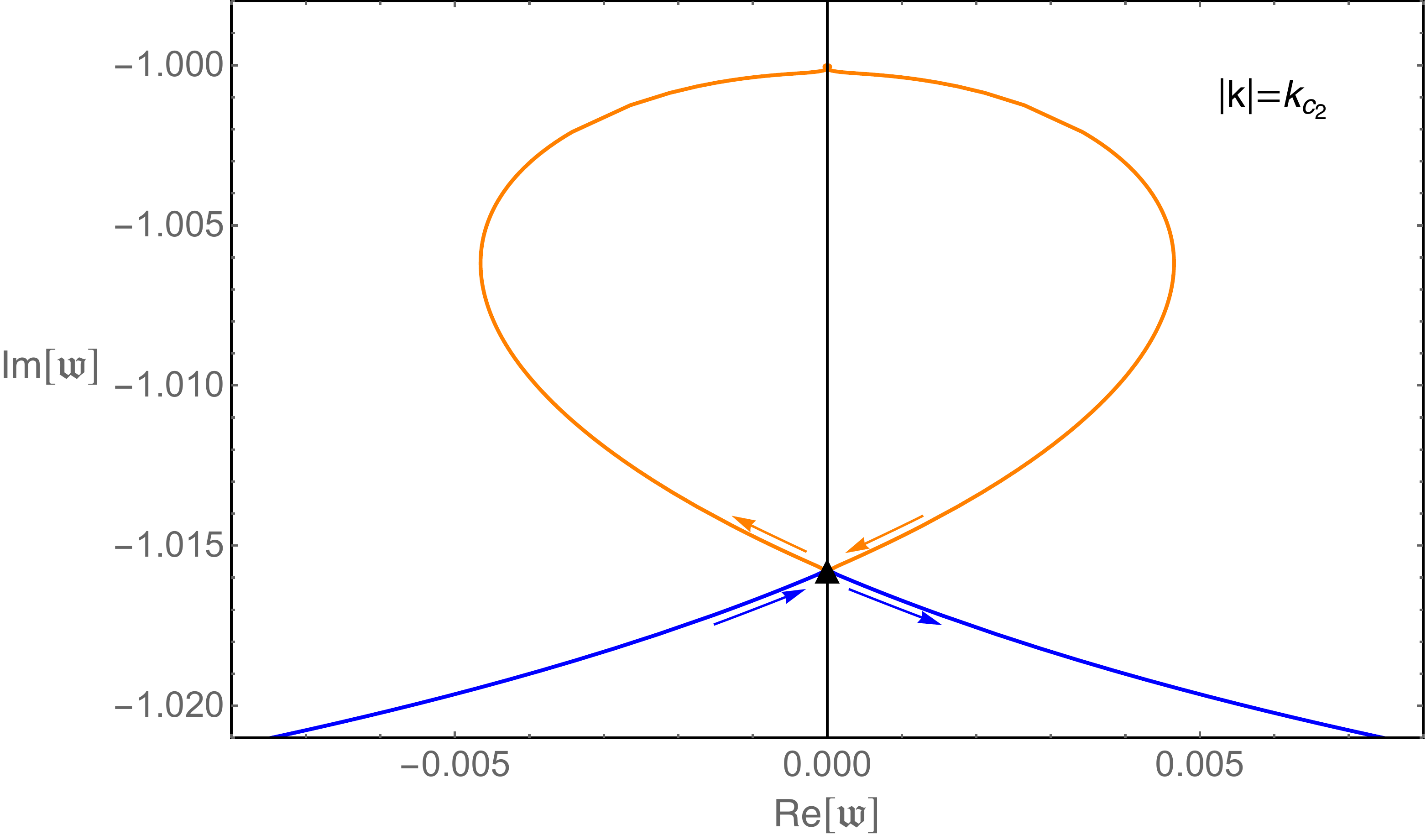}
\vspace{2mm}
\\
\includegraphics[width=1\linewidth]{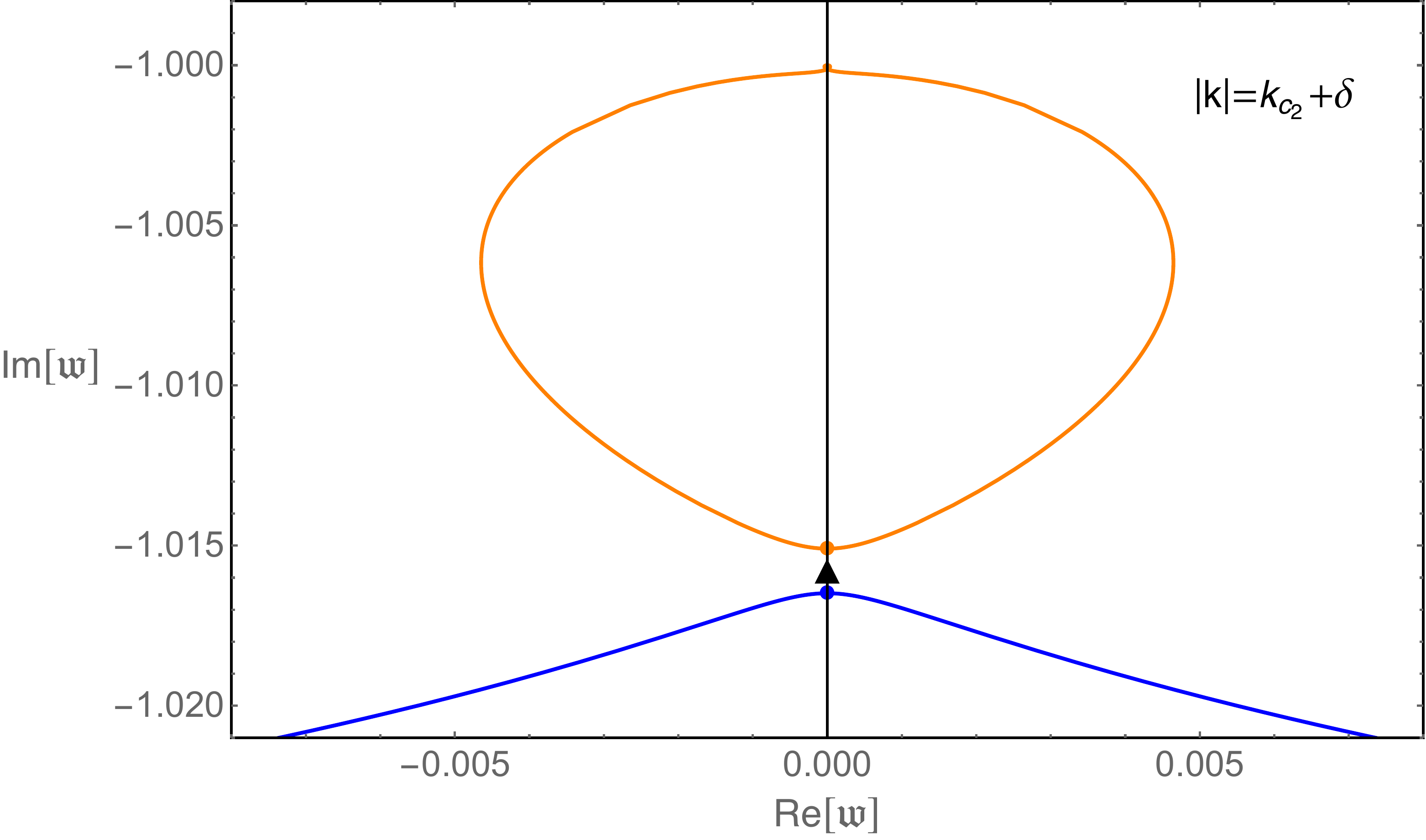}
\end{minipage}}
  \caption{\small Frequencies of the hydrodynamic and the first non-hydrodynamic modes by tuning the phase of $k$ with fixing   $|k|$ close to $k_{eq}$ for $\alpha=2$ ({\em left}) and $\alpha=0$ ({\em right}) at low temperature. 
  }
  \label{fig:complexq}
\end{figure}

The pole collision for $\alpha=2$ with complex momentum close the collision point is shown in the left pole in Fig. \ref{fig:complexq}, while for $\alpha=0$ it is shown in the right pole in Fig. \ref{fig:complexq}. The underlying non-hydrodynamic modes are different in these two cases, however, they show quite similar behavior in the complex momentum space. Before the collision the hydrodynamic mode and non-hydrodynamic mode are of topological $S^1$ separately for a fixed module of the complex momentum close to the collision point. The QNMs start from the locations of the values at real $k$ and move anticlockwise when we increase the phase from $0$ to $\pi$. When the poles collide, they connect. After the collision they become a single closed curve. There is a topological change between $S^1\times S^1 $ and $S^1$ during the pole collision. These behaviors are quite similar to the studies in \cite{Grozdanov:2019kge, Grozdanov:2019uhi} where the pole collisions occur at complex momentum. 

\newpage
 
\end{document}